# The 2-Token Theorem:
# Recognising History-Deterministic Parity Automata Efficiently

Karoliina Lehtinen
Aix-Marseille Université, Marseille, France
`karoliina.lehtinen@lis-lab.fr`

Aditya Prakash
University of Warwick, Coventry, UK
`aditya.prakash@warwick.ac.uk`

## Abstract

History-determinism is a restricted notion of nondeterminism in automata, where the nondeterminism can be successfully resolved based solely on the prefix read so far. History-deterministic automata still allow for exponential succinctness in automata over infinite words compared to deterministic automata (Kuperberg and Skrzypczak, 2015), allow for canonical forms unlike deterministic automata (Abu Radi and Kupferman, 2019 and 2020; Ehlers and Schewe, 2022), and retain some of the algorithmic properties of deterministic automata, for example for reactive synthesis (Henzinger and Piterman, 2006; Ehlers and Khalimov, 2024).

Despite the topic of history-determinism having received a lot of attention over the last decade, the complexity of deciding whether a parity automaton is history-deterministic has, up till now, remained open. We show that history-determinism for a parity automaton with a fixed parity index can be checked in PTIME, thus improving upon the naive EXPTIME upper bound of Henzinger and Piterman that has stood since 2006. More precisely, we show that the so-called 2-token game, which can be solved in PTIME for parity automata with a fixed parity index, characterises history-determinism for parity automata. This game was introduced by Bagnol and Kuperberg in 2018, who showed that to decide if a Büchi automaton is history-deterministic, it suffices to find the winner of the 2-token game on it. They conjectured that this 2-token game based characterisation extends to parity automata. Boker, Kuperberg, Lehtinen, and Skrzypcak showed in 2020 that this conjecture holds for coBüchi automata as well. We prove Bagnol and Kuperberg's conjecture that the winner of the 2-token game characterises history-determinism on parity automata.

As consequences of our result, we also get the 2-token game based characterisation for history-determinism in alternating parity automata and $\omega$-regular automata with any acceptance condition, thus yielding efficient algorithms for deciding their history-determinism. Additionally, we also obtain an efficient algorithm for deciding the good-enough-synthesis (Almagor and Kupferman, 2020) for specifications given by a deterministic parity automata.

# Contents

# 1 Introduction

Automata over infinite words can describe systems that run indefinitely, by describing properties such as liveness (all processes eventually make progress) and safety (free from errors), for example. The class of $\omega$-regular automata, characterised by a finite state space, is a particularly important formalism for verification problems such as model-checking of modal and temporal logics, e.g., the modal $\mu$-calculus [SE89] and linear temporal logic [Pnu77, PR89]. They are also a key formalism in reactive synthesis, also known as Church synthesis [Chu57, CHVB18], where the goal is to generate the model of a system guaranteed to satisfy a specification in all possible environments.

Within the class of $\omega$-regular automata, we are particularly interested in Büchi, coBüchi and parity automata. Büchi automata, in which a run is accepting if it visits an accepting transition[1] infinitely often, stand out since nondeterministic Büchi automata are as expressive $\omega$-regular automata with parity, Muller, Rabin, Streett, or Emerson-Lei acceptance conditions, as well as monadic second order logic over words (S1S) [Rab69] and $\omega$-regular expressions. CoBüchi automata have the dual acceptance condition: a run is accepting if it visits rejecting transitions only finitely many times. Nondeterministic coBüchi automata are weaker than nondeterministic Büchi automata. Parity automata generalise both Büchi and coBüchi automata. These are automata in which every transition is labelled by a natural number, which is called the priority of that transition. A run in a parity automaton is accepting if the *least priority that occurs infinitely often in that run is even. Deterministic* parity automata are as expressive as nondeterministic Büchi automata, while deterministic Büchi automata are much weaker.

This already suggest the important role of nondeterminism in automata. Nondeterminism can express languages of infinite words using simpler acceptance conditions, and are also exponentially more succinct than deterministic automata (as is the case over finite words); but this comes at a cost, since deterministic automata have better algorithmic properties. In particular, language inclusion, which is the core question in model-checking, is **NL**-complete for deterministic parity automata [AF20, Theorem 5] but **PSPACE**-complete for nondeterministic parity automata.

History-deterministic automata are an intermediate model between nondeterministic and deterministic automata that allows some nondeterminism, as long as it can be resolved on-the-fly, without knowledge of the future of the word. History-deterministic parity automata are exponentially more succinct than deterministic parity automata [KS15] and, at the same time, enjoy good algorithmic properties similar to deterministic parity automata: the problem of language inclusion reduces to deciding simulation (polynomial time if the number of priorities is fixed [Pra24]), while Church's reactive synthesis problem for specifications given by history-deterministic automata can be reduced to solving parity games [HP06]—which is feasible in quasi-polynomial time [CJK+22]. History-deterministic automata are also used in constructing canonical representations of $\omega$-regular languages [AK22, ES22]. Furthermore, the problem of deciding whether an automaton is history-deterministic, which we tackle in this paper, is **LOGSPACE**-interreducible with the good-enough synthesis problem ([AK20] and [BL23b, Page 23]), a variant of the Church synthesis problem.

Thanks to its appeal for synthesis and model-checking, history-determinism has received significant attention over the last decade. In the $\omega$-regular setting, questions of minimisation [RK19, AK22, Cas22, CIK+25], canonicity [AK22, ES22], and memory [CCL22] have been studied, while beyond the regular setting, it has been used to solve synthesis problems for more complex automata models, such as pushdown automata [LZ20, LZ22, GJLZ22], vector addition systems with states [PT23, BPT23], timed automata [BHL+24] and quantitative automata [BL21, BL23a]. For a more comprehensive survey, we refer the reader to [BL23b].

---

[1] While acceptance has historically mostly been defined on states, we use transition-based definitions, as they are mathematically more elegant and versatile, see for example, [AK22] and [Cas23, Chapter VI].

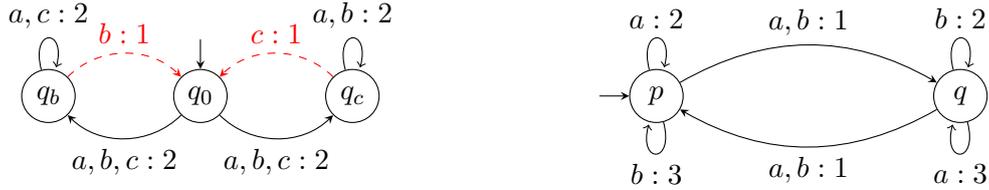

Figure 1: On the left: an example of a history-deterministic coBüchi automaton. On the right: an example of a parity automaton that is not history-deterministic.

Despite this attention, the complexity of deciding history-determinism for parity automata, has, up till now, remained stubbornly open. Let us first explain why this is a tricky question. History-determinism of an automaton is characterised by the winner of the HD game on that automaton. A play of this game starts with a token in the initial state, and proceeds in infinitely many rounds. In each round, Adam plays a letter from the input alphabet and Eve responds by moving the token along a transition of the automaton over that letter. In the limit of this play, Adam constructs an infinite word and Eve constructs a run on that word. Eve wins the play is either Adam's word is *not* in the language of the automaton, or her run is an accepting run.

An automaton is history-deterministic if Eve has a winning strategy in the HD game. In other words, there must be a unified strategy to build accepting runs for all words of the language, transition by transition, without knowledge of the future of the word.

A naive algorithm to decide the history-determinism of parity automata solves the HD game directly and takes exponential time, since it involves determinising the automaton [HP06, Section 4]. In contrast, the best known lower bound is only of **NP**-hardness [Pra24]. There are polynomial time algorithms for deciding history-determinism for certain subclasses of parity automata, however, which has led to the conjecture that this problem is in **P** when the number of priorities in the parity automata are fixed.

In 2015, Kuperberg and Skrzypczak [KS15] gave a polynomial-time algorithm for coBüchi automata, which first constructs a language-equivalent history-deterministic automaton, provided Eve wins the so-called Joker game (which Eve must win if the automaton is history-deterministic). Then deciding the history-determinism of the original automaton reduces to checking whether the original automaton simulates this constructed HD automaton.

---

**Example 1.** *Consider the coBüchi automaton on the left in Fig. 1, which we argue is history-deterministic. This automaton has nondeterminism on letters $a, b, c$ at the initial state $q_0$, and recognises the language of words that either contain finitely many $b$'s or finitely many $c$'s. Eve's winning strategy in the HD game on this automaton is to alternate between moving her token to the state $q_b$ and $q_c$ whenever her token is at the initial state $q_0$. If Adam's word is in the language, then her token only takes the dashed priority-1 transitions finitely many times, and hence her run is accepting.*

*Now consider the parity automaton on the right in Fig. 1, which has its priorities in $[1, 3]$. We will argue that this automaton is not history-deterministic. The language accepted by this automaton is the set of words that either see infinitely many $a$'s (and an accepting run eventually stays in $p$), or infinitely many $b$'s (and an accepting run eventually stays in $q$), and hence the automaton accepts all words in $\{a, b\}^\omega$. Adam has the following strategy to win the HD game: If Eve's token is at state $p$ (respectively $q$), then Adam picks the letter $a$ (respectively $b$). This results in Eve's token never taking a transition of even priority, and thus her run is rejecting.*

---

This was followed, in 2018, by Bagnol and Kuperberg giving a polynomial-time algorithm to decide history-determinism for Büchi automata [BK18]—these are parity automata whose priorities are all 0 or 1. Their algorithm relied on solving the so-called 2-token game. This game is similar to the HD game, where Adam constructs a word letter-by-letter and Eve constructs a run on her token transition-by-transition, but in addition, Adam has two distinguishable tokens on which he also constructs runs transition-by-transition. In each round of the 2-token game, Adam selects a letter, then Eve selects a transition on her token, and then Adam selects a transition on each of his two tokens. In the limit of an infinite play, Adam builds a word, Eve builds a run on her token, and Adam builds a run on each of his two tokens, all on Adam's word. Eve wins this play if the following condition holds: if either of Adam's runs on his tokens is accepting, then the run of Eve's token is accepting.

Bagnol and Kuperberg showed that for every Büchi automaton, Eve has a strategy to win the 2-token game on it if and only if it is HD. Thus, to decide if a Büchi automaton is HD, it suffices to solve the 2-token game on it, and doing so takes polynomial time. Bagnol and Kuperberg conjectured that this 2-token game based characterisation of history-determinism is true for all of parity automata: we call this the 2-*token conjecture*. For parity automata with a fixed number of priorities, the 2-token game can be solved in **P**, and in **PSPACE** if the number of priorities is not fixed.

Boker, Kuperberg, Lehtinen, and Skrzypczak proved that this conjecture is also true for coBüchi automata [BKLS20], by a more involved proof than the Büchi case, and which, like the proof for Büchi automata [BK18], is not constructive, that is, does not provide a strategy for Eve in the history-determinism game. Crucially, neither their argument not the argument of Bagnol and Kuperberg generalised to parity automata. Our main contribution is a proof of the 2-token conjecture.

**The 2-Token Theorem.** *For every nondeterministic parity automaton $\mathcal{A}$, Eve wins the 2-token game on $\mathcal{A}$ if and only if $\mathcal{A}$ is history-deterministic. Thus, the problem of deciding history-determinism is in **P** for parity automata with a fixed number of priorities, and in **PSPACE** if the number of priorities is part of the input.*

As corollaries, we obtain that the above upper bounds for deciding history-determinism also applies to alternating parity automata [BKLS20, Page 11], and that the 2-token game characterises history-determinism of $\omega$-regular automata due to [CIK+25, Lemma 36]. As a further corollary, we also obtain a 2-token game based algorithm for deciding the good-enough-synthesis for specifications given by deterministic parity automata [AK20], since this problem is logspace-interreducible[2] to deciding history-determinism [BL23b].

Our proof for the 2-token theorem is constructive, i.e., we are able to construct a winning strategy in the HD game based on a strategy in the 2-token game. As a result, we also get alternative constructive proofs of the 2-token theorem for Büchi and coBüchi automata. Our proof for the coBüchi case, in particular, is simpler than that of [BKLS20].

## 2 History-Deterministic Automata and Token Games

We use $\mathbb{N}$ to denote the set of natural numbers $\{0, 1, 2, \dots\}$. For two natural numbers $i < j$, we will use $[i, j]$ to denote the set of numbers $\{i, i+1, i+2, \dots, j\}$. An *alphabet* $\Sigma$ is a finite set of *letters*. We use $\Sigma^*$ and $\Sigma^\omega$ to denote the set of words of finite and countably infinite length over $\Sigma$, respectively. A language is a subset of $\Sigma^\omega$. For a finite word $u$ and a language $L$, we use $u^{-1}L$ to denote the language $\{w \mid uw \in L\}$.

**Parity Automata.** An $[i, j]$ nondeterministic parity automaton (or $[i, j]$ automaton, for short) $\mathcal{A} = (Q, \Sigma, q_0, \Delta)$ is a directed graph with a finite set of *states* $Q$, and edges that are each labelled by a *letter* from a finite set of alphabet $\Sigma$ and a *priority* in $[i, j]$. These edges are called *transitions*, and a transition $\delta$ from the state $q$ to $q'$ on the letter $a$ with a priority $c$ is denoted with $q \xrightarrow{a:c} q'$. We write that $\delta$ is a *c-transition* to indicate that the priority of $\delta$ is $c$.

---

[2]Boker and Lehtinen show that these two problems are polynomially interreducible, but logspace interreducibility is clear from their discussion as well.

A *run* of the automaton $\mathcal{A}$ on an infinite word $w$ is an infinite path that starts at the *initial state* $q_0$, and follows transitions consisting of the letters of $w$ in sequence. We say that this run is *accepting* if the *least priority occurring infinitely often* in the labels of its transitions is even. A word $w$ is *accepted* by $\mathcal{A}$ if $w$ has an accepting run, and the language of $\mathcal{A}$, denoted $L(\mathcal{A})$, is the set of words accepted by $\mathcal{A}$. For a state $q$ in $\mathcal{A}$, we use $(\mathcal{A}, q)$ to denote the automaton $\mathcal{A}$ with $q$ as the initial state. A parity automaton is said to be *deterministic* if for each state and letter, there is at most one transition outgoing from that state on that letter.

A Büchi (resp. coBüchi) automaton is a $[0, 1]$(resp. $[1, 2]$) automaton. A safety (resp. reachability) automaton is a $[1, 2]$ automaton that has a unique rejecting (resp. accepting) sink state, on which we only have outgoing transitions with priority 1 (resp. 2) for each letter, and all other transitions in the automaton have priority 2 (resp. 1). We note that safety and reachability automata can be equivalently viewed as a $[0, 1]$ automata, by changing the priorities of 2-transitions to be 0; this operation does not change the acceptance or rejection of any run.

We assume all our automata to be *complete*, i.e., from every state and letter, there is at least one outgoing transition from that state on that letter. We say that a parity automaton $\mathcal{B}$ is a *subautomaton* of a parity automaton $\mathcal{A}$ if $\mathcal{B}$ can be obtained by deleting some transitions from $\mathcal{A}$.

**Games.** An *arena* is a directed graph $G = (V, E)$ with vertices partitioned into $V_\forall$ and $V_\exists$ between two players Adam and Eve, respectively. Additionally, a vertex $v_0 \in V_\forall$ is designated as the initial vertex. We say that vertices in $V_\exists$ (resp. $V_\forall$) are owned by Eve (resp. Adam).

A *play* on this arena is an infinite path starting at $v_0$ and is formed as follows. A play starts with a token at $v_0$ and it proceeds for infinitely many rounds. At each round, the player who owns the vertex on which the token is currently placed chooses an outgoing edge, and the token is moved along this edge to the next vertex for another round of play. This creates an infinite path in the arena, which we call a play of $G$.

A game $\mathcal{G}$ consists of an arena $(V, E)$ and a winning objective $L \subseteq E^\omega$. A *play* of $\mathcal{G}$ starts with a token at a designated start vertex $v$ and proceeds in infinitely many rounds. In each round, the player who owns the vertex the token is at moves the token along an outgoing edge from that vertex to the target of that edge, from where the next round starts. The edges taken by the token constitutes an infinite path, which we call a *play* of $\mathcal{G}$. Eve wins a play if that play is in $L$, and Adam wins otherwise.

A *finite play* is a prefix of an infinite play. A *strategy* for Eve in such a game $\mathcal{G}$ is a function from the set of finite plays that end at an Eve's vertex to an outgoing edge from that vertex. Eve's strategy is said to be *winning* if any play produced while she chooses edges according to this strategy is winning for her. Strategies and winning strategies for Adam and defined analogously. We say Eve (resp. Adam) wins $\mathcal{G}$ if she (resp. he) has a winning strategy in $\mathcal{G}$. We say that Eve (resp. Adam) wins from $v$ in $\mathcal{G}$ if she (resp. he) wins the game $\mathcal{G}$ with its initial vertex at $v$. The set of vertices which Eve (resp. Adam) wins from is called the *winning region* of Eve (resp. Adam).

We say that a strategy (for either player) is *positional* if it is a strategy that only depends on the current position the token is at. We say that a strategy for a player is *uniform* if it is a positional strategy using which that player can win from all vertices in their winning region.

We will deal with $\omega$-regular games in this work—games whose winning objective $L$ is recognised by a nondeterministic Büchi automaton. These games are determined, i.e, they have a *winner* [GH82].

**Parity games.** In an $[i, j]$ *parity game* $\mathcal{G} = (V, E)$, or $[i, j]$ *game* for short, the edges $E$ are each labelled by a positive natural number from the interval $[i, j]$, which we call the *priority* of that edge. The winning objective for Eve consists of plays in which the least priority occurring infinitely often is even. Parity games are known to be *positionally determined*, that is, there is a winner and the winner can win using a uniform strategy [FBB+23, Theorem 15].

**History determinism** The history-determinism game (HD game) of an automaton is a two player turn-based game between Adam and Eve, who take alternating turns to select a letter and a transition in the automaton (on that letter), respectively. After the game ends, the sequence of Adam's choices of letters is an infinite word, and the sequence of Eve's choices of transitions is a run on that word. Eve wins the game if her run is accepting or Adam's word is rejecting.

**Definition 2.1** (History-determinism game). *Given a parity automaton $\mathcal{A} = (Q, \Sigma, q_0, \Delta)$, the history-determinism (HD) game of $\mathcal{A}$ is defined between the players Adam and Eve as follows, with positions in $Q$. The game starts with Eve's token at $q_0$ and proceeds for infinitely many rounds. For each $i \in \mathbb{N}$, round $i$ starts at a position $q_i \in Q$, and proceeds as follows.*

1. *Adam selects a letter $a_i \in \Sigma$.*

2. *Eve moves her token along a transition $q_i \xrightarrow{a_i : c_i} q_{i+1} \in \Delta$.*

*Eve's token then is at $q_{i+1}$ from where the round $(i + 1)$ is played. Thus, the play of a HD game can be seen as Adam constructing a word letter-by-letter, and Eve constructing a run transition-by-transition on her token on the same word. Eve wins this play if the following condition holds: if Adam's word is in $\mathcal{L}(\mathcal{A})$, then the run of Eve's token is accepting.*

We say that an automaton is history-deterministic (HD) if Eve has a winning strategy in the HD game on $\mathcal{A}$. A direct approach to decide the history-determinism of an automaton that involves directly solving the HD game involves determinising the automaton $\mathcal{A}$, and thus is algorithmically expensive since determinisation takes exponential time [HP06, Section 4].

**Token games** Token games were introduced by Bagnol and Kuperberg as a potential alternative for deciding history-determinism, which avoids the bottleneck of determinisation [BK18].

**Definition 2.2** ($k$-token game). *For a natural number $k \geq 1$, the $k$-token game on parity automata $\mathcal{A} = (Q, \Sigma, q_0, \Delta)$, and $\mathcal{B}_l = (P^l, \Sigma, p_0^l, \Delta_l)$ for each $l \in [k]$, denoted $Gk(\mathcal{A}; \mathcal{B}_1, \mathcal{B}_2, \dots, \mathcal{B}_k)$ is defined between the players Adam and Eve as follows, with positions in $Q \times P^1 \times \dots P^k$. The game starts with an Eve's token in the initial state of $\mathcal{A}$, and $k$ tokens of Adam that are placed in the initial states of the $k$ automata $\mathcal{B}_1, \dots, \mathcal{B}_l$, respectively. For each $i \in \mathbb{N}$, the round $i$ starts at a position $(q_i, (p_i^1, p_i^2, \dots, p_i^k)) \in Q \times P^1 \times \dots \times P^k$, and proceeds as follows.*

1. *Adam selects a letter $a_i \in \Sigma$.*

2. *Eve selects a transition on her token $q_i \xrightarrow{a_i : c} q_{i+1} \in \Delta$.*

3. *Adam selects $k$ transitions $p_i^l \xrightarrow{a_i : c_i^l} p_{i+1}^l \in \Delta_l$ on $a_i$ for each of his $k$ tokens.*

*The new position is $(q_{i+1}, (p_{i+1}^1, p_{i+1}^2, \dots, p_{i+1}^k))$, from where round $(i + 1)$ begins.*

Thus, in a play of the $k$-token game on $\mathcal{A}, \mathcal{B}_1, \mathcal{B}_2, \dots, \mathcal{B}_k$, Eve constructs a run on her token and Adam a run on each of his $k$ tokens, all on the same word. Eve wins this play if the following condition holds: if at least one of Adam's $k$ runs on his tokens are accepting, then Eve's run on her token is accepting.

If Eve (resp. Adam) has a winning strategy in the above game, we say Eve (resp. Adam) wins $Gk(\mathcal{A}; \mathcal{B}_1, \ldots, \mathcal{B}_k)$. We will often care about the $k$-token game where both Eve and Adam's tokens are all in the same automaton $\mathcal{A}$ but start at different states. We will then write $Gk(q; p_1, p_2, \ldots, p_k)$ in $\mathcal{A}$ to denote the game

$$Gk((\mathcal{A}, q); (\mathcal{A}, p_1), \ldots, (\mathcal{A}, p_k)).$$

Finally, we use $Gk(\mathcal{A})$ to denote the game where Eve's token and Adam's $k$ tokens all start at the initial state of $\mathcal{A}$, and we call this game as the $k$-token game on $\mathcal{A}$. The crucial and elegant insight of Bagnol and Kuperberg's work is that the 2-token game on $\mathcal{A}$ have the same winner as the $k$-token on $\mathcal{A}$ for any $k$ larger than 2.

**Lemma 2.3** ([BK18])**.** *Eve wins the* 2*-token game on* $\mathcal{A}$ *if and only if Eve wins the* $k$*-token game on* $\mathcal{A}$ *for all* $k \geq 1$.

We show that if Eve wins the 2-token game then she wins the 3-token game, using which our reader should be able to prove Lemma 2.3. Fix a strategy $\sigma_{G2}$ for Eve in the 2-token game, and consider the following strategy for Eve in the 3-token game. Eve stores in her memory, an additional token, in which she selects transitions by playing the 2-token game against Adam's second and third token using $\sigma_{G2}$. Eve chooses transitions on her token by playing the 2-token game against Adam's first token and her memory token using $\sigma_{G2}$. In any play of the 3-token game on $\mathcal{A}$ where Eve is playing according to this strategy, if either of Adam's tokens produces an accepting run, then either Adam's first token or Eve's memory token produces an accepting run, and hence Eve's token produces an accepting run.

The following result on transitivity of $G1$ games will be useful for us.

**Lemma 2.4.** *Let* $\mathcal{A}, \mathcal{B},$ *and* $\mathcal{C}$ *be parity automata such that Eve wins* $G1(\mathcal{A}; \mathcal{B})$ *and* $G1(\mathcal{B}; \mathcal{C})$. *Then Eve wins* $G1(\mathcal{A}; \mathcal{C})$.

Additionally, we observe that if Eve wins $G1(\mathcal{A}; \mathcal{B})$ then Eve has a positional winning strategy. Indeed, the winning condition for Eve in $G1$ is a disjunction of two parity objectives: either Adam's run is rejecting, or Eve's run is accepting. Disjunctions of parity conditions can be expressed as a Rabin objective [CHP07, Page 6], and Eve has positional winning strategies on games with Rabin objectives [Eme85]. Additionally, Eve has an uniform strategy in such games, i.e., a positional strategy using which Eve wins from all vertices in her winning region.

**Lemma 2.5** ([Pra25, Proposition 3.16])**.** *If Eve wins* $G1(\mathcal{A}; \mathcal{B})$ *for some parity automata* $\mathcal{A}$ *and* $\mathcal{B}$, *then Eve has a uniform strategy in* $G1(\mathcal{A}; \mathcal{B})$.

While 1-token games are not enough to characterise history-determinism already on Büchi or coBüchi automata [BK18, Lemma 8], they characterise history-determinism on safety and reachability automata.

**Lemma 2.6** ([BKS17, Theorem 17],[BL23a, Theorem 4.5 and 4.8])**.** *Let* $\mathcal{A}$ *be a safety or reachability automaton. Then the following statements are equivalent.*

1. *Eve wins the* 1*-token game on* $\mathcal{A}$.

2. $\mathcal{A}$ *is history-deterministic.*

3. *Eve has a positional winning strategy in the HD game on* $\mathcal{A}$.

4. $\mathcal{A}$ *has a language-equivalent deterministic subautomaton* $\mathcal{D}$.

**Simulation**  For parity automata $\mathcal{A}$ and $\mathcal{B}$, the (fair) simulation game [HKR02] of $\mathcal{B}$ by $\mathcal{A}$ is similar to $G1(\mathcal{A}; \mathcal{B})$, except that in each round Adam moves his token before Eve moves her token.

**Definition 2.7** (Simulation game). *Given two parity automata $\mathcal{A} = (P, \Sigma, p_0, \Delta_A)$ and $\mathcal{B} = (Q, \Sigma, q_0, \Delta_B)$, the* simulation game *of $\mathcal{B}$ by $\mathcal{A}$ is a two player game played between Eve and Adam as follows, with positions in $Q \times P$. The game starts at $(q_0, p_0)$, and proceeds for infinitely many rounds. For each $i \geq 0$, round $i$ starts at position $(q_i, p_i)$ and proceeds as follows:*

1. *Adam selects a letter $a_i \in \Sigma$;*

2. *Adam selects a transition $q_i \xrightarrow{a_i : c'} q_{i+1}$ in $\mathcal{B}$;*

3. *Eve selects a transition $p_i \xrightarrow{a_i : c} p_{i+1}$ in $\mathcal{A}$.*

*In the limit of an infinite play, Adam and Eve have each construct a run on the same word in $\mathcal{B}$ and $\mathcal{A}$, respectively. We say that Eve wins the play if either her run is accepting or Adam's run is rejecting.*

If Eve has a winning strategy in the simulation game of $\mathcal{B}$ by $\mathcal{A}$, then we say that $\mathcal{A}$ simulates $\mathcal{B}$. We say that $\mathcal{A}$ and $\mathcal{B}$ are simulation-equivalent if $\mathcal{A}$ simulates $\mathcal{B}$ and $\mathcal{B}$ simulates $\mathcal{A}$. Observe that if $\mathcal{A}$ simulates $\mathcal{B}$, then $L(\mathcal{B}) \subseteq L(\mathcal{A})$.

**Lemma 2.8** ([AJP24, Corollary 10]). *If $\mathcal{A}$ and $\mathcal{B}$ are two parity automata that are simulation-equivalent, then $\mathcal{A}$ is HD if and only if $\mathcal{B}$ is HD.*

# 3   Overview of Techniques

Our proof of the 2-token theorem involves induction on the *parity-index hierarchy*. For an $[i, j]$ parity automaton $\mathcal{A}$, note that shifting all priorities by 2 in the same direction does not change the acceptance of each run in the automaton, and hence we can assume that $i = 0$ or $1$. If this is the case, then $[i, j]$ is said to be the *parity index* of the automaton. Parity automata thus have an alternating hierarchy with respect to its parity indices as shown in Fig. 2, where automata with larger parity indices are (syntactically) more powerful than those with smaller parity indices.

Our inductive proof of the 2-token theorem consists of the following two steps.

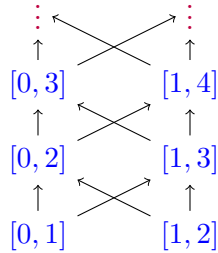

Figure 2: The parity index hierarchy. Büchi and coBüchi automata correspond to $[0, 1]$ and $[1, 2]$ automata respectively, and are at the bottom of the index hierarchy.

**Even-to-odd induction step.** We show that if the 2-token theorem holds for $[0, K]$ automata (equivalently $[2, K + 2]$ automata) for some $K \geq 1$, then the 2-token theorem also holds for $[1, K + 2]$ automata.

**Odd-to-even induction step.** We show that if the 2-token theorem holds for $[1, K]$ automata for some $K \geq 2$, then the 2-token theorem also holds for $[0, K]$ automata.

Before presenting each of the induction steps, we will prove the 2-token theorem for coBüchi and coBüchi automata as easier versions of the induction step, to prepare the reader. We then present the two induction steps, which will prove the 2-token theorem.

Observe that one direction of the 2-token theorem holds easily: if an automaton $\mathcal{A}$ is history-deterministic, then Eve has a winning strategy in the 2-token game on $\mathcal{A}$, where she chooses her transitions according to a winning strategy in the HD game on $\mathcal{A}$, ignoring Adam's tokens. Thus, the more interesting direction is showing that if Eve wins the 2-token game on an automaton, then that automaton is HD.

A recurring pattern in the arguments used in our proofs is the following: starting from an automaton on which Eve wins the 2-token game, we will make certain modifications to get, as an intermediate step, a *simulation-equivalent* automaton on which Eve still wins the 2-token game, and which have some nice additional properties. Due to these properties, it will be easier to show that the new automaton is HD, and it will follow from simulation-equivalence that the automaton we started with is HD as well.

In both induction steps, we use an important intermediate property. For the even-to-odd induction step, we call this intermediate property *1-safe coverage*, while for the odd-to-even induction step, we call the analogous property *0-reach covering*. For the even-to-odd induction step, the structural modification to obtain 1-safe coverage involves relabelling of priorities while preserving the acceptance of each run, and uses a refinement of classical techniques that have been used in [KS15, BKLS20]. The odd-to-even induction step is more involved, and uses modifications which involve the concepts of ranks on parity games [Jur00, Wal02] and Zielonka trees [DJW97].

Another crucial modification we will use throughout allows us to assume—in the context of showing that an automaton is HD if Eve wins the 2-token game on it—that *Eve wins the $k$-token game from everywhere for any $k \geq 1$*, i.e., Eve wins the $k$-token game from every configuration of states that can be reached in the $k$-token game from the initial configuration where all tokens start at the initial state.

To make this more precise, let us call states $q, p_1, p_2, \ldots, p_k$ as coreachable if there is a finite word $u$ on which there are finite runs from the initial state to the states $q, p_1, p_2, \ldots, p_k$. We say that Eve wins the $k$-token game from everywhere in $\mathcal{A}$ if she wins $Gk(q; p_1, p_2, \ldots, p_k)$ for all $(k+1)$ states $q, p_1, p_2, \ldots, p_k$ in $\mathcal{A}$.

We can strengthen this more. Let us call the transitive closure of the binary coreachability relation *weak coreachability* and denote the relation as $\mathsf{CR}^*$. Note that weak coreachability is an equivalence relation between the states of the automaton. We say that states $p_1, p_2, \ldots, p_k$ are weakly coreachable in $\mathcal{A}$ if they are pairwise weakly coreachable in $\mathcal{A}$.

Then, if Eve wins the $k$-token game from everywhere in $\mathcal{A}$ then she wins $Gk(q; p_1, p_2, \ldots, p_k)$ in $\mathcal{A}$ from all $(k+1)$ states $(q, p_1, p_2, \ldots, p_k)$ that are weakly coreachable in $\mathcal{A}$. The above discussion is summed up in the following theorem.

**Theorem 3.1.** *Let $\mathcal{A}$ be a parity automaton on which Eve wins the 2-token game. Then, there is a simulation-equivalent subautomaton $\mathcal{B}$ of $\mathcal{A}$ such that the following two conditions hold.*

1. *Eve wins $Gk(q; p_1, p_2, \ldots, p_k)$ in $\mathcal{B}$ for all $(k+1)$ states $q, p_1, p_2, \ldots, p_k$ that are weakly coreachable in $\mathcal{B}$.*

2. *If $\mathcal{B}$ is history-deterministic, then so is $\mathcal{A}$.*

Since automata that are simulation-equivalent are either all HD or all not HD (Lemma 2.8), we note that the second item in Theorem 3.1 above is a consequence of simulation-equivalence of $\mathcal{A}$ and $\mathcal{B}$. When trying to prove that a parity automaton $\mathcal{A}$ on which Eve wins the 2-token game is HD, we will assume due to Theorem 3.1 that Eve wins the 2-token game from everywhere on $\mathcal{A}$.

Before continuing, let us set some more notation that we will use throughout. For a state $p$, we will use $\mathsf{CR}^*(\mathcal{A}, p)$ to denote the set of states that are weakly coreachable to $p$ in $\mathcal{A}$. For a finite word $u$, we will use $\mathsf{CR}^*(\mathcal{A}, u)$ to refer to the set $\mathsf{CR}^*(\mathcal{A}, p)$, where $p$ is a state such that there is a finite run from the initial state of $\mathcal{A}$ to $p$ on $u$. Note that the choice of $p$ does not matter here.

The rest of this section is organised as follows: in Section 3.2 and Section 3.4, we will prove our even-to-odd and our odd-to-even induction steps, respectively. We precede these induction steps with proofs of the 2-token theorem for coBüchi automata in Section 3.1 and for Büchi automata in Section 3.3 as warm-ups for the even-to-odd and odd-to-even induction steps, respectively.

## 3.1 Even to Odd Warm-up: CoBüchi Automata

Before detailing our even-to-odd induction step, we prepare the reader by showing that the 2-token game characterises history-determinism for coBüchi automata, with a novel, simpler proof than that of [BKLS20, Theorem 28]. It is also more constructive, in the sense that we obtain a strategy for Eve in the HD game based on a strategy in the 2-token game. Our arguments are then adapted to work for the even-to-odd induction step. We will show the following result.

**Theorem 3.2.** *Let $\mathcal{A}$ be a coBüchi automaton on which Eve wins the 1-token game from everywhere. Then, $\mathcal{A}$ is history-deterministic.*

The 2-token theorem for coBüchi automata (Corollary 3.3) then follows from Theorem 3.1.

**Corollary 3.3.** ([BKLS20].) *For every coBüchi automaton $\mathcal{A}$, Eve wins the 2-token game on $\mathcal{A}$ if and only if $\mathcal{A}$ is HD.*

We first do a *priority-normalisation* procedure: For a coBüchi automaton $\mathcal{A}$, consider the graph $G_{>1}$ of $\mathcal{A}$ consisting of only priority 2 transitions. For a priority 2 transition in $\mathcal{A}$ that is not part of any SCC in $G_{>1}$, we change its priority to 1 to obtain $\mathcal{A}'$; Other transitions of $\mathcal{A}$ are kept as is in $\mathcal{A}'$. Then, a run in $\mathcal{A}$ is accepting if and only if a run in $\mathcal{A}'$ is, and therefore, Eve wins the HD game (resp. the 1-token game from everywhere) on $\mathcal{A}$ if and only if the same holds for $\mathcal{A}'$ (Proposition 6.1).

We then define the safety automaton $\mathcal{A}_{\mathtt{safe}}$ obtained by preserving all priority 2 transitions in $\mathcal{A}$ and having priority 1 transitions lead to a rejecting sink state. The crucial property required for our proof, a version of which appeared first in [KS15, Lemma 56 in full version], is of safe coverage.

**Safe coverage.** We say that an automaton $\mathcal{A}$ has *safe coverage* if for each state $p$ there is a state $q$ in $\mathsf{CR}^*(\mathcal{A}, p)$, such that Eve wins $G1(q; p)$ in $\mathcal{A}_{\mathtt{safe}}$.

**Lemma 3.4.** *Every coBüchi automaton on which Eve wins the 1-token game from everywhere and that is priority-normalised has safe coverage.*

*Proof.* Suppose that $\mathcal{A}$ is priority-normalised and that $\mathcal{A}$ does not have safe coverage. We will show that Eve does not win the 1-token game from everywhere in $\mathcal{A}$.

Since $\mathcal{A}$ does not have safe coverage, there is a state $q$ such that Adam wins $G1(p; q)$ in $\mathcal{A}_{\mathtt{safe}}$ for all states $p \in \mathsf{CR}^*(\mathcal{A}, q)$. We describe a winning strategy of Adam for $G1(q; p)$ in $\mathcal{A}$.

Adam, in $G1(q; q)$ in $\mathcal{A}$, chooses letters and transitions on his token according to his strategy in $G1(q; q)$ in $\mathcal{A}_{\mathtt{safe}}$. This ensures that Eve is eventually forced to take a 1-transition in $\mathcal{A}$, while Adam's token has reached some state $q'$ via only priority 2 transitions. Since $\mathcal{A}$ is priority-normalised, and Adam has not changed SCC of $G_{>1}$, there is a finite run back from $q'$ to $q$ via only priority 2 transitions, and so Adam plays letters and picks transitions on his token in the 1-token game according to this finite run. When Adam's token is back at state $q$, Eve's token is at some state $p'$ that is weakly coreachable to $q$ in $\mathcal{A}$ (Proposition 5.3). By assumption, Adam wins $G1(p'; q)$ in $\mathcal{A}_{\mathtt{safe}}$, and Adam thus repeats his strategy as above, ensuring that the run on his token takes only priority 2 transitions and hence, is accepting, while the run on Eve's token takes infinitely many priority 1 transitions and hence, is rejecting. □

We then construct a winning strategy for Eve in the HD game on a coBüchi automaton that has safe coverage and on which Eve wins the 1-token game from everywhere, proving Theorem 3.2.

*Sketch of Eve's strategy.* Due to the transitivity of $G1$ games (Lemma 2.4) and the finiteness of states in $\mathcal{A}$, the property of safe coverage implies the following statement: for every state $p$, there is a state $q$ in $\mathsf{CR}^*(\mathcal{A}, p)$, such that Eve wins $G1(q; p)$ and $G1(q; q)$ in $\mathcal{A}_{\mathtt{safe}}$. Note, from Lemma 2.6, that if Eve wins $G1(q; q)$ in $\mathcal{A}_{\mathtt{safe}}$, then $(\mathcal{A}_{\mathtt{safe}}, q)$ is HD and determinisable-by-pruning: we call such states $q$ safe-deterministic. We will use these safe-deterministic states to build a strategy for Eve in the HD game on $\mathcal{A}$.

Fix $\sigma_{\mathtt{safe}}$ to be a positional HD strategy in $\mathcal{A}_{\mathtt{safe}}$ for all safe-deterministic states, and a positional winning strategy $\sigma_{G1}$ for Eve in the 1-token game in $\mathcal{A}$ from all pairs of weakly coreachable states (Lemma 2.5). We describe the following winning strategy for Eve in the HD game on $\mathcal{A}$. At all points, when Eve's token is at $p$, Eve's token will *track* a memory token at a safe-deterministic state $q$ that is in $\mathsf{CR}^*(\mathcal{A}, p)$. Eve will choose transitions on her token according to $\sigma_{G1}$ against her memory token and her memory token will take transitions according to $\sigma_{\mathtt{safe}}$, until her memory token takes a transition of priority 1. When this happens, suppose Eve's token is at the state $p'$. Eve then resets her memory token to some other safe-deterministic state $q' \in \mathsf{CR}^*(\mathcal{A}, p)$.

The choice of this $q'$ is crucial and is made as follows: if $u$ is the finite word read so far in the HD game, then consider the longest suffix $v$ of $u$ such that there is a run $\rho_v$ in $\mathcal{A}_{\mathtt{safe}}$ only consisting of transitions from $\sigma_{\mathtt{safe}}$ ending at some state in $\mathsf{CR}^*(\mathcal{A}, p)$. We pick $q'$ to be the endpoint of $\rho_v$; note that $q'$ is safe-deterministic, since it was reached by $\sigma_{\mathtt{safe}}$. This concludes the description our strategy. We show in Section 6 that this is a winning strategy for Eve. □

## 3.2 The Even-to-Odd Induction Step

In this section, we will show the following result:

**Theorem 3.5.** *Let $K \geq 1$ be a natural number, such that for every $[0, K]$ automaton $\mathcal{A}$, Eve wins the 2-token game on $\mathcal{A}$ if and only if $\mathcal{A}$ is HD. Then, for every $[1, K + 2]$ automaton $\mathcal{A}$, Eve wins the 2-token game on $\mathcal{A}$ if and only if $\mathcal{A}$ is HD.*

Observe that if $\mathcal{A}$ is HD, then Eve wins the 2-token game on it trivially. Towards proving Theorem 3.5, for the rest of this section, we assume that the 2-token theorem holds for $[0, K]$ automata, or equivalently, on $[2, K + 2]$ automata.

We extend the *priority-normalisation* procedure from coBüchi automata to $[1, K + 2]$ automata, calling it 2-priority reduction. On every $[1, K+2]$ automaton $\mathcal{A}$, the 2-priority reduction procedure relabels the priorities of $\mathcal{A}$ while preserving the (in)acceptance of each run so that the following condition holds: for the graph $\mathcal{G}_{>1}$ consisting of transitions of priority at least 2 in $\mathcal{A}$, all transitions in $\mathcal{G}_{>1}$ belong to an SCC, and each SCC contains at least one transition of priority 2 (Lemma 7.3). Similar to $\mathcal{A}_{\mathtt{safe}}$ for coBüchi automata, we define the 2-approximation of $\mathcal{A}$, denoted by $\mathcal{A}_{>1}$ as the automaton obtained by the following modifications on $\mathcal{A}$: we preserve transitions of priority at least 2 in $\mathcal{A}_{>1}$ and transitions of priority 1 are redirected towards a rejecting sink state and have their priority changed to 3. The rejecting sink state has a self loop of priority 3 on every letter. Observe that $\mathcal{A}_{>1}$ is a $[2, K + 2]$ automaton.

**1-safe coverage**  Similar to how we used safe coverage as an intermediate property to show the 2-token theorem for coBüchi automata, we will use the property of *1-safe coverage* as an intermediate step for showing the even-to-odd induction step. We say $\mathcal{A}$ has 1-safe coverage if for each state $p$ there is a state $q \in \mathsf{CR}^*(\mathcal{A}, p)$ such that Eve wins $G2(q; p, p)$ in $\mathcal{A}_{>1}$. Note that with our proof of the 2-token theorem for coBüchi automata in mind, this definition is well-motivated. The proof of the induction step then consists of first showing that if Eve wins the 2-token game from everywhere and is priority reduced, then $\mathcal{A}$ has 1-safe coverage, and then showing that an automaton with 1-safe coverage and on which Eve wins the 2-token game is history-deterministic. In this latter part, we will use our induction hypothesis of the 2-token theorem for $[2, K + 2]$ automata, analogously to how we used the 1-token characterisation of history-determinism Lemma 2.6 on safety automata to show the 2-token theorem for coBüchi automata.

Analogous to Lemma 3.4 for coBüchi automata, we now show that $\mathcal{A}$ has 1-safe coverage. To do so, we first define a safety automaton $\mathcal{A}_{\mathtt{safe}}$, as we had for coBüchi automata: transitions of priority at least 2 in $\mathcal{A}$ become safe, that is, priority 2 transitions in $\mathcal{A}_{\mathtt{safe}}$, while transitions of priority 1 lead to a rejecting sink state. We can also define safe coverage and argue that $\mathcal{A}$ has safe coverage similarly to coBüchi automata.

**Lemma 3.6.** *Every $[1, K + 2]$ automaton $\mathcal{A}$ that is 2-priority reduced and on which Eve wins the 1-token game from everywhere has safe coverage, i.e., for every state $p$ in $\mathcal{A}$ there is a state $q$ weakly coreachable to $p$ in $\mathcal{A}$, such that Eve wins $G1(q; p)$ in $\mathcal{A}_{\mathtt{safe}}$.*

**Lemma 3.7.** *Every $[1, K + 2]$ automaton that is 2-priority reduced and on which Eve wins the 2-token game from everywhere has 1-safe coverage.*

*Proof sketch.* Suppose towards a contradiction that a $[1, K + 2]$ automaton $\mathcal{A}$ is 2-priority reduced and Eve wins the 2-token game from everywhere but that there is a state $p$, such that Adam wins $G2(q; p, p)$ in $\mathcal{A}_{>1}$ for all $q \in \mathsf{CR}^*(\mathcal{A}, p)$. Since $\mathcal{A}$ has safe coverage (Lemma 3.6), let $s$ be a safe-deterministic state in $\mathsf{CR}^*(\mathcal{A}, p)$, such that Eve wins $G1(s; p)$ in $\mathcal{A}_{\mathtt{safe}}$: an existence of such an $s$ is shown similarly to the case for coBüchi automata.

We will construct a winning strategy for Adam in the game $\mathcal{G} = G3(p; p, p, s)$ in $\mathcal{A}$, which contradicts the fact that Eve wins the 2-token game from everywhere in $\mathcal{A}$ (Lemma 5.7). Adam's winning strategy will use two memory tokens $m_1$ and $m_2$ that take transitions in $\mathcal{A}_{>1}$ and that are at the state $p$ initially and at each *reset*. Until Eve's token takes a priority 1 transition, Adam's first two tokens will take transitions according to a winning strategy (for Eve) in $G1$ on $\mathcal{A}$ against these memory tokens, respectively. Adam chooses letters and transitions on his memory tokens according to his winning strategy in $G2(p; p, p)$ in $\mathcal{A}_{>1}$; Adam chooses transition on his third token according to the HD strategy for Eve in $(\mathcal{A}_{\mathtt{safe}}, s)$. When Eve's token takes a priority 1 transition, Adam moves his third token along priority 2 transitions back to $s$, and he resets his memory tokens to be both at $p$. This is possible since $\mathcal{A}$ is 2-priority reduced. Eve's token is then at some state $q$ weakly coreachable to $p$. Since Adam wins $G2(q; p, p)$ in $\mathcal{A}_{>1}$, Adam can then repeat this strategy.

In any play where Adam plays with this strategy, one of the following two cases occur.

1. Eve's token takes infinitely many priority 1 transitions while Adam's third token takes infinitely many priority 2 transitions and no priority 1 transition.

2. Eve's token does not take a priority 1 transition after some point, and so Eve's token produces a rejecting run (since Adam plays according to a winning strategy in $G2$ on $\mathcal{A}_{>1}$), Adam's memory tokens are eventually not reset and one of them produces an accepting run, implying that the corresponding run of Adam's token is accepting as well.

We note that Eve loses in both cases, as desired. □

If Eve wins $G2(q; p, p)$ and $G2(r; q, q)$ in $\mathcal{A}_{>1}$ for some states $q$ and $r$, then we can argue that Eve wins $G2(r; p, p)$ in $\mathcal{A}_{>1}$ too (Lemma 5.1.4). Using the finiteness of states, we obtain that if $\mathcal{A}$ has 1-safe coverage, then for every state $q$ in $\mathcal{A}$, there is a state $p$ in $\mathsf{CR}^*(\mathcal{A}, q)$ such that Eve wins $G2(p; q, q)$ and $G2(p; p, p)$ in $\mathcal{A}_{>1}$(Lemma 7.7). Using the hypothesis of the 2-token theorem for $[0, K]$ automata, we note that states $q$ as above are such that $(\mathcal{A}_{>1}, q)$ is HD: we call such states as *1-safe HD* states.

We then show that an automaton $\mathcal{A}$ with 1-safe coverage on which Eve wins the 2-token game from everywhere is HD. First, we note that a naive adaptation of the proof for coBüchi automata fails. A run in $\mathcal{A}_{>1}$ that does not end up at the rejecting sink state need not be accepting, unlike for $\mathcal{A}_{\mathtt{safe}}$ when $\mathcal{A}$ is a coBüchi automaton. This is a problem since we then cannot just play the 1-token game against some 1-safe HD state. If Eve uses sufficiently many tokens instead of just one, however, then we will show that she can play the 1-token game against all weakly coreachable 1-safe HD states in a way that she constructs an accepting run in at least one of her tokens if the word is the language.

This corresponds to $\mathcal{A}$ being *explorable* [HK23]. A nondeterministic parity automaton $\mathcal{B}$ is *k-explorable* if Eve wins the $k$-HD game on it, where she has $k$ tokens. In each round, Adam selects a letter, and Eve responds with a transition on each of her $k$ tokens on that letter. Eve wins a play in this game if one of the runs on her $k$ tokens is accepting, or Adam's word is not in the language of $\mathcal{B}$. An automaton is explorable if it is $k$-explorable for some finite $k$.

We can then combine explorability of $\mathcal{A}$ and the fact that Eve wins the 2-token game to show that $\mathcal{A}$ is HD ([HK23, Section 2.4]): if an automaton is $k$-explorable for some $k$, then Eve in the HD game can construct an accepting run on any accepting word by playing the $k$-token game against $k$ tokens that are following a winning strategy in the $k$-HD game. Thus, we deduce that $\mathcal{A}$ is HD, and this completes our even-to-odd induction step.

## 3.3 Odd to Even Warm-up: Büchi Automata

We proceed towards the second half our our argument for the 2-token theorem, which is the odd-to-even induction step. Like in the even-to-odd case, we start by elaborating on the simpler case of Büchi automata first. Analogous to Theorem 3.2 for coBüchi automata, we will show the following result, which implies the 2-token theorem for Büchi automata (Corollary 3.10).

**Theorem 3.8.** *For every Büchi automaton $\mathcal{A}$, if Eve wins the 1-token game from everywhere on $\mathcal{A}$, then $\mathcal{A}$ is HD.*

We note that the above result also follows due to a recent result of Acharya, Jurdziński, and Prakash [AJP24, Theorem B]. We here give a proof of the above where we explicitly construct an Eve strategy in the HD game, which instead borrows from the proof of the correctness of the polynomial-time determinisation procedure that they gave for HD Büchi automata [AJP24, Theorem C].

Our proof relies on playing the 1-token games in approximations of Büchi automata. For a Büchi automaton $\mathcal{A}$, we define the 1-*approximation* of $\mathcal{A}$, denoted by $\mathcal{A}_{>0}$, as the automaton that has the same states as $\mathcal{A}$ along with an additional *accepting sink state* $q_\top$. The transitions of priority 1 in $\mathcal{A}$ are added to $\mathcal{A}_{>0}$, while transitions of priority 0 in $\mathcal{A}$ are redirected to the accepting sink state $q_\top$ in $\mathcal{A}_{>0}$ and have priority 2. We have self-loops of priority 2 on $q_\top$ on all letters in $\Sigma$. Note that $\mathcal{A}_{>0}$ is a reachability automaton.

**Reach covering.** We say that a Büchi $\mathcal{A}$ has *reach covering* if for every state $q$ there is a state $p$ in $\mathsf{CR}^*(\mathcal{A}, q)$, such that Eve wins $G1(q; p)$ in $\mathcal{A}_{>0}$. Reach covering is analogous to safe coverage for coBüchi automata, but observe that the $G1$ relation between states $q$ and $p$ here are in the opposite order from that of safe coverage.

**Lemma 3.9.** *Every Büchi automaton $\mathcal{A}$ on which Eve wins the 1-token game from everywhere and that has reach covering, is HD.*

*Sketch of Eve's strategy.* If $\mathcal{A}$ has reach covering then, due to the transitivity of $G1$ (Lemma 2.4) and the finiteness of states of $\mathcal{A}$, the following statement is true: for each state $q$ there is state $p \in \mathsf{CR}^*(\mathcal{A}, q)$, such that Eve wins $G1(q; p)$ and $G1(p; p)$ in $\mathcal{A}_{>0}$. Observer that if Eve wins $G1(p; p)$ in $\mathcal{A}_{>0}$ then Eve has a positional winning strategy in the HD game on $(\mathcal{A}_{>0}, p)$ (Lemma 2.6), and we thus call such states $p$ as *reach-deterministic*.

We fix a positional winning strategy $\tau_{G1}$ for Eve in the 1-token game on $\mathcal{A}_{>0}$ from all pairs $q, p$ of states in $\mathcal{A}_{>0}$ such that Eve wins $G1(q; p)$ in $\mathcal{A}_{>0}$, as well as a positional winning strategy $\sigma_{>0}$ for Eve in the HD game on $\mathcal{A}_{>0}$ from states $q$ that are reach-deterministic.

We describe a winning strategy for Eve in the HD game on $\mathcal{A}$ as follows. When Eve's token is at a state $q$, she stores a memory token at a state $p$ in $\mathcal{A}_{>0}$ such that $p$ is reach-deterministic, Eve wins $G1(q; p)$ in $\mathcal{A}_{>0}$, and $p$ is either the accepting sink state in $\mathcal{A}_{>0}$ or $p$ is in $\mathsf{CR}^*(\mathcal{A}, q)$. When Adam picks a letter $a$, let $\delta = q \xrightarrow{a:c} q'$ be the transition given by Eve's strategy $\tau_{G1}$ against Eve's memory token. If $q'$ is not the accepting sink state, then she takes the transition $\delta$, and Eve updates her memory token to take the transition on $a$ given by $\sigma_{>0}$.

Otherwise, there is an outgoing transition of priority 0 in $\mathcal{A}$ from $q$ to $\bar{q}$, and she picks this transition on her token. She resets her memory token to be at some state $\bar{p} \in \mathsf{CR}^*(\mathcal{A}, \bar{q})$ such that Eve wins $G1(\bar{q}; \bar{p})$ in $\mathcal{A}_{>0}$ and $\bar{p}$ is reach-deterministic. $\square$

Due to Lemma 3.9 above, it suffices to show (for proving Theorem 3.8) that every Büchi automaton $\mathcal{A}$ on which Eve wins the 1-token game from everywhere can be modified to a simulation-equivalent automaton that additionally has reach covering. This is more technical than for coBüchi automata, and uses (a restriction of) the concept of progress measures on parity games [Jur00], equivalently dubbed as signatures [Wal02] or ranks [Büc83].

For a $[0, P]$ parity game $\mathcal{G}$, where Eve wins from every vertex $v$, we define $\mathrm{rank}(v)$ as the largest number $k \in \mathbb{N} \cup \{\infty\}$ such that Adam has a strategy $\sigma$ in $\mathcal{G}$ that satisfies the following condition: in every play $\rho$ starting from $v$ where Adam is playing according to $\sigma$, at least $k$ many edges of priority 1 occur in $\rho$ before an edge of priority 0. Then dually, Eve has a positional winning strategy $\tau$ in $\mathcal{G}$ that ensures, for every vertex $v$, any play from $v$ where Eve is playing according to $\tau$ does not contain more that $\mathrm{rank}(v)$ many priority 1 edges before a priority 0 edge [Wal02, Lemma 8]. We will call such a strategy the *optimal strategy* in $\mathcal{G}$.

For a Büchi automaton $\mathcal{A}$ on which Eve wins the 1-token game from everywhere, the 1-token game on $\mathcal{A}$ can be explicitly represented as a $[0, 2]$ parity game $\mathcal{G}_1(\mathcal{A})$. In $\mathcal{G}_1(\mathcal{A})$, a priority 0 edge corresponds to Eve's token taking a 0-transition, a priority 1 edge corresponds to Adam's and Eve's token taking a 0-transition and 1-transition respectively, and a priority 2 edge corresponds to both Eve's and Adam's token taking 1-transitions. We denote the position of the 1-token game where Eve's and Adam's tokens are at $q$ and $p$, respectively, as $(q, p)$. Observe that if $\mathrm{rank}(q, p) = 0$ for weakly coreachable states $p$ and $q$, then Eve wins $G1(q; p)$ in $\mathcal{A}_{>0}$ using an optimal strategy in $\mathcal{G}_1(\mathcal{A})$. For each state $q$, we define its optimal rank as

$$\mathrm{opt}(q) = \min\{\mathrm{rank}(q, p) \mid q \text{ and } p \text{ are weakly coreachable in } \mathcal{A}.\}$$

Thus, if $\mathrm{opt}(q) = 0$ for all states $q$, then $\mathcal{A}$ has reach covering. This is the goal of the rank-reduction procedure below.

**Rank-reduction.** Set $\mathcal{A}_0 = \mathcal{A}$. For each $i \geq 0$, we perform the following three steps on $\mathcal{A}_i$ until $\mathcal{A}_{i+1} = \mathcal{A}_i$.

Step 1 For each state state $q$ in $\mathcal{A}_i$, compute the optimal rank of $q$ in $\mathcal{G}_1(\mathcal{A}_i)$, which we denote $\mathrm{opt}_i(q)$.

Step 2 Obtain $\mathcal{A}_i'$ from $\mathcal{A}_i$ by removing all transitions $q \xrightarrow{a:1} q'$ with $\mathrm{opt}_i(q) < \mathrm{opt}_i(q')$.

Step 3 Obtain $\mathcal{A}_{i+1}$ from $\mathcal{A}_i'$ by changing priorities of transitions $q \xrightarrow{a:1} q'$ with $\mathrm{opt}_i(q) > \mathrm{opt}_i(q')$ to be 0.

The above procedure stabilises since each step either deletes transitions or decreases the priority of certain transitions. We prove that the automaton $\mathcal{A}_I$ obtained at the end of this procedure is such that Eve wins the 1-token game from everywhere on it, $\mathcal{A}_I$ has $\mathsf{opt}(q) = 0$ for all states $q$, and $\mathcal{A}_I$ is simulation-equivalent to $\mathcal{A}$. It follows then that $\mathcal{A}_I$ has reach covering and hence is HD (Lemma 3.9), and from Lemma 2.8, so is $\mathcal{A}$. This concludes the proof of Theorem 3.8. We note, due to Theorem 3.1, we obtain the 2-token theorem for Büchi automata.

**Corollary 3.10.** *For every Büchi automaton $\mathcal{A}$, Eve wins the 2-token game on $\mathcal{A}$ if and only if $\mathcal{A}$ is history-deterministic.*

### 3.4 The Odd-to-Even Induction Step

We now build upon our proof of Theorem 3.8 to show the odd-to-even induction step, completing our proof of the 2-token theorem.

**Theorem 3.11.** *Let $K > 1$ be a natural number such that for every $[1, K]$ automaton $\mathcal{A}$, Eve wins the 2-token game on $\mathcal{A}$ if and only if $\mathcal{A}$ is HD. Then, for every $[0, K]$ automaton $\mathcal{A}$, Eve wins the 2-token game on $\mathcal{A}$ if and only if $\mathcal{A}$ is HD.*

We assume that the 2-token theorem holds for $[1, K]$ automata for some fixed $K \geq 2$ in the rest of this section. We will show, using this assumption, that every $[0, K]$ automaton on which Eve wins the 2-token game is HD.

We start by extending reach covering as a relation based on 2-token games on approximations of $[0, K]$ automata, similar to how we extended safe coverage to 1-safe coverage. More concretely, for a $[0, K]$ automaton $\mathcal{A}$, we define $\mathcal{A}_{>0}$ as the automaton in which all transitions of $\mathcal{A}$ of priority at least 1 are kept as is, and transitions of priority 0 are redirected to an accepting sink state and have their priority changed to 2. We have self-loops of priority 2 on each letter on the accepting sink state. Note that $\mathcal{A}_{>0}$ is a $[1, K]$ automaton.

**0-Reach covering.** We say that a $[0, K]$ automaton $\mathcal{A}$ has *0-reach covering* if for every state $q$ there is a state $p \in \mathsf{CR}^*(\mathcal{A}, q)$, such that Eve wins $G2(q; p, p)$ in $\mathcal{A}_{>0}$.

We can show, nearly identically to Lemma 3.9 for Büchi automata, that $[0, K]$ automata that have 0-reach covering and on which Eve wins the 2-token game from everywhere are HD.

**Lemma 3.12.** *Every $[0, K]$ automaton on which Eve wins the 2-token game from everywhere and that has 0-reach covering is history-deterministic.*

The challenge lies in modifying $[0, K]$ automata on which Eve wins the 2-token game from everywhere to automata that also have 0-reach covering. Unlike for Büchi automata, however, the 2-token games on $[0, K]$ automata are not parity games, but instead *Muller games*. A Muller game is a two-player game where every edge is labelled by a colour in $C$, and the winning condition is given by $\mathcal{F} \subseteq \mathcal{P}(C)$, a set of nonempty subsets of $C$. Eve wins an infinite play in this game if the set of colours visited infinitely often in that play is a set in $\mathcal{F}$. For the 2-token game, we have $C = [0, K] \times [0, K] \times [0, K]$ where the first component represents priorities of transitions on Eve's token, while the second and third component represent that of Adam's token. The set $\mathcal{F}$ consists of subsets of $S$ for which the following holds: if the lowest priority occurring in the $i^{th}$ component amongst elements of $S$ is even for $i = 2$ or $i = 3$, then this is also the case for $i = 1$.

Every Muller game $\mathcal{M}$ can be converted to a parity game $\mathcal{G}$ by taking product with the Zielonka tree $\mathcal{Z}_{C, \mathcal{F}}$ of its winning condition [DJW97, CCF21]. This is a rooted ordered tree with its root labelled by $C$. For a node labelled $X$, its children are nodes labelled by distinct maximal nonempty subsets $Y$ of $X$ such that $X$ is in $\mathcal{F}$ if and only if $Y$ is not in $\mathcal{F}$. The vertices of the parity game $\mathcal{G}$ are pairs $(m, \beta)$ where $m$ is a vertex in $\mathcal{M}$ and $\beta$ is a branch in $\mathcal{Z}_{C, \mathcal{F}}$, i.e., paths from the root to a leaf in $\mathcal{Z}_{C, \mathcal{F}}$.

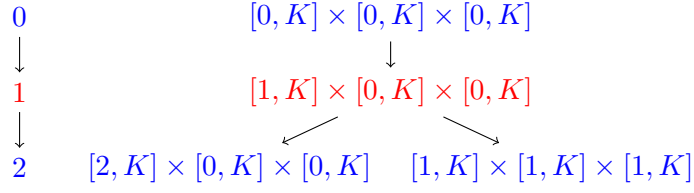

Figure 3: The first three layers of the Zielonka tree $\mathcal{Z}_{[0,K]}$

For our purposes, we only require an understanding of the first three levels of the Zielonka tree $\mathcal{Z}_{[0,K]}$ for the 2-token game on $[0,K]$ automaton, which are as shown in Fig. 3. In this tree, we will call a branch $\beta$ a right branch (resp. left branch) if it contains the node labelled $[1,K] \times [1,K] \times [1,K]$ (resp. $[2,K] \times [0,K] \times [0,K]$).

For a $[0,K]$ automaton $\mathcal{A}$ on which Eve wins the 2-token game from everywhere, we use $\mathcal{G}_2(\mathcal{A})$ to denote the parity game obtained by taking product of the 2-token game on $\mathcal{A}$ with $\mathcal{Z}_{[0,K]}$. The vertices of this game are $(q,p,r,\beta)$ where $q,p,r$ are weakly coreachable in $\mathcal{A}$ and $\beta$ is a branch of $\mathcal{Z}_{[0,K]}$. We leave the precise details of this game to Section 9 and instead state the following observation that follows from the structure of $\mathcal{Z}_{[0,K]}$.

**Proposition 3.13.** *Suppose* $rank(q,p,r,\beta) = 0$ *in* $\mathcal{G}_2(\mathcal{A})$ *for some weakly coreachable states* $q,p,r$ *in* $\mathcal{A}$ *and a branch* $\beta$ *of the Zielonka tree. Fix* $\sigma$ *to be an optimal winning strategy for Eve in* $\mathcal{G}_2(\mathcal{A})$. *Then the following two statements are true.*

1. *If* $\beta$ *is a left-branch, then the strategy* $\sigma$ *from* $\mathcal{G}_2(q,p,r,\beta)$ *in* $\mathcal{A}$ *ensures that a priority 0 transition occurs before the first priority 1 transition in the run of Eve's token.*

2. *If* $\beta$ *is a right-branch, then the strategy* $\sigma$ *from* $\mathcal{G}_2(q,p,r,\beta)$ *in* $\mathcal{A}$ *ensures that a 0-priority transition in Eve's token occurs earlier or in the same round as the first occurrence of a 0-priority transition in any of Adam's tokens.*

Let us define the optimal rank of state $q$ as

$$\text{opt}(q) = \min\{rank(q,p,r,\beta) \mid q,p,r \text{ are weakly coreachable and } \beta \text{ is a branch is } \mathcal{Z}_{[0,K]}\}.$$

If $\text{opt}(q) = rank(q,p,r,\beta) = 0$ for some right branch $\beta$ and $p,r \in \mathsf{CR}^*(\mathcal{A},q)$, then we say that $q$ is a *right state*. Otherwise, if $\text{opt}(q) = 0$ but $q$ is not a right state, then we say that $q$ is a *non-right* state. Note that Proposition 3.13 tells us that if $rank(q,p,r,\beta) = 0$ for some right branch $\beta$, then Eve wins $G2(q;p,r)$ in $\mathcal{A}_{>0}$ by choosing transitions using $\sigma$. Therefore, if all states are right states in $\mathcal{A}$, then for every state $q$, there are $p,r \in \mathsf{CR}^*(\mathcal{A},q)$, such that Eve wins $G2(q;p,r)$ in $\mathcal{A}_{>0}$. We can then show that $\mathcal{A}$ will have 0-reach covering in this case. Thus, the goal of the normalisation procedure we present below is to make all states right.

**The normalisation procedure.** Set $\mathcal{A}_0 = \mathcal{A}$. For each $i \geq 0$, we do the following three steps on $\mathcal{A}_i$ until $\mathcal{A}_i = \mathcal{A}_{i+1}$.

1. *Rank-reduction.* We modify the automaton $\mathcal{A}_i$ into a simulation equivalent automaton $\mathcal{B}_i$ so that $\text{opt}(q) = 0$ for each state $q$ in $\mathcal{G}_2(\mathcal{A}_i)$. This is similar to the rank-reduction procedure for Büchi automata.

2. *Branch-separation.* For the automaton $\mathcal{B}_i$, we remove the following transitions to obtain $\mathcal{C}_i$.

   (a) Transitions $q \xrightarrow{a:c} q'$ from a right state $q$ to a non-right state $q'$ in which the priority $c$ is at least 1.

   (b) Transitions of priority 1 that are outgoing from non-right states.

3. *Priority-modification.* The branch separation step might change the fact that all states have their optimal rank as 0. We nevertheless use the right (resp. non-right) states of $\mathcal{C}_i$ to refer to the states that were originally right (resp. non-right) in $\mathcal{B}_i$. For any transition $p \xrightarrow{a:c} p'$ in $\mathcal{C}_i$ where $p$ is an non-right state, if the priority $c$ is not 0 or 1, then we decrease the priority of that transition by 2. We let $\mathcal{A}_{i+1}$ be the automaton thus obtained.

The above iterative normalisation procedure terminates, since in each step we either delete transitions or decrease the priorities of certain transitions. We show that the automaton $\mathcal{A}_N$ obtained from this procedure is simulation-equivalent to $\mathcal{A}$, by showing that each of the above steps preserve simulation-equivalence to $\mathcal{A}$. Furthermore, we also show that all states in $\mathcal{A}_N$ are right states in $\mathcal{A}$, and thus, $\mathcal{A}_N$ has 0-reach covering. It follows that $\mathcal{A}_N$ is HD, and hence so if $\mathcal{A}$. This proves our odd-to-even induction step.

# 4 Applications and Concluding Remarks

Combining Theorem 3.5, Theorem 3.11, and Corollary 3.3 with a simple inductive argument proves the 2-token theorem.

**The 2-Token Theorem.** *For every nondeterministic parity automaton $\mathcal{A}$, Eve wins the 2-token game on $\mathcal{A}$ if and only if $\mathcal{A}$ is history-deterministic. Thus, the problem of deciding history-determinism is in $\mathbf{P}$ for parity automata with a fixed number of priorities, and in $\mathbf{PSPACE}$ if the number of priorities is part of the input.*

2-token games can be solved in polynomial time for a parity automaton with fixed parity index [BKLS20, Proposition 24]. If the parity index is not fixed, then the 2-token games can be solved in $\mathbf{PSPACE}$ since the winning condition of the 2-token game can be represented as an Emerson-Lei condition [HD05]. We refer the reader to the PhD thesis of Prakash for details on how to solve the 2-token game in $\mathbf{PSPACE}$ [Pra25, Section 3.5]. Due to [Pra25, Theorem 3.39], we also obtain the following upper bound for deciding history-determinism of parity automata.

**Theorem 4.1.** *There is an algorithm to decide history-determinism for $[0, d]$ parity automata $\mathcal{A}$ in time*

$$d \cdot (2^{3d} \cdot m^3)^{1+o(1)},$$

*where $m$ is the number of transitions in $\mathcal{A}$.*

Boker, Kuperberg, Lehtinen, and Skrzypczak have shown that if the 2-token conjecture is true for nondeterministic parity automata, then the 2-token games can also be used to decide history-determinism in alternating parity automata [BKLS20, Page 11]. Therefore, the above upper bounds also apply to alternating parity automata.

**Corollary 4.2.** *History-determinism of nondeterministic and alternating parity automata with a fixed parity index can be decided in $\mathbf{P}$, and in $\mathbf{PSPACE}$ in general.*

Casares, Idir, Kuperberg, Mascle, and Prakash [CIK+25, Lemma 36 in the full version] observe that if the 2-token conjecture is true for parity automata, then it is also true for $\omega$-regular parity automata: these are automata where each transition is labelled by a colour from a finite set of colours $C$. The acceptance condition for infinite runs is specified by a language $L \subseteq C^\omega$ that can be recognised by a nondeterministic Büchi automaton.

**Theorem 4.3.** *For every $\omega$-regular automaton $\mathcal{A}$, Eve wins the 2-token game on $\mathcal{A}$ if and only if $\mathcal{A}$ is history-deterministic.*

The exact complexity of solving 2-token games for an $\omega$-regular automata depends on the representation of the acceptance condition, but it is almost always more efficient to decide history-determinism via solving the 2-token game than via a direct approach which would involve determinising the automaton.

Prakash, in 2024, proved that checking history-determinism is **NP**-hard for parity automata, when the parity index is not fixed. Since our result only gives an upper bound of **PSPACE**, we still have a complexity gap.

**Open Problem 1.** *What is the exact complexity of deciding history-determinism for nondeterministic parity automata whose parity index is not fixed?*

We conjecture that the problem is **NP**-complete. More specifically, we conjecture that for every history-deterministic automaton $\mathcal{H}$, there is another equivalent history-deterministic automaton $\mathcal{H}'$ with as many or fewer states which satisfies certain properties that makes it easier to verify history-determinism of $\mathcal{H}'$. An **NP**-upper bound would then involve, to check the history-determinism of $\mathcal{H}$, guessing such an $\mathcal{H}'$ and the polynomial time certificate required to verify history-determinism of $\mathcal{H}'$, as well as positional strategies in simulation games between $\mathcal{H}$ and $\mathcal{H}'$.

Our work is a step towards such an approach, and in general, towards understanding history-deterministic parity automata. Indeed, since our proofs shows that every $[0, K]$ (resp. $[1, K]$) history-deterministic automaton $\mathcal{A}$ has a normal form in which the automata $\mathcal{A}_{>0}$ (resp. $\mathcal{A}_{>1}$) are history-deterministic from certain states, this leads us to believe that history-deterministic automata and strategies for Eve in the HD game are not too wild.

**Church synthesis and Good-enough synthesis** Church synthesis [Chu57, CHVB18], also known as reactive synthesis, has as ambition to produce a system that is correct by construction directly from a specification. It represents the interaction of a system with its environment as a game, in which, at each turn $i$, the environment produces an input letter $a_i$ from an input alphabet $I$, to which the system must respond with a letter from an $b_i$ from an output alphabet $O$. Then, the goal of the system is to guarantee that in the limit, the pair of words $(a_0 a_1 \ldots, b_0 b_1 \ldots)$ that describes the result of this interaction is in the specification $S \subset I^\omega \times O^\omega$. For example, in the synthesis of a scheduler whose task is to allocate resources to clients, the inputs might be requests for resources, the outputs might be resource allocations, and the specification defines what is a satisfactory schedule, with conditions such as every request being eventually answered and priorities being respected and so on. A winning strategy in this games corresponds to a system that guarantees the specification, whatever the behaviour of its environment. The *realisability problem* asks, given a specification $S$, whether such a system strategy exists. The synthesis problem also asks to produce said strategy when it exists.

One issue with Church synthesis is that a specification $S \in I^\omega \times O^\omega$ can only be realisable if for each input sequence $w_I \in I^\omega$, there is an output sequence $w_O \in O^\omega$ such that $(w_I, w_O) \in S$. Concretely, the specification of a safety critical coffee machine that requires it to produce a coffee whenever a user presses the button might turn out to be unrealisable because it is possible for users to refuse to refill the water canister. Then, no system behaviour could guarantee the satisfaction of the specification for *all* environment behaviours, and therefore we can not synthesise a coffee machine. Instead of giving up, we might like to synthesise a coffee machine that is guaranteed to work *as long as* the users are being reasonable, i.e. refill the water canister when required to do so. This is captured by a version of the Church synthesis problem, called *good-enough synthesis* [AK20], in which the system is only required to guarantee the specification for input words in the projection of $S$ onto the first component. As with Church synthesis, good-enough realisability is the decision problem asking whether there is such a system, while the good-enough synthesis problem is also interested in producing the corresponding strategy. A naive solution is to encode the condition on the environment into the specification, but this might make the specification exponentially larger.

The complexity of the good-enough realisability and synthesis problems from an LTL specification, as studied in [AK20], is dominated by the double-exponential determinisation procedure. However, if the specification is given directly as a deterministic parity automaton, then the complexity of the good-enough realisability problem is exactly the complexity of deciding whether a non-deterministic parity automaton is history-deterministic [BL23a]. Our results therefore directly imply a novel procedure to decide the good-enough realisability problem of deterministic parity automata.

**Corollary 4.4.** *Deciding whether the language of a deterministic parity automaton is good-enough realisable is in* **P** *for each fixed number of priorities and in* **PSPACE** *in general.*

# 5  Winning Token Games from Everywhere

We now start by giving the detailed proof of our result that allows us to assume without loss of generality, in the context of proving the 2-token theorem, that if Eve wins the 2-token game on $\mathcal{A}$, then Eve wins the 2-token game (or in general, the $k$-token game) from everywhere in $\mathcal{A}$.

**Theorem 3.1.** *Let $\mathcal{A}$ be a parity automaton on which Eve wins the 2-token game. Then, there is a simulation-equivalent subautomaton $\mathcal{B}$ of $\mathcal{A}$ such that the following two conditions hold.*

  1. *Eve wins $Gk(q; p_1, p_2, \ldots, p_k)$ in $\mathcal{B}$ for all $(k+1)$ states $q, p_1, p_2, \ldots, p_k$ that are weakly coreachable in $\mathcal{B}$.*

  2. *If $\mathcal{B}$ is history-deterministic, then so is $\mathcal{A}$.*

We remind the reader that we say that Eve wins the $k$-token game from everywhere in $\mathcal{B}$ if every tuple of states that can be reached via some play of the $k$-token game on $\mathcal{B}$ is winning for Eve, where $q_0$ is the initial state of $\mathcal{B}$.

## 5.1  Properties of token games

Recall that for parity automata $\mathcal{A}$ and $\mathcal{B}$, we write $G1(\mathcal{A}; \mathcal{B})$ if Eve wins $G1(\mathcal{A}; \mathcal{B})$. We show that the $G1$ games are transitive.

**Lemma 2.4.** *Let $\mathcal{A}, \mathcal{B},$ and $\mathcal{C}$ be parity automata such that Eve wins $G1(\mathcal{A}; \mathcal{B})$ and $G1(\mathcal{B}; \mathcal{C})$. Then Eve wins $G1(\mathcal{A}; \mathcal{C})$.*

*Proof.* Fix a winning strategy $\sigma_{AB}$ for Eve in $G1(\mathcal{A}; \mathcal{B})$, and a winning strategy $\sigma_{BC}$ for Eve in $G1(\mathcal{B}; \mathcal{C})$. We describe a winning strategy $\sigma_{AC}$ for Eve in $G1(\mathcal{A}; \mathcal{C})$ as follows. Eve stores in her memory, a token that takes transitions in $\mathcal{B}$. In $G1(\mathcal{A}; \mathcal{C})$, she will choose transitions her memory tokens according to her strategy $\sigma_{BC}$ against Adam's token in $\mathcal{C}$, and choose transitions on her token in $\mathcal{A}$ against her memory token via $\sigma_{AB}$. Then, if Adam's run in $\mathcal{C}$ is accepting, the run produced by her memory token in $\mathcal{B}$ is accepting since $\sigma_{BC}$ was a winning strategy for Eve in $G1(\mathcal{B}; \mathcal{C})$. Since $\sigma_{AB}$ is a winning strategy for Eve in $G1(\mathcal{A}; \mathcal{B})$, it follows that the run on her token is accepting as well. □

We continue with the following results on multiple token games, which follow easily or can be proved similarly to Lemma 2.4.

**Lemma 5.1.** *Let $\mathcal{A}, \mathcal{B}, \mathcal{B}', \mathcal{C}, \mathcal{C}'$ be nondeterministic parity automata. Then the following statements hold.*

  1. *If Eve wins $G2(\mathcal{A}; \mathcal{B}, \mathcal{C})$, then Eve wins $G2(\mathcal{A}; \mathcal{C}, \mathcal{B})$.*

  2. *If Eve wins $G2(\mathcal{A}; \mathcal{B}, \mathcal{C})$, then Eve wins $G1(\mathcal{A}; \mathcal{B})$ and Eve wins $G1(\mathcal{A}; \mathcal{C})$.*

*3. If Eve wins $G2(\mathcal{A}; \mathcal{B}, \mathcal{C})$ and $G1(\mathcal{B}; \mathcal{B}')$, then Eve wins $G2(\mathcal{A}; \mathcal{B}', \mathcal{C})$.*

*4. If Eve wins $G1(\mathcal{A}; \mathcal{A}')$ and $G2(\mathcal{A}'; \mathcal{B}, \mathcal{C})$, then Eve wins $G2(\mathcal{A}; \mathcal{B}, \mathcal{C})$.*

*5. If Eve wins $G1(\mathcal{A}; \mathcal{B})$, then $\mathcal{A}$ simulates $\mathcal{B}$.*

*Proof.* **(1)** This follows from symmetry.

**(2)** If Eve wins $G2(\mathcal{A}; \mathcal{B}, \mathcal{C})$ using strategy $\sigma$, then Eve can play $G1(\mathcal{A}; \mathcal{B})$ using $\sigma$ as if she is playing $G2(\mathcal{A}; \mathcal{B}, \mathcal{C})$ where Adam is picking transitions arbitrarily in $\mathcal{C}$. Then, if Adam's run on his token in $\mathcal{B}$ is accepting, so is the run of Eve's token in $\mathcal{A}$.

**(3)** Fix a winning strategy $\sigma_{ABC}$ for Eve in $G2(\mathcal{A}; \mathcal{B}, \mathcal{C})$ and a winning strategy $\sigma_{BB'}$ for Eve in $G1(\mathcal{B}; \mathcal{B}')$.

We describe a winning strategy $\sigma$ for Eve in $G2(\mathcal{A}; \mathcal{B}', \mathcal{C})$ as follows. Eve will keep in her memory a token in $\mathcal{B}$, on which she builds a run on Adam's word simultaneously with the run of her token. She chooses transitions on her memory token in $\mathcal{B}$ by using $\sigma_{BB'}$ against Adam's token in $\mathcal{B}'$, while she chooses transition on her token in $\mathcal{A}$ by playing $\sigma_{ABC}$ against her memory token in $\mathcal{B}$ and Adam's token in $\mathcal{C}$.

then if Adam's first token in $\mathcal{B}'$ constructs an accepting run, so does Eve's memory token in $\mathcal{B}$ (due to $\sigma_{BB'}$) and hence so does her token in $\mathcal{A}$ (due to $\sigma_{ABC}$). If Adam's token in $\mathcal{C}$ is accepting, then so does Eve's token in $\mathcal{A}$, due to $\sigma_{ABC}$.

**(4)** Fix a winning strategy $\sigma_{AA'}$ for Eve in $G1(\mathcal{A}; \mathcal{A}')$ and a winning strategy $\sigma_{A'BC}$ for Eve in $G2(\mathcal{A}'; \mathcal{B}, \mathcal{C})$.

We describe a winning strategy for Eve in $G2(\mathcal{A}; \mathcal{B}, \mathcal{C})$ as follows. Eve keeps in her memory a token in $\mathcal{A}'$ on which she builds runs on Adam's word simultaneously with the run of her token. She chooses transitions on her memory token in $\mathcal{A}'$ by playing $\sigma_{A'BC}$ against Adam's tokens in $\mathcal{B}$ and $\mathcal{C}$, and Eve chooses transitions on her token in $\mathcal{A}$ by playing $\sigma_{AA'}$ against her memory token in $\mathcal{A}'$. Then if the runs of Adam on his token in $\mathcal{B}$ or $\mathcal{C}$ is accepting, so is the run of Eve's memory token in $\mathcal{A}'$, and hence so is the run of Eve's token in $\mathcal{A}$.

**(5)** Eve can use her winning strategy in $G1(\mathcal{A}; \mathcal{B})$ to play in the simulation game between $\mathcal{A}$ and $\mathcal{B}$, where she ignores the additional information she gets by Adam picking a transition on his token earlier than Eve. $\qquad\square$

## 5.2 Coreachability and a useful pruning

Recall that we write $Gk(q; p_1, p_2, \ldots, p_k)$ in $\mathcal{A}$, where $\mathcal{A}$ is a parity automaton and $q, p_1, p_2, \ldots, p_k$ are states in $\mathcal{A}$, as a shorthand for the game $Gk((\mathcal{A}, q); (\mathcal{A}, p_1), (\mathcal{A}, p_2), \ldots, (\mathcal{A}, p_k))$. We will deal with token games on the same automata where the tokens are at *coreachable states* or *weakly-coreachable states*. We refer the reader to Section 2 for the definitions of coreachability and weak coreachability, and make the following two observations.

**Proposition 5.2.** *Let $p$ and $q$ be two coreachable states in a parity automaton $\mathcal{A}$. If $p'$ and $q'$ are states such that there are finite runs from $p$ to $p'$ and from $q$ to $q'$ on a finite word $u$, then $p'$ and $q'$ are coreachable in $\mathcal{A}$.*

*Proof.* Let $v$ be a finite word such that there are runs from the initial state of $\mathcal{A}$ to $p$ and $q$ on $v$. Then there are runs from the initial state of $\mathcal{A}$ to $p'$ and $q'$ on the word $vu$, and thus $p'$ and $q'$ are coreachable in $\mathcal{A}$. $\qquad\square$

We can extend the above result to weak coreachability.

**Proposition 5.3.** *Let $p$ and $q$ be two weakly coreachable states in a parity automaton $\mathcal{A}$. If $p'$ and $q'$ are states such that there are finite runs from $p$ to $p'$ and from $q$ to $q'$ on a finite word $u$ then $p'$ and $q'$ are weakly coreachable in $\mathcal{A}$.*

*Proof.* Let $q_1, q_2, \ldots, q_k$ be states such that $p$ and $q_1$ are coreachable, $q_i$ and $q_{i+1}$ are coreachable for every $i \in [1, k-1]$, and $q_k$ and $q$ are coreachable. For each $i \in [1, k]$, let $q_i'$ be a state such that there is a run from $q_i$ to $q_i'$ on $u$; note that such a state exists since we assume our automata to be complete.

then from Proposition 5.2, we know that $p'$ is coreachable to $q_1'$, $q_i'$ is coreachable to $q_{i+1}'$ for each $i \in [1, k-1]$, and $q_k'$ and $q'$ are coreachable. Thus, $p'$ and $q'$ are weakly coreachable. $\qquad \square$

**Remark 1.** *Suppose we call a binary relation $R$ between the states of an automaton $\mathcal{A}$ as* successor-closed *if the following two conditions hold.*

1. *$(q_0, q_0)$ is in $R$, where $q_0$ is the initial state of $\mathcal{A}$.*

2. *For every $(p, q) \in R$ and a letter $a$, all pairs $(p', q')$ such that there are transitions on $a$ from $p$ and $q$ to $p'$ and $q'$ respectively are also in $R$.*

*Then note that coreachability is the smallest relation that is successor-closed. Similarly, weak-coreachability is the smallest equivalence relation that is successor-closed.*

We now proceed towards proving Theorem 3.1. We start with the following result, which is the crux in our proof.

**Lemma 5.4.** *If Eve wins the 2-token game on $\mathcal{A}$, then there is a simulation equivalent subautomaton $\mathcal{B}$ of $\mathcal{A}$ on which Eve wins the 2-token game from everywhere.*

*Proof.* Since Eve wins the 2-token game on $\mathcal{A}$, she wins the 3-token game on $\mathcal{A}$, say using a winning strategy $\sigma_3$. Consider the strategy $\sigma_2$ for Eve in the 2-token game obtained from $\sigma_3$, in which she plays using $\sigma_3$ supposing Adam's third token is copying her token. Observe that $\sigma_2$ is a winning strategy for Eve in the 2-token game. Let $\Delta'$ be the set of all transitions $\delta$ for which there is a play $\rho$ where Eve is playing according to $\sigma_2$ such that Eve's token takes the transition $\delta$ at least once in $\rho$. We let $\mathcal{B}$ be the subautomaton of $\mathcal{A}$ consisting of the transitions in $\Delta'$.

We will show that $\mathcal{B}$ is the subautomaton that satisfies the desired properties. Let $q_0$ be the initial state of $\mathcal{A}$. We start by proving the following claim.

**Claim 1.** *For every finite word $u$ and states $p, r_1, r_2$ such that there are runs on $u$ from $q_0$ to $p$ in $\mathcal{B}$ and from $q_0$ to $r_1$ and to $r_2$ in $\mathcal{A}$, Eve wins $G3((\mathcal{B}, p); (\mathcal{A}, r_1), (\mathcal{A}, r_2), (\mathcal{A}, p))$.*

We will show the claim by induction on the length of $u$. For the base case when $|u| = 0$, that is, $u = \varepsilon$, we need to show that Eve wins $G3((\mathcal{B}, q_0); (\mathcal{A}, q_0), (\mathcal{A}, q_0), (\mathcal{A}, q_0))$. We know that Eve wins $G2((\mathcal{B}, q_0); (\mathcal{A}, q_0), (\mathcal{A}, q_0))$ using the strategy $\sigma_2$; note that $\sigma_2$ only takes transitions in $\mathcal{B}$. Thus, in particular, Eve wins $G1((\mathcal{B}, q_0); (\mathcal{A}, q_0))$ (Lemma 5.1.2). Since Eve wins $G3(q_0; q_0, q_0, q_0)$ in $\mathcal{A}$, we get using an extension of Lemma 5.1.4 to 3-token games that Eve wins $G3((\mathcal{B}, q_0); (\mathcal{A}, q_0), (\mathcal{A}, q_0), (\mathcal{A}, q_0))$, as desired.

Thus, assume that the claimed statement holds for words $u_k$ of length $k$. We will show that the claimed statement also holds for all words $u_{k+1}$ of length $k+1$. Let $u_{k+1} = u_k \cdot a$ such that length of $u_k$ is $k$ and $a$ is a letter in $\Sigma$. Let $p, r_1, r_2$ be states to which there is a run on $u_{k+1}$ in automata $\mathcal{B}, \mathcal{A},$ and $\mathcal{A}$ respectively from $q_0$. Then there are states $\text{pre}(p), \text{pre}(r_1), \text{pre}(r_2)$ such that $\text{pre}(p)$ (resp. $\text{pre}(r_1), \text{pre}(r_2)$) is reachable in $\mathcal{B}$ (resp. $\mathcal{A}$) on the word $u_k$, and $\text{pre}(p) \xrightarrow{a} p$ (resp. $\text{pre}(r_i) \xrightarrow{a} r_i$ for $i = 1, 2$) is a transition in $\mathcal{B}$ (resp. $\mathcal{A}$).

We know using the induction hypothesis that

$$\text{Eve wins } G3((\mathcal{B}, \text{pre}(p)); (\mathcal{A}, \text{pre}(r_1)), (\mathcal{A}, \text{pre}(r_2)), (\mathcal{A}, \text{pre}(p))). \tag{1}$$

Thus, there is a state $p'$ such that there is a transition from $\text{pre}(p)$ to $p'$ in $\mathcal{B}$ on $a$ and

$$\text{Eve wins } G3((\mathcal{B}, p'); (\mathcal{A}, r_1), (\mathcal{A}, r_2), (\mathcal{A}, p)). \tag{2}$$

Further, since $\mathrm{pre}(p) \xrightarrow{a} p$ is a transition in $\mathcal{B}$, we know there are states $q_1$ and $q_2$ that are weakly coreachable to $\mathrm{pre}(p)$ in $\mathcal{A}$ such that

$$\text{Eve wins } G3((\mathcal{B}, \mathrm{pre}(p)); (\mathcal{A}, q_1), (\mathcal{A}, q_2), (\mathcal{A}, \mathrm{pre}(p)))$$

and the strategy $\sigma_2$ chooses the transition $\mathrm{pre}(p) \xrightarrow{a} p$ on Eve's token. Since $\sigma_2$ is a winning strategy, we note that

$$\text{Eve wins } G3((\mathcal{B}, p); (\mathcal{A}, \mathrm{suc}(q_1)), (\mathcal{A}, \mathrm{suc}(q_2)), (\mathcal{A}, \mathrm{suc}(\mathrm{pre}(p)))) \tag{3}$$

for any states $\mathrm{suc}(q_1), \mathrm{suc}(q_2)$, and $\mathrm{suc}(\mathrm{pre}(p))$ that can be reached upon reading the letter $a$ from $q_1, q_2$ and $\mathrm{pre}(p)$ respectively in $\mathcal{A}$. In particular, Eve wins $G1((\mathcal{B}, p); (\mathcal{A}, p'))$. Since $\mathcal{B}$ is a subautomaton of $\mathcal{A}$, we note that Eve also wins $G1((\mathcal{B}, p); (\mathcal{B}, p'))$. Combining this with Eq. (2) using the extension of Lemma 5.1.4 to 3-token games, we get that

$$\text{Eve wins } G3((\mathcal{B}, p); (\mathcal{A}, r_1), (\mathcal{A}, r_2), (\mathcal{A}, p)), \tag{4}$$

thus completing our induction step. This completes the proof of our claim.

Note that since $\mathcal{B}$ is a subautomaton of $\mathcal{A}$, we know that $\mathcal{A}$ simulates $\mathcal{B}$. Furthermore, since Eve wins $G1((\mathcal{B}, q_0); (\mathcal{A}, q_0))$ (due to construction of $\mathcal{B}$), we note that $\mathcal{B}$ simulates $\mathcal{A}$ (Lemma 5.1.5). Therefore, $\mathcal{A}$ and $\mathcal{B}$ are simulation-equivalent.

The above claim also implies that Eve wins the 2-token game from everywhere in $\mathcal{B}$. Indeed, if $p, r_1, r_2$ are states that can be reached upon reading the same word from $q_0$ in $\mathcal{B}$, then the same holds for $\mathcal{A}$. Claim 1 then tells us that Eve wins $G2((\mathcal{B}, p); (\mathcal{A}, r_1), (\mathcal{A}, r_2))$. Since $\mathcal{B}$ is a subautomaton of $\mathcal{A}$, we obtain that Eve wins $G2((\mathcal{B}, p); (\mathcal{B}, r_1), (\mathcal{B}, r_2))$, as desired. $\qquad\square$

For $k \geq 1$ and a parity automaton $\mathcal{A}$, recall that winning the $k$-token game from everywhere is equivalent to winning the $k$-token game from all $(k+1)$ states $q, p_1, p_2, \ldots, p_k$ that are coreachable in $\mathcal{A}$. In the next three lemmas, we will extend the above result to winning $k$-token games from all weakly coreachable state for every $k \geq 1$. which will prove Theorem 3.1. We start with 1-token games.

**Lemma 5.5.** *Let $\mathcal{B}$ be an automaton on which Eve wins the 1-token game from everywhere. Then Eve wins the 1-token game from all pairs of weakly coreachable states in $\mathcal{B}$.*

*Proof.* This follows from the transitivity of $G1$-games and the fact that weak-coreachability is the transitive closure of coreachability. $\qquad\square$

We next extend the above to 2-token games

**Lemma 5.6.** *Let $\mathcal{B}$ be an automaton such that Eve wins the 2-token game from everywhere in $\mathcal{B}$. Then, Eve wins the 2-token game from all configurations of weakly coreachable states in $\mathcal{B}$.*

*Proof.* Let $p, q, r$ be weakly coreachable states in $\mathcal{B}$. Since Eve wins the 2-token game from everywhere in $\mathcal{B}$, she also wins the 1-token game from everywhere. Thus Eve wins $G2(p; p, p)$, and due to Lemma 5.5, Eve wins $G1(p; q)$ and Eve wins $G1(p; r)$. Using Lemma 5.1.3, we obtain that Eve wins $G2(p; q, r)$ in $\mathcal{B}$, as desired. $\qquad\square$

We then extend this to the $k$-token games for every $k \geq 1$, by combining it with Lemma 2.3.

**Lemma 5.7.** *Let $\mathcal{B}$ be an automaton such that Eve wins the 2-token game from everywhere in $\mathcal{B}$. Then, Eve wins the $k$-token game from all tuples $(q; p_1, p_2, \ldots, p_k)$ in $\mathcal{B}$, where $(q, p_1, \ldots, p_k)$ are weakly coreachable states in $\mathcal{B}$.*

*Proof.* Since Eve wins $G2(q; q, q)$ in $\mathcal{B}$, Eve also wins $Gk(q; q, q, \ldots, q)$ in $\mathcal{B}$ .(Lemma 2.3). Lemma 5.5 then tells us Eve wins $G1(q; p_i)$ for each $i \in [1, k]$. Combining this with an (extension of) Lemma 5.1.3, we obtain that Eve wins $Gk(q; p_1, p_2, \ldots, p_k)$ in $\mathcal{B}$. $\qquad \square$

Lemmas 2.8, 5.4 and 5.6 together then prove Theorem 3.1.

**Theorem 3.1.** *Let $\mathcal{A}$ be a parity automaton on which Eve wins the 2-token game. Then, there is a simulation-equivalent subautomaton $\mathcal{B}$ of $\mathcal{A}$ such that the following two conditions hold.*

1. *Eve wins $Gk(q; p_1, p_2, \ldots, p_k)$ in $\mathcal{B}$ for all $(k + 1)$ states $q, p_1, p_2, \ldots, p_k$ that are weakly coreachable in $\mathcal{B}$.*

2. *If $\mathcal{B}$ is history-deterministic, then so is $\mathcal{A}$.*

**Semantic-determinism**

We note that if $\mathcal{A}$ is an automaton such that Eve wins the 1-token game on $\mathcal{A}$ from everywhere, then $\mathcal{A}$ is *semantically-deterministic* [ARK23]. Semantically-deterministic automata are nondeterministic automata where from each state and each letter, all outgoing transitions from that state on that letter lead to language-equivalent states (we say states $p$ and $q$ are language-equivalent if $L(\mathcal{A}, p) = L(\mathcal{A}, q)$). This can easily be extended to finite words: if an automaton $\mathcal{A}$ is semantically deterministic, then all states that can be reached from $p$ upon reading $u$ are language-equivalent, for any state $p$ and finite word $u$.

**Lemma 5.8.** *For a parity automaton $\mathcal{A}$, if Eve wins the 1-token game from everywhere in $\mathcal{A}$, then $\mathcal{A}$ is semantically-deterministic. Additionally, if $p$ and $q$ are weakly coreachable states, then they are language-equivalent.*

*Proof.* Observe that if Eve wins $G1((\mathcal{A}, p); (\mathcal{A}, q))$ for some states $p$ and $q$, then $(\mathcal{A}, p)$ simulates $(\mathcal{A}, q)$ and hence $L(\mathcal{A}, q) \subseteq L(\mathcal{A}, p)$. Thus, if states $p$ and $q$ are weakly coreachable, then we get that $L(\mathcal{A}, p) = L(\mathcal{A}, q)$ since Eve wins $G1(p; q)$ and $G1(q; p)$ in $\mathcal{A}$ (Lemma 5.5). It follows that $\mathcal{A}$ is semantically-deterministic. $\qquad \square$

# 6 CoBüchi Automata

Boker, Kuperberg, Lehtinen, and Skrzypczak showed in 2020 that given a coBüchi or $[1, 2]$ automaton $\mathcal{A}$, Eve wins the 2-token game on $\mathcal{A}$ if and only if $\mathcal{A}$ is history-deterministic [BKLS20]. Their argument is fairly involved, and uses insights from the polynomial time algorithm Kuperberg and Skrzypczak gave in 2015 [KS15] and Bagnol and Kuperberg's result of 2018 [BK18] that shows the 2-token game characterisation of history-determinism on Büchi automata.

Their argument roughly goes as follows. They show that if Eve wins the 2-token game on an automaton $\mathcal{A}$, then $\mathcal{A}$ satisfies a certain property, (a simplification of) which we call safe coverage below. Towards a contradiction, they then assume that $\mathcal{A}$ is not history-deterministic, and hence Adam has a finite memory strategy in the HD game. They then show that Adam, using the safe coverage of $\mathcal{A}$ and his finite memory, can win the $N$-token game, where $N$ is doubly exponential in the size of $\mathcal{A}$. However, since, 2-token games and $N$-token games on $\mathcal{A}$ have the same winner, this is a contradiction.

In this section, we give a simpler proof for the 2-token game characterisation of history-determinism for coBüchi automata and show the following stronger statement.

**Theorem 3.2.** *Let $\mathcal{A}$ be a coBüchi automaton on which Eve wins the 1-token game from everywhere. Then, $\mathcal{A}$ is history-deterministic.*

We will then extend the ideas used in our proof of Theorem 3.2 to show our even-to-odd inductive step of the 2-token theorem: if 2-token game characterises history-determinism on $[0, K]$ automata for some $K \geq 1$, then 2-token game characterises history-determinism on $[1, K + 2]$ automata (Theorem 3.5).

Observe that Theorem 3.2 implies the 2-token game characterisation of history-determinism on coBüchi automata: if Eve wins the 2-token game on a coBüchi automaton $\mathcal{A}$, Theorem 3.1 allows us to assume without loss of generality that Eve wins the 2-token game from everywhere on $\mathcal{A}$, and hence also the 1-token game from everywhere on $\mathcal{A}$.

**Corollary 3.3.** ([BKLS20]**.**) *For every coBüchi automaton $\mathcal{A}$, Eve wins the 2-token game on $\mathcal{A}$ if and only if $\mathcal{A}$ is HD.*

We proceed towards proving Theorem 3.2. We first describe a *priority normalisation* procedure, which preserves the (non-)acceptance of infinite runs. Let $G_{>1}$ be the graph consisting of the 2-transitions of $\mathcal{A}$; let the 2-SCCs of $\mathcal{A}$ be the SCCs of $G_{>1}$. That is, a 2-SCC $S$ in $\mathcal{A}$ is a maximal subautomaton of $\mathcal{A}$ such that (i) all transitions in $S$ have priority 2, and (ii) for every two (not necessarily distinct) states $p, q$ in $S$, there is a path from $p$ to $q$ in $S$.

We say that a coBüchi automaton is *priority-normalised* if every 2-transition in that automaton occurs in a 2-SCC of that automaton.

Consider the coBüchi automaton $\mathcal{B}$ that is obtained from $\mathcal{A}$ by making 1 the priority of every 2-transition in $\mathcal{A}$ that is not in any 2-SCC of $\mathcal{A}$. The priorities of other transitions in $\mathcal{B}$ is the same as their respective priorities in $\mathcal{A}$ (see Fig. 4 for an illustration).

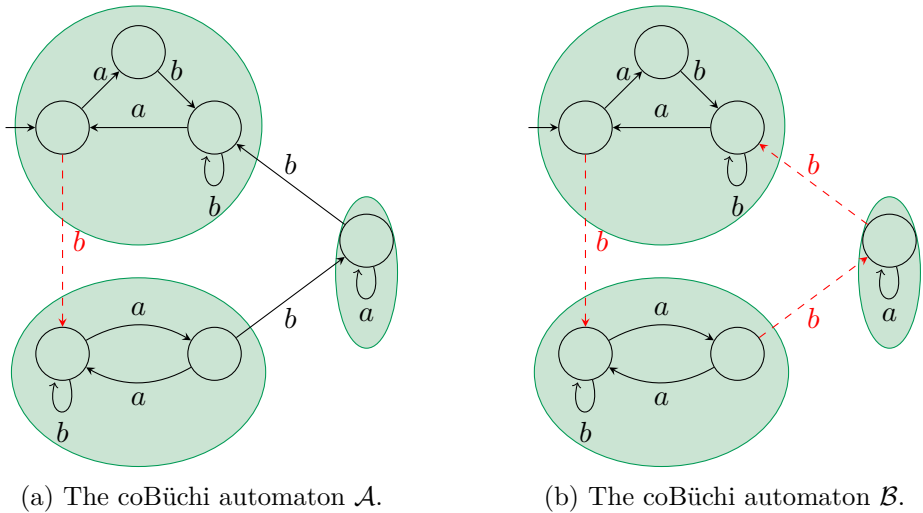

(a) The coBüchi automaton $\mathcal{A}$.   (b) The coBüchi automaton $\mathcal{B}$.

Figure 4: An illustration of the priority relabelling. The priority 2 transitions are represented by solid edges, while priority 1 transitions are represented by dashed edges. The 2-SCCs are represented by green ellipses. Note that in $\mathcal{B}$, all transitions outside 2-SCCs have priority 1.

It is clear that $\mathcal{B}$ is *priority-normalised*. We claim that this modification preserves the acceptance of runs. This follows from the following sequence of equivalent statements, where the equivalence of consecutive statements are easy to verify.

1. A run $\rho$ in $\mathcal{A}$ is accepting.

2. $\rho$ contains finitely many 1-transitions in $\mathcal{A}$.

3. $\rho$ eventually stays in a 2-SCC of $\mathcal{A}$.

4. $\rho$ contains finitely many 1-transitions in $\mathcal{B}$.

5. $\rho$ is an accepting run in $\mathcal{B}$.

In particular, if Eve wins the 1-token game from everywhere in $\mathcal{A}$, then Eve also wins the 1-token game from everywhere in $\mathcal{B}$.

**Proposition 6.1.** *For every coBüchi automaton $\mathcal{A}$, there is a simulation-equivalent automaton $\mathcal{B}$ that is priority-normalised.*

Let $\mathcal{A}_{\texttt{safe}}$ be the safety automaton obtained from a co Büchi $\mathcal{A}$ as follows. The automaton $\mathcal{A}_{\texttt{safe}}$ has all the states of $\mathcal{A}$ along with an additional rejecting sink state. The transitions of $\mathcal{A}$ that have priority 2 are preserved in $\mathcal{A}_{\texttt{safe}}$, and transitions of priority 1 are redirected to the rejecting sink state (see Fig. 5).

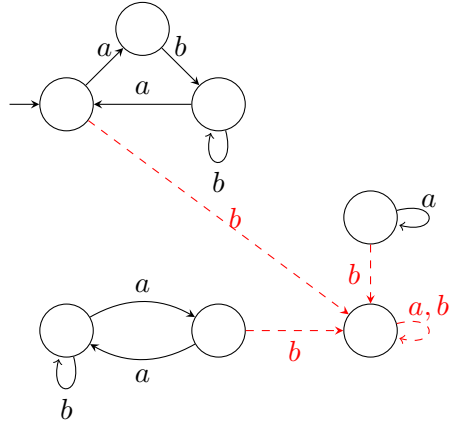

Figure 5: The safety automata $\mathcal{B}_{\texttt{safe}}$ for the automaton $\mathcal{B}$ in Fig. 4. Solid transitions have priority 2, while dashed transitions have priority 1.

We write that the automaton $\mathcal{A}$ has safe coverage if, for each state $p$, there is a state $q$ weakly coreachable to $p$ in $\mathcal{A}$ such that Eve wins $G1(q;p)$ in $\mathcal{A}_{\texttt{safe}}$. Recall that we showed in Section 3.1 that $\mathcal{A}$ has safe coverage.

**Lemma 3.4.** *Every coBüchi automaton on which Eve wins the 1-token game from everywhere and that is priority-normalised has safe coverage.*

The following observation on coBüchi automata with safe coverage follows from the transitivity of $G1$ games (Lemma 2.4) and the finiteness of the state-space.

**Lemma 6.2.** *If a coBüchi automaton $\mathcal{A}$ has safe coverage, then for every state $q$ in $\mathcal{A}$, there is a state $p$ in $\mathsf{CR}^*(\mathcal{A}, q)$ such that Eve wins $G1(p;q)$ in $\mathcal{A}_{\texttt{safe}}$ and $G1(p;p)$ in $\mathcal{A}_{\texttt{safe}}$.*

Recall that Eve wins the 1-token game on a safety automaton if and only it is determinisable-by-pruning, i.e., contains a language-equivalent deterministic subautomaton (Lemma 2.6). Thus, for a coBüchi automaton $\mathcal{A}$, we call a state $p$ of $\mathcal{A}$ *safe-deterministic* if Eve wins $G1(p;p)$ in $\mathcal{A}_{\texttt{safe}}$. The above lemma can then be restated as the following result.

**Lemma 6.3.** *For each state $p$ in $\mathcal{A}$, there is a state $q$ in $\mathsf{CR}^*(\mathcal{A}, p)$ such that Eve wins $G1(q;p)$ in $\mathcal{A}_{\texttt{safe}}$ and $q$ is safe-deterministic.*

Using Lemma 6.3, we now prove that a coBüchi automaton $\mathcal{A}$ on which Eve wins the 1-token game from everywhere is history-deterministic.

*Proof of Theorem 3.2.* We assume, without loss of generality due to Proposition 6.1, that $\mathcal{A}$ is priority-normalised. Fix $\sigma$ to be a uniform winning strategy for Eve in the HD game on $\mathcal{A}_{\texttt{safe}}$ from states that are safe-deterministic (Lemma 2.6). Let $\mathcal{D}_{\texttt{safe}}$ be the deterministic safety automaton obtained as the subautomaton of $\mathcal{A}_{\mathtt{safe}}$ consisting of states that are safe-deterministic and transitions that are in $\sigma$. Note that $L(\mathcal{D}_{\mathtt{safe}}, p) = L(\mathcal{A}_{\mathtt{safe}}, p)$ for every state $p$ that is safe-deterministic.

For a finite word $u$, we call a state $p$ *active* on $u$ if $p$ is in $\mathsf{CR}^*(\mathcal{A}, u)$ and $p$ is safe-deterministic. Note that from Lemmas 3.4 and 6.3, for every finite word $u$, there is some state that is active on $u$.

Let $w$ be a (finite or infinite) word. Consider then the DAG $D_w$ consisting of states of the form $(u, q)$, where $u$ is a finite prefix of $w$ and $q$ is a state in $\mathcal{A}$ that is active on $u$. We add an edge from $(u, q)$ to $(ua, q')$ in $D_w$ if $a \in \Sigma$, $u$ and $ua$ are prefixes of $w$, and $q \xrightarrow{a} q'$ is a transition in $\mathcal{D}_{\mathtt{safe}}$ such that $q'$ is not the rejecting sink state. Observe that each vertex in the DAG $D_w$ has outdegree at most 1 since $\mathcal{D}_{\mathtt{safe}}$ is deterministic.

We claim that if $w$ is an infinite word in $L(\mathcal{A})$ then the DAG $D_w$ contains an infinite path. Indeed, let $\rho$ be an accepting run of $\mathcal{A}$ on $w = uw'$ such that $\rho$ ends in some state $p$ upon reading the word $u$, after which $\rho$ does not contain a 1-transition. Then note that $w' \in L(\mathcal{A}_{\mathtt{safe}}, p)$. Let $q \in \mathsf{CR}^*(\mathcal{A}, p)$ be a state such that Eve wins $G1(q; p)$ in $\mathcal{A}_{\mathtt{safe}}$ and $q$ is safe-deterministic. Then

$$w' \in L(\mathcal{A}_{\mathtt{safe}}, p) \subseteq L(\mathcal{A}_{\mathtt{safe}}, q) = L(\mathcal{D}_{\mathtt{safe}}, q).$$

It is then clear that in DAG $D_w$, the unique path from $(u, q)$ is of infinite length.

Recall that if Eve wins the 1-token game from everywhere on $\mathcal{A}$, then she wins the 1-token game from all pairs of weakly coreachable states too (Lemma 5.5). We thus fix a uniform strategy $\sigma_{G1}$ for Eve in the 1-token game on $\mathcal{A}$ from all pairs of weakly coreachable states. Eve then has the following winning strategy in the HD game on $\mathcal{A}$.

At the start of the HD game on $\mathcal{A}$, Eve *tracks* a state $r_0$ in $\mathsf{CR}^*(\mathcal{A}, q_0)$ such that $r_0$ is safe-deterministic. In general, after Adam has played $u$ in the HD game on $\mathcal{A}$, Eve will be at some state $q_u$, and she will be *tracking* some state $r_u$ that is weakly coreachable to $q_u$ in $\mathcal{A}$ and is safe-deterministic. After Adam plays the letter $a$, Eve will then choose the $a$-transition on her token given by the strategy $\sigma_{G1}$ from the position of the 1-token game on $\mathcal{A}$ where Eve's token is at $q_u$ and Adam's token is at $r_u$. If $r_u$ has a transition to a state $r_{ua}$ in $\mathcal{D}_{\mathtt{safe}}$ that is not the rejecting sink state, or equivalently, there is an edge from $(u, r)$ to $(ua, r_{ua})$ in $D_{ua}$, then we let Eve track $r_{ua}$ for the next round. If there is no outgoing edge from $(u, r)$ in $D_{ua}$, then Eve *resets* to instead track a state $r'_{ua}$ such that the following holds: the path ending at $(ua, r'_{ua})$ has the longest length amongst the paths ending at vertices of the form $(ua, s)$ in $D_{ua}$ where $s$ is an active state of $ua$.

We claim that the strategy thus described is a winning strategy for Eve in the HD game on $\mathcal{A}$. Indeed, if Adam produces a word $w$ that is in $L(\mathcal{A})$, then the DAG $D_w$ contains an infinite path. Thus, after some point, Eve never resets the states she is tracking, and the states she is tracking constitutes an accepting run in $\mathcal{A}$. Since $\sigma_{G1}$ is a winning strategy for Eve from all pairs of weakly coreachable states, it follows that the run on Eve's token in the HD game is accepting run as well. This concludes our proof. $\qquad\square$

This concludes the proof that the 2-token game characterises history-determinism for coBüchi automata.

# 7 Climbing from Even to Odd

We proceed to show that if the 2-token game characterises history-determinism for $[0, K]$ automata, or equivalently $[2, K + 2]$ automata, then we can add a most important odd priority and extend the 2-token game characterisation of history-determinism to $[1, K + 2]$ automata.

**Theorem 3.5.** *Let $K \geq 1$ be a natural number, such that for every $[0,K]$ automaton $\mathcal{A}$, Eve wins the 2-token game on $\mathcal{A}$ if and only if $\mathcal{A}$ is HD. Then, for every $[1, K+2]$ automaton $\mathcal{A}$, Eve wins the 2-token game on $\mathcal{A}$ if and only if $\mathcal{A}$ is HD.*

Note that if an automaton is HD, then Eve wins the 2-token game on it, since Eve can pick transitions in the 2-token game using her strategy in the HD game, ignoring transitions of Adam's tokens. We need to show, that if Eve wins the 2-token game on a $[1, K+2]$ automaton, then that automaton is HD, provided that the 2-token game characterisation holds for $[0,K]$ automata. We proceed in the rest of this section with this hypothesis , for some fixed natural number $K \geq 1$.

**Hypothesis 7.1.** *For every $[0, K]$ automaton $\mathcal{A}$, Eve wins the 2-token game on $\mathcal{A}$ if and only if $\mathcal{A}$ is HD.*

We will prove Theorem 3.5 by using a similar template to the coBüchi case. We will start in Section 7.1 by introducing a convenient relabelling of priorities for $[1, K+2]$ automata, so that the automata we deal with are '2-priority reduced'. This relabelling of priorities will preserve the acceptance and rejection of each run.

In Section 7.2, we will introduce a property called 1-safe coverage, which is an extension of the property of safe coverage we introduced for coBüchi automata. We will show that every $[1, K+2]$ automaton on which Eve wins the 2-token game from everywhere and is 2-priority reduced has 1-safe coverage (Lemma 3.7). We will then show in Section 7.3 that every $[1, K+2]$ automaton with 1-safe coverage and on which Eve wins the 2-token game is history-deterministic (assuming Hypothesis 7.1), thus concluding the proof of Theorem 3.5.

## 7.1 2-priority reduction

In this section, we introduce a priority-normalisation procedure for $[0, K+2]$ automata, which is similar to the priority-normalisation procedure for coBüchi automata (Section 6).

For every $[1, K+2]$ automaton $\mathcal{A}$, we define the *2-SCCs* of $\mathcal{A}$ as the maximal strongly connected components in $\mathcal{A}$ consisting of transitions with priority at least 2. More precisely, a 2-SCC $\mathcal{S}$ of $\mathcal{A}$ is a maximal subautomaton of $\mathcal{A}$ satisfying the following two conditions.

1. All transitions in $\mathcal{S}$ have priority at least 2.

2. For every two (not necessarily distinct) states $p, q$ in $\mathcal{S}$, there is a path from $p$ to $q$ in $\mathcal{S}$.

Observe that the 2-SCCs of a $[0, K+2]$ automata are similarly defined to that for coBüchi automata in Section 6, where we considered SCCs that do not contain a priority 1 transition.

We say that a $[0, K+2]$ automaton $\mathcal{A}$ is *2-priority reduced* if the following two conditions are satisfied.

1. All transitions of priority at least 2 in $\mathcal{A}$ occur in some 2-SCC of $\mathcal{A}$.

2. Every 2-SCC has at least one transition of priority 2.

We refer the reader to Fig. 6 for an example of a 2-priority reduced automaton.

We now show that we can relabel the priorities of transitions of every $[1, K+2]$ automaton while preserving the acceptance and rejection of each run so that the new automaton is 2-priority reduced. This relabelling of priorities is done via the following iterative procedure.

**2-priority reduction.** Let $\mathcal{A}$ be a $[1, K+2]$ automaton, and let $\mathcal{A}_0 = \mathcal{A}$. For each $i \geq 0$, we perform the following two steps until $\mathcal{A}_{i+1} = \mathcal{A}_i$.

1. For every transition in $\mathcal{A}_i$ that does not appear in a 2-SCC of $\mathcal{A}_i$ we change its priority to 1, to obtain $\mathcal{A}'_i$.

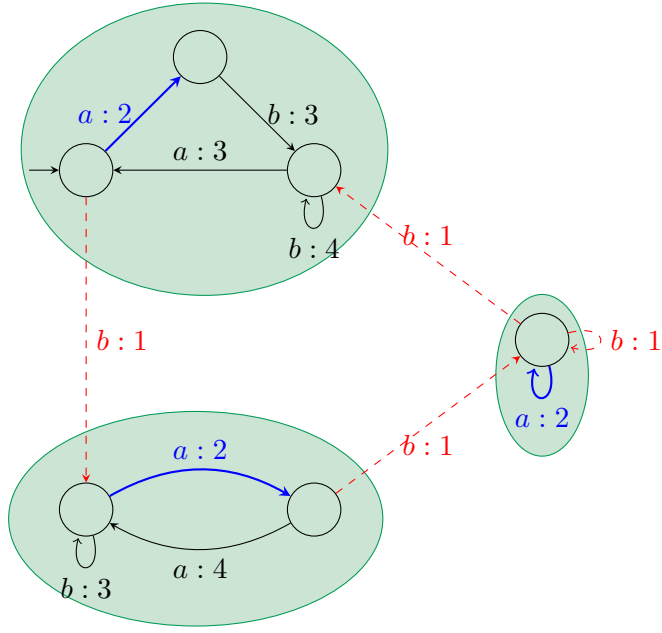

Figure 6: A 2-priority reduced automaton. Note that all transitions outside 2-SCCs (represented by green ellipses) have priority 1, and each 2-SCC has at least one transition of priority 2.

2. For every 2-SCC of $\mathcal{A}'_i$ in which the priorities of all transitions in it are strictly greater than 2, we decrease the priority of every transition in that 2-SCC by 2. We let $\mathcal{A}_{i+1}$ be the resulting $[1, K+2]$ automaton.

Observe that in each iteration of 2-priority reduction, we are only decreasing the priorities of certain transitions. Hence, the above procedure terminates. We first show that the 2-priority reduction procedure preserves the acceptance of each run, by showing that each iteration of the 2-priority reduction procedure preserves the acceptance of each run.

**Lemma 7.2.** *For each $i \geq 0$, each infinite run $\rho$ in $\mathcal{A}_i$ is accepting if and only if $\rho$ is accepting in $\mathcal{A}_{i+1}$.*

*Proof.* Let $\rho$ be a run in $\mathcal{A}_i$. We make a case distinction according to whether there are finitely of infinitely many priority 1 transitions occurring in $\rho$.

If $\rho$ contains finitely many transitions of priority 1 in $\mathcal{A}_i$, then $\rho$ eventually only visits transitions that are all in a single 2-SCC of $\mathcal{A}_i$. Since the priorities of transitions in this strongly connected component are either decreased for all transitions by 2 or unchanged in $\mathcal{A}_{i+1}$, the lowest priority appearing infinitely often in $\rho$ in $\mathcal{A}_i$ is even if and only if the lowest priority appearing infinitely often in $\rho$ in $\mathcal{A}_{i+1}$ is even. It follows that in this case, $\rho$ in an accepting run in $\mathcal{A}_i$ if and only if $\rho$ is an accepting run in $\mathcal{A}_{i+1}$.

If $\rho$ contains infinitely many transitions of priority 1 in $\mathcal{A}_i$, then it does the same in $\mathcal{A}_{i+1}$, since the priority 1 transitions remain unchanged. Since 1 is the lowest priority occurring in $\rho$ in both $\mathcal{A}_i$ and $\mathcal{A}_{i+1}$, $\rho$ is a rejecting run in both automata. $\qquad\square$

For a $[1, K+2]$ automaton $\mathcal{A}$, we use $\mathcal{A}{\Downarrow}$ to denote the automaton obtained via the 2-priority reduction procedure on $\mathcal{A}$. It then follows from Lemma 7.2 that every run $\rho$ is accepting in $\mathcal{A}$ if and only if $\rho$ is accepting in $\mathcal{A}{\Downarrow}$. We next show that $\mathcal{A}{\Downarrow}$ is 2-priority reduced, and that it suffices to reason with 2-priority reduced automata in this section.

**Lemma 7.3.** *Let $\mathcal{A}$ be $[1, K+2]$ automaton. Then:*

1. *$\mathcal{A}{\Downarrow}$ is 2-priority reduced, and*

2. *$\mathcal{A}$ is HD if and only if $\mathcal{A}\Downarrow$ is, and Eve wins the 2-token game from everywhere in $\mathcal{A}\Downarrow$ if and only if Eve wins the 2-token game from everywhere in $\mathcal{A}$.*

*Proof.* Since an iteration of the 2-priority reduction procedure has no effect on $\mathcal{A}\Downarrow$, $\mathcal{A}\Downarrow$ is 2-priority reduced. The second statement follows from the fact that a run is accepting in $\mathcal{A}$ if and only if that run is accepting in $\mathcal{A}\Downarrow$. □

## 7.2 1-safe coverage

Similar to how we defined safe coverage for coBüchi automata, we define an analogous property of 1-safe coverage for $[1, K+2]$ automata, which is based on the 2-token game relations on an approximation $\mathcal{A}$. Concretely, for a $[1, K+2]$ automaton $\mathcal{A}$, we define the 2-approximation of $\mathcal{A}$, denoted by $\mathcal{A}_{>1}$, as the automaton obtained by modifying $\mathcal{A}$, so that transitions of priority at least 2 are preserved, while transitions of priority 1 are redirected to a rejecting sink state $q_{\perp}$. The transitions leading to the rejecting sink state $q_{\perp}$ in $\mathcal{A}_{>1}$ (including the self-loops from $q_{\perp}$) have priority 3. Note that $\mathcal{A}_{>1}$ is a $[2, K+2]$ automaton.

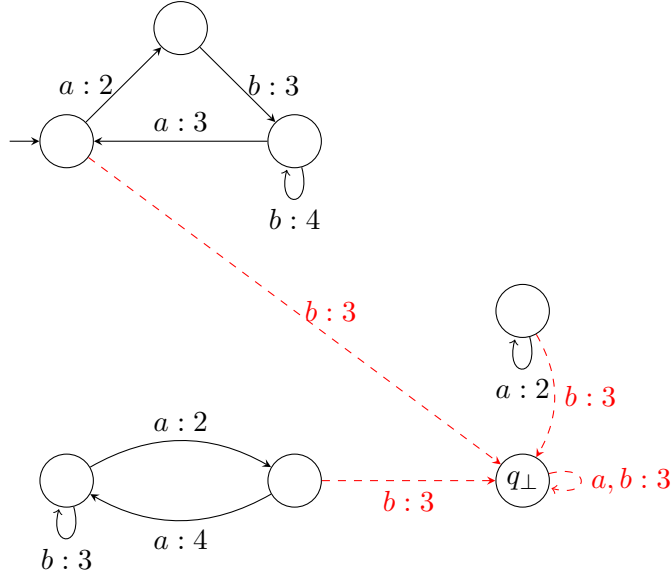

Figure 7: The 2-approximation of the automaton in Fig. 6.

For a $[1, K+2]$ automaton $\mathcal{A}$, we say that $\mathcal{A}$ has *1-safe coverage* if for every state $q$ in $\mathcal{A}$, there is a state $p$ weakly coreachable to $q$ in $\mathcal{A}$ such that Eve wins $G2(p; q, q)$ in $\mathcal{A}_{>1}$.

In the rest of this section, we prove the following result.

**Lemma 3.7.** *Every $[1, K+2]$ automaton that is 2-priority reduced and on which Eve wins the 2-token game from everywhere has 1-safe coverage.*

To show Lemma 3.7, we will also use the property of safe coverage that we had defined for coBüchi automata, and which we now define for $[1, K+2]$ automata. For a $[1, K+2]$ automaton $\mathcal{A}$, define the automaton $\mathcal{A}_{\texttt{safe}}$ as the safety automaton in which all transitions of $\mathcal{A}$ with priority at least 2 are made safe, i.e., relabelled to have priority 2, while transitions of priority 1 in $\mathcal{A}$ are redirected to a rejecting sink state in $\mathcal{A}_{\texttt{safe}}$, where the self loops on the rejecting sink state have priority 1. (For coBüchi automata $\mathcal{A}_{\texttt{safe}}$ and $\mathcal{A}_{>1}$ coincide.)

The following observation is easy to see.

**Proposition 7.4.** *For every $[1, K+2]$ automaton $\mathcal{C}$, $L(\mathcal{C}_{>1}) \subseteq L(\mathcal{C}_{\texttt{safe}})$ and $L(\mathcal{C}_{>1}) \subseteq L(\mathcal{C})$.*

*Proof.* If $\rho$ is an accepting run over some word $w$ in $\mathcal{C}_{>1}$, then the same run is also an accepting run in $\mathcal{C}_{\texttt{safe}}$ and $\mathcal{C}$. □

We say that a $[1, K+2]$ automaton $\mathcal{A}$ has *safe coverage* if for each state $q$ in $\mathcal{A}$, there is a state $p$ weakly coreachable to $q$ in $\mathcal{A}$ such that Eve wins $G1(p; q)$ in $\mathcal{A}_{\mathtt{safe}}$.

**Remark 2.** *Let us compare the definitions of safe coverage and 1-safe coverage. Firstly, the former is based on 1-token games in safety automata, and the latter on 2-token games in $[2, K+2]$ or $[0, K]$ automata. The reasoning behind going from safety to $[0, K]$ automata and to involve 2 tokens instead of 1-token games for our even-to-odd induction step is natural: similar to how we used the 1-token game characterisation of history-determinism on safety automata to obtain Theorem 3.2 for coBüchi automata, we would like to use the 2-token game characterisation of history-determinism on $[0, K]$ automata (Hypothesis 7.1) to show the 2-token game characterisation of history-determinism on $[1, K+2]$ automata.*

We next prove that every 2-priority reduced $[1, K+2]$ automaton on which Eve wins the 1-token game from everywhere has safe coverage. This is analogous to Lemma 3.4 for coBüchi automata, and is proved similarly: if $q$ is a state in $\mathcal{A}$ such that Adam wins $G1(p; q)$ in $\mathcal{A}_{\mathtt{safe}}$ for all $p$ in $\mathsf{CR}^*(\mathcal{A})$, then we show that Adam wins $G1(q; q)$ in $\mathcal{A}$. The only difference with the proof of Lemma 3.4 is in the construction of Adam's strategy in $G1(q; q)$ in $\mathcal{A}$, where whenever Eve's token takes a priority 1 transition and his token has not seen a 1-transition, Adam's token returns to $q$ via a 2-transition without ever taking a 1-transition. This is possible because $\mathcal{A}$ is 2-priority reduced.

**Lemma 3.6.** *Every $[1, K+2]$ automaton $\mathcal{A}$ that is 2-priority reduced and on which Eve wins the 1-token game from everywhere has safe coverage, i.e., for every state $p$ in $\mathcal{A}$ there is a state $q$ weakly coreachable to $p$ in $\mathcal{A}$, such that Eve wins $G1(q; p)$ in $\mathcal{A}_{\mathtt{safe}}$.*

*Proof.* Let $q$ be a state such that Adam wins $G1(p; q)$ in $\mathcal{A}_{\mathtt{safe}}$ for all states $p$ that are weakly coreachable to $q$ in $\mathcal{A}$. We will show that Adam wins $G1(q; q)$ in $\mathcal{A}$ by building a winning strategy $\sigma$ for Adam, which contradicts the fact that Eve wins the 1-token game from everywhere in $\mathcal{A}$.

From the position $(q, q)$ in $G1$ on $\mathcal{A}$, Adam chooses letters and transitions on his token according to his winning strategy for $G1(q; q)$ in $\mathcal{A}_{\mathtt{safe}}$, which will cause Eve's token to eventually take a transition of priority 1 in $\mathcal{A}$, while all priorities taken by Adam's token have priority at least 2. Due to $\mathcal{A}$ being 2-priority reduced, we know that the current position of Adam's token is in the same 2-SCC of $\mathcal{A}$ as $q$, and this 2-SCC contains a priority 2 transition. Then, after Eve's token sees 1, Adam picks letters and transitions on his token so that his token first takes a priority 2 transition and then returns to $q$, all while not seeing any priority 1 transition. Eve's token is then at some state $p$ that is weakly coreachable to $q$, and Adam's token is at $q$. Adam can then repeat the same strategy as above, since Adam wins $G1(p; q)$ in $\mathcal{A}_{\mathtt{safe}}$. Repeating this strategy infinitely many times causes the run of Eve's token to contain infinitely many priority 1 transitions, implying that the run on her token is rejecting. The lowest priority occurring infinitely often for Adam's token is 2 and hence Adam wins $G1(q; q)$ in $\mathcal{A}$. □

The next result follows from the transitivity of 1-token games (Lemma 2.4), the fact that there are finitely many states in $\mathcal{A}$, and the 1-token game characterisation of history-determinism on safety automata (Lemma 2.6).

**Lemma 7.5.** *Let $\mathcal{A}$ be a $[1, K+2]$ automaton with safe coverage. Then for every state $q$, there is a state $p$ that is weakly coreachable to $q$ in $\mathcal{A}$ such that $(\mathcal{A}_{\mathtt{safe}}, p)$ is HD and Eve wins $G1(p; q)$ in $\mathcal{A}_{\mathtt{safe}}$.*

*Proof.* We start by showing that for each $q$ in $\mathcal{A}$, there is a state $p$ in $\mathsf{CR}^*(\mathcal{A}, p)$ such that Eve wins $G1(p; p)$ in $\mathcal{A}_{\mathtt{safe}}$ and Eve wins $G1(p; q)$ in $\mathcal{A}_{\mathtt{safe}}$. The result then follows from the fact that safety automata on which Eve wins the 1-token game are history-deterministic (Lemma 2.6).

Let $q$ be a state in $\mathcal{A}$, and consider the directed graph $G$ whose vertices consist of the states in $\mathsf{CR}^*(\mathcal{A}, q)$. There is an edge in $G$ from vertices $r$ to $s$ if Eve wins $G1(s; r)$ in $\mathcal{A}_{\mathtt{safe}}$.

Since $\mathcal{A}$ has safe coverage, every vertex in $G$ has outdegree at least 1. Thus, for every vertex $r$, there is a vertex $s$ such that there is a path from $r$ to $s$ and a cycle consisting of the vertex $s$ in $G$. It then follows from transitivity of $G1$ (Lemma 2.4) that Eve wins $G1(s;r)$ in $\mathcal{A}_{\mathtt{safe}}$ and $G1(s;s)$ in $\mathcal{A}_{\mathtt{safe}}$. This concludes the proof of Lemma 7.5. $\qquad\square$

We now prove Lemma 3.7.

**Lemma 3.7.** *Every $[1, K+2]$ automaton that is 2-priority reduced and on which Eve wins the 2-token game from everywhere has 1-safe coverage.*

*Proof.* Let $\mathcal{A}$ be a $[1, K+2]$ automaton that is 2-priority reduced, on which Eve wins the 2-token game from everywhere, and suppose towards a contradiction that $\mathcal{A}$ does not have 1-safe coverage. We will show that Eve does not win the 3-token game from everywhere in $\mathcal{A}$, which will imply that Eve does not win the 2-token game from everywhere (due to Lemma 5.7). This is a contradiction, and thus will prove Lemma 3.7.

Since $\mathcal{A}$ does not have 1-safe coverage, there is a state $q$ in $\mathcal{A}$ such that for all states $p$ in $\mathsf{CR}^*(\mathcal{A}, q)$, Adam wins $G2(p; q, q)$ in $\mathcal{A}_{>1}$. Let $r$ be a state in $\mathsf{CR}^*(\mathcal{A}, q)$ such that $(\mathcal{A}_{\mathtt{safe}}, r)$ is HD and Eve wins $G1(r; q)$ in $\mathcal{A}_{\mathtt{safe}}$; we know that such an $r$ exists due to Lemma 7.5. Note that this implies that $(\mathcal{A}_{\mathtt{safe}}, r)$ simulates $(\mathcal{A}_{\mathtt{safe}}, q)$ (Lemma 5.1.5) and hence $L(\mathcal{A}_{\mathtt{safe}}, q) \subseteq L(\mathcal{A}_{\mathtt{safe}}, r)$.

Since Eve wins 1-token game from everywhere in $\mathcal{A}$, she wins the 1-token game from all pairs of weakly coreachable states in $\mathcal{A}$, due to the transitivity of 1-token games (Lemma 2.4). Thus, we can fix a uniform strategy $\sigma_{G1}$ for Eve in the 1-token game on $\mathcal{A}$ from all pairs of weakly coreachable states in $\mathcal{A}$, which we know exists due to Lemmas 2.5 and 5.5. We also fix a positional winning strategy $\sigma_{\mathtt{safe}}$ in the HD game on $(\mathcal{A}_{\mathtt{safe}}, r)$, which we know exists due to Lemma 2.6. We will show that Adam wins $G3(p; q, q, r)$ in $\mathcal{A}$, which implies that Eve does not win the 2-token game from everywhere in $\mathcal{A}$, as desired (Lemma 5.7).

**Claim 2.** *Adam wins $G3(p; q, q, r)$ in $\mathcal{A}$.*

We will prove the claim by describe a winning strategy $\sigma$ for Adam in $G3(q; q, q, r)$ in $\mathcal{A}$, where Adam stores two virtual tokens $m_1$ and $m_2$ in his memory, which take transitions in $\mathcal{A}_{>1}$. At the start of each round, Eve's token and Adam's tokens will all be at states that are weakly coreachable in $\mathcal{A}$. Adam's virtual tokens will be at either the rejecting sink state in $\mathcal{A}_{>1}$ or at some state in $\mathcal{A}_{>1}$ that is weakly coreachable to the state of Eve's token in $\mathcal{A}$.

Adam's tokens $t_1$ and $t_2$ will play in $G1$ against the virtual tokens $m_1$ and $m_2$, respectively, using $\sigma_{G1}$, whenever the corresponding virtual token is not in the rejecting sink state $q_\perp$ of $\mathcal{A}_{>1}$. If the token $m_i$ for $i = 1$ or 2 is in the rejecting sink state of $\mathcal{A}_{>1}$, then Adam will pick transition on token $t_i$ arbitrarily. We now describe how Adam chooses letters, the behaviour of these two virtual tokens, and the transitions on his third token.

Initially and at each *reset*, the first two virtual tokens $m_1$ and $m_2$ will be at the state $q$ in $\mathcal{A}_{>1}$, Eve's token will be at a state $p$ in $\mathsf{CR}^*(\mathcal{A}, p)$ while Adam's third token will be at the state $r$. Adam's first and second tokens will also be at some states $q_1$ and $q_2$ that are in $\mathsf{CR}^*(\mathcal{A}, p)$.

Until Eve's token takes a priority 1 transition in $\mathcal{A}$, Adam chooses transitions on his virtual tokens and letters according to a winning strategy for Adam in $G2(p; q, q)$ in $\mathcal{A}_{>1}$ against Eve's token. Adam chooses transitions on his third token $t_3$ using the strategy $\sigma_{\mathtt{safe}}$. This guarantees that the run on his third token does not take a 1-transition in $\mathcal{A}$: recall that $L(\mathcal{A}_{\mathtt{safe}}, q) \subseteq L(\mathcal{A}_{\mathtt{safe}}, r)$ since Eve wins $G1(r; q)$ in $\mathcal{A}_{\mathtt{safe}}$, and the finite word played by Adam so far must be the prefix of a word in $L(\mathcal{A}_{>1}, q) \subseteq L(\mathcal{A}_{\mathtt{safe}}, q)$ (Proposition 7.4), since Adam is choosing letters and transitions on his token according to his winning strategy in $G2(p; q, q)$ in $\mathcal{A}_{>1}$.

Whenever Eve's token takes a 1-transition, Adam changes his strategy and plays a word that allows $t_3$ to return to $r$ via a priority 2 transition and without seeing a priority 1 transition; this is possible due to the fact that $\mathcal{A}$ is 2-priority reduced. Then his strategy *resets* his virtual tokens $m_1$ and $m_2$ to be at $q$. Note that since $r$ and $p$ are weakly coreachable in $\mathcal{A}$, Eve's token must also at this point be in some state $p'$ such that is in $\mathsf{CR}^*(\mathcal{A}, q)$, which allows Adam to again play a winning strategy in $G2(p'; q, q)$ in $\mathcal{A}_{>1}$ at each reset. Adam's original tokens will also be in some states $q_1, q_2$ that are in $\mathsf{CR}^*(\mathcal{A}, q)$, and hence Adam can again choose transitions on his token $t_1$ and $t_2$ by playing according to $\sigma_{G1}$ against his virtual tokens $m_1$ and $m_2$ respectively.

This concludes the description of Adam's strategy. In every play where Adam is playing according to the above strategy, if the run on Eve's token contains infinitely many priority 1 transitions and therefore is rejecting, then Adam's token $t_3$ contains infinitely many priority 2 transitions and no priority 1 transition, and hence is accepting. Otherwise, if Eve's run eventually does not contain a priority 1 transition, then finitely many resets occur. Since Adam is playing according to a winning strategy in $G2(\mathcal{A}_{>1})$ on his memory tokens against Eve's token, the run on Eve's token is rejecting, while one of the memory tokens $m_i$ of Adam eventually produces an accepting run in $\mathcal{A}_{>1}$. The corresponding token of Adam where Adam is choosing transitions using $\sigma_{G1}$ against this memory token then constructs an accepting run in $\mathcal{A}$. Thus, Adam wins every play of $G3(q; q, q, r)$ in $\mathcal{A}$ where Adam is playing according to the above strategy, as desired. □

### 7.3 Automata with 1-safe coverage are HD

We prove in this section that every $[0, K+2]$ automaton that has 1-safe coverage and on which Eve wins the 2-token game from everywhere is history-deterministic.

**Lemma 7.6.** *Every $[1, K+2]$ automaton that has 1-safe coverage and on which Eve wins the 2-token game from everywhere is history-deterministic.*

Lemma 3.7 together with Lemma 7.6 will then imply Theorem 3.5. We start by proving the following result using Lemma 3.7, that is analogous to Lemma 7.5 for safe coverage, and is proved similarly.

**Lemma 7.7.** *Let $\mathcal{A}$ be a $[1, K+2]$ automaton that has 1-safe coverage. Then for every state $q$ in $\mathcal{A}$ there is a state $r$ weakly coreachable to $q$ in $\mathcal{A}$ such that Eve wins $G2(r; q, q)$ in $\mathcal{A}_{>1}$ and that $(\mathcal{A}_{>1}, r)$ is HD.*

*Proof.* Note that if Eve wins $G2(\mathcal{A}_1; \mathcal{A}_2, \mathcal{A}_2)$ and $G2(\mathcal{A}_2; \mathcal{A}_3, \mathcal{A}_3)$ for some parity automata $\mathcal{A}_1, \mathcal{A}_2, \mathcal{A}_3$, then Eve also wins $G2(\mathcal{A}_1; \mathcal{A}_3, \mathcal{A}_3)$ (Lemma 5.1.4). Thus, it follows, from the facts that $\mathcal{A}$ has finitely many states and that $\mathcal{A}$ has 1-safe coverage, that for every state $q$ in $\mathcal{A}$, there is a state $r$ in $\mathsf{CR}^*(\mathcal{A}, q)$ such that Eve wins $G2(r; q, q)$ in $\mathcal{A}_{>1}$ and that Eve wins $G2(r; r, r)$ in $\mathcal{A}_{>1}$. Hypothesis 7.1 then implies that $(\mathcal{A}_{>1}, r)$ is HD, from which the lemma follows. □

For a $[1, K+2]$ automaton $\mathcal{A}$ that has 1-safe coverage, call a state $q$ in $\mathcal{A}$ *1-safe HD* if $(\mathcal{A}_{>1}, q)$ is HD. We will use the winning strategies for Eve in the HD game on $(\mathcal{A}_{>1}, q)$ from states $q$ that are 1-safe HD to construct a winning strategy for Eve in the HD game on $\mathcal{A}$.

For coBüchi automata, recall that we had constructed, in the proof of Theorem 3.2, a winning strategy for Eve in the HD game by playing the 1-token game against safe-deterministic states. This approach does not work immediately: we do not have the luxury to play the 1-token game against a memory token following the HD strategy in $\mathcal{A}_{>1}$ from a 1-safe HD state that we 'track' until a 1-transition, since, unlike for coBüchi automaton, a run in $\mathcal{A}_{>1}$ for a $[1, K+2]$ automaton $\mathcal{A}$ that does not end at a rejecting sink state can still be rejecting.

We will therefore use *explorability*, introduced by Hazard and Kuperberg [HK23], as an intermediate step. For a natural number $k \geq 1$, a nondeterministic parity automaton is $k$-explorable if Eve wins the following $k$-HD game on it, where she has $k$ tokens instead of one in the HD game, with which she constructs $k$ runs. Her objective is to produce an accepting run on at least one of her tokens if Adam's word is in the language of the automaton.

**Definition 7.8** (*k*-Explorability). *For a nondeterministic parity automaton $\mathcal{B}$ and $k \geq 1$ a natural number, the $k$-HD game on $\mathcal{B}$ is played with $k$ tokens of Eve in $\mathcal{B}$. At the start of the game, Eve has $k$ tokens $t_1, t_2, \ldots, t_k$ in the initial state of $\mathcal{B}$. In round $i$ for each $i \in \mathbb{N}$,*

1. *Adam selects a letter $a_i$;*

2. *Eve moves each of her $l$ tokens along an $a_i$-transition from the tokens' current positions.*

*Eve wins a play of the $k$-HD game if and only if the following condition holds: if Adam's word is in $L(\mathcal{B})$, then at least one of the runs on Eve's $k$ tokens are accepting. We say that $\mathcal{B}$ is $k$-explorable if Eve has a winning strategy in the $k$-HD game on it. We say that $\mathcal{B}$ is explorable if $\mathcal{B}$ is $m$-explorable for some positive natural number $m$.*

Note that an automaton is HD if and only if it is 1-explorable. Hazard and Kuperberg observed that 2-token game characterises history-determinism on explorable automata, which was also their original motivation to study explorable automata: in the hope that it helps towards the resolution of the 2-token conjecture.

**Lemma 7.9** ([HK23, Pages 8-9])**.** *Let $\mathcal{B}$ be an explorable parity automaton. If Eve wins the 2-token game on $\mathcal{B}$, then $\mathcal{B}$ is HD.*

*Proof.* Let $\mathcal{B}$ be $k$-explorable for some $k \geq 1$. Since Eve wins the 2-token game on $\mathcal{B}$, Eve wins the $k$-token game on $\mathcal{B}$ as well (Lemma 2.3). Eve then has the following winning strategy in the HD game on $\mathcal{B}$. She stores in her memory $k$ tokens that are following a winning strategy in the $k$-HD game on $\mathcal{B}$, and constructs a run on her token in the HD game on $\mathcal{B}$ by choosing transitions according to a winning strategy in the $k$-token game on $\mathcal{B}$ against these $k$-memory tokens. □

Then, to prove Lemma 7.6, it suffices to prove the following.

**Lemma 7.10.** *Let $\mathcal{A}$ be a $[1, K+2]$ automaton that has 1-safe coverage and on which Eve wins the 2-token game from everywhere. Then $\mathcal{A}$ is explorable.*

*Proof.* For each state $q$ in $\mathcal{A}$, we know, due to Lemma 7.7, that there is a state $p$ in $\mathsf{CR}^*(\mathcal{A}, q)$ such that $(\mathcal{A}_{>1}, p)$ is HD and Eve wins $G2(p; q, q)$ in $\mathcal{A}_{>1}$.

Let $\sigma$ be a finite-memory strategy for Eve in the HD game of $\mathcal{A}_{>1}$, which is winning from all states $p$ that are HD in $\mathcal{A}_{>1}$, and let $\mathcal{M}$ be the memory of $\sigma$. Let $\mathcal{D}_{>1}$ be the deterministic automaton that is obtained by taking the product of $\mathcal{A}_{>1}$ with the strategy $\sigma$. That is, the states of $\mathcal{D}_{>1}$ are pairs $(p, m)$ such that $p$ is a state in $\mathcal{A}_{>1}$ such that $(\mathcal{A}_{>1}, p)$ is HD, and $m$ is a state in $\mathcal{M}$. There is a transition $(p, m) \xrightarrow{a:c} (p', m')$ in $\mathcal{D}_{>1}$ if in the HD game on $\mathcal{A}_{>1}$, when Eve's token is at $p$, her memory is $m$, and Adam plays the letter $a$, then $\sigma$ takes the transition $q \xrightarrow{a:c} q'$ in $\mathcal{A}_{>1}$ and updates its memory to $m'$.

We say that a state $(p, m)$ in $\mathcal{D}_{>1}$ is *active* on $u$ if $p$ is a state in $\mathsf{CR}^*(\mathcal{A}, u)$. Note that if $(p, m)$ is an active state in $\mathcal{D}_{>1}$, then $p$ is 1-safe HD.

Fix a positional winning strategy $\sigma_{G1}$ for Eve in the 1-token game on $\mathcal{A}$ from all pairs of weakly coreachable states. We will show that $\mathcal{A}$ is $k$-explorable, where $k$ is the number of states in $\mathcal{D}_{>1}$, by describing a winning strategy for Eve in the $k$-HD game on $\mathcal{A}$. At a high-level, each of Eve's tokens in the $k$-HD game on $\mathcal{A}$ will play in the 1-token game on $\mathcal{A}$ according to $\sigma_{G1}$, against a corresponding memory token that is taking transitions in $\mathcal{A}$ according to the projection of a corresponding run in $\mathcal{D}_{>1}$ on the $\mathcal{A}$ component.

In more details, let us denote Eve's tokens in this game by $t_1, t_2, \ldots, t_k$, and we suppose that her $k$-tokens $t_1 < t_2 < \cdots < t_k$ are ordered. We describe her strategy in the $k$-HD game on $\mathcal{A}$ inductively as the rounds proceed.

After Adam has read a finite word $u$ in the $k$-HD game on $\mathcal{A}$, we will have exactly one token of Eve *tracking* every state $(p, m)$ of $\mathcal{D}_{>1}$ such that $(p, m)$ is active on $u$. The tokens that are not tracking any active states are said to be *idle*.

When Adam plays the letter $a \in \Sigma$, Eve's tokens take transitions as follows:

1. If Eve's token $t$ at state $q$ is tracking $(p, m)$, then Eve, on the token $t$, takes the transition given by $\sigma_{G1}(q; p)$ in $G1(q; p)$ in $\mathcal{A}$ when Adam plays the letter $a$. Eve then updates the state that the token $t$ is tracking to be $(p', m')$, where $(p, m) \xrightarrow{a} (p', m')$ is the unique transition from $(p, m)$ on $\mathcal{A}$ in $\mathcal{D}_{>1}$. If $p'$ is the rejecting sink state in $\mathcal{A}_{>1}$, then token $t$ becomes idle and stops tracking $(p', m')$.

2. Otherwise if Eve's token $t$ is idle, then Eve takes an arbitrary outgoing transition on $a$ from the current state of $t$.

Eve then further updates the states of $\mathcal{D}_{>1}$ that her tokens are tracking as follows.

1. If $(p', m')$ is an active state on $ua$ in $\mathcal{D}_{>1}$ such that there are multiple tokens tracking $(p', m')$, we make all but the smallest token idle.

2. If there are active states of $\mathcal{D}_{>1}$ on $ua$ that are not tracked by any token, we pick one idle token for each of them and let them *track* those active states. Note that this is possible because Eve has as many tokens as there are states in $\mathcal{D}_{>1}$.

Thus, at the end of each round, exactly one Eve's token tracks an active state on $ua$. This concludes the description of Eve's strategy in the $k$-HD game on $\mathcal{A}$.

We claim that this strategy is winning for Eve in the $k$-HD game on $\mathcal{A}$. To prove this, suppose that Adam plays a word $w$ in $L(\mathcal{A})$ in a play of the $k$-HD game on $\mathcal{A}$ where Eve is playing according to the above strategy. Then, there is a run of $\mathcal{A}$ over $w$ that contains finitely many priority 1 transitions. Let $w = uw'$ be a decomposition of $w$ such that there is state $r$ of $\mathcal{A}$ reachable by $u$ from the initial state of $\mathcal{A}$ and $w' \in L(\mathcal{A}_{>1}, r)$. Then, by Lemma 7.7, there is a state $p$ in $\mathsf{CR}^*(u)$ such that Eve wins $G1(p; r)$ in $\mathcal{A}_{>1}$ and $(\mathcal{A}_{>1}, p)$ is HD. Since $L(\mathcal{A}_{>1}, p) \supseteq L(\mathcal{A}_{>1}, r)$, $w' \in L(\mathcal{A}_{>1}, p)$. Thus, from some state $(p, m)$ in $\mathcal{D}_{>1}$, the run $\rho_D$ of $\mathcal{D}_{>1}$ on $w'$ is accepting, and $(p, m)$ is active after $u$. Furthermore, for all prefixes $v$ of $w'$, the state $(p_v, m_v)$ that $\rho_D$ reaches after the prefix $v$ is not of the form $(q_\perp, m)$ where $q_\perp$ is the sink state in $\mathcal{A}_{>1}$, and thus $(p_v, m_v)$ is active.

Since the tokens are ordered, and whenever more than one tokens are tracking the same state in $\mathcal{D}_{>1}$ we let the smallest token continue to track that state, we note that eventually the same token $t$ uninterruptedly tracks states occurring in $\rho_D$. Then, since $\rho_D$ is accepting, the run on $t$ is also accepting because $\sigma_{G1}$ is a winning strategy. It follows that the strategy we have described is a winning strategy for Eve in the $k$-HD game on $\mathcal{A}$ and, therefore, $\mathcal{A}$ is explorable. $\qquad\square$

Lemma 7.6 follows, since 2-token games characterise history-determinism on explorable automata.

**Lemma 7.6.** *Every $[1, K + 2]$ automaton that has 1-safe coverage and on which Eve wins the 2-token game from everywhere is history-deterministic.*

We conclude with the proof of Theorem 3.5.

**Theorem 3.5.** *Let $K \geq 1$ be a natural number, such that for every $[0, K]$ automaton $\mathcal{A}$, Eve wins the 2-token game on $\mathcal{A}$ if and only if $\mathcal{A}$ is HD. Then, for every $[1, K + 2]$ automaton $\mathcal{A}$, Eve wins the 2-token game on $\mathcal{A}$ if and only if $\mathcal{A}$ is HD.*

*Proof.* We will show that if Eve wins the 2-token game on a $[1, K + 2]$ automaton $\mathcal{A}$ then $\mathcal{A}$ is history-deterministic, under the assumption that the same holds for $[0, K]$ automata (Hypothesis 7.1). By Theorem 3.1, we know that $\mathcal{A}$ has a simulation-equivalent subautomaton $\mathcal{B}$ such that Eve wins the 2-token game from everywhere in $\mathcal{A}$. Then Eve wins the 2-token game from everywhere in $\mathcal{B}{\Downarrow}$ (Lemma 7.3). By Lemma 3.7, $\mathcal{B}{\Downarrow}$ has 1-safe coverage, and it thus follows from Lemma 7.6 that $\mathcal{B}{\Downarrow}$ is HD. Thus, $\mathcal{B}$ is HD (Lemma 7.3) and due to the simulation-equivalence of $\mathcal{A}$ and $\mathcal{B}$, $\mathcal{A}$ is HD as well (Lemma 2.8). $\qquad\square$

## 8 Büchi Automata

As for coBüchi automata (Theorem 3.2), we shall show that any Büchi automaton on which Eve wins the 1-token game from everywhere is history-deterministic (Theorem 3.8). This follows from the 1-token game characterisation of history-determinism for semantically-deterministic Büchi automata shown by Acharya, Jurdziński, and Prakash in 2024 [AJP24, Theorem B], and Lemma 5.8. This proof is non-constructive, and is in similar spirit to the previous non-constructive proofs of the 2-token theorem for Büchi and coBüchi automata [BK18, BKLS20].

We will give a constructive proof of Theorem 3.8, i.e., starting from a strategy for Eve to win the 1-token game from all pairs of weakly-coreachable states in a Büchi automata, we construct a strategy for Eve in the HD game on that automaton. Our proof of this result borrows heavily from the work of Acharya, Jurdziński, and Prakash in 2024, where they showed that every history-deterministic Büchi automaton can be converted to a language-equivalent deterministic Büchi automaton with a quadratic blow up in state space [AJP24, Theorem C]. The arguments used in our proof serve as a warm-up for our proof of the odd-to-even induction step of the 2-token theorem (Theorem 3.11).

**Theorem 3.8.** *For every Büchi automaton $\mathcal{A}$, if Eve wins the 1-token game from everywhere on $\mathcal{A}$, then $\mathcal{A}$ is HD.*

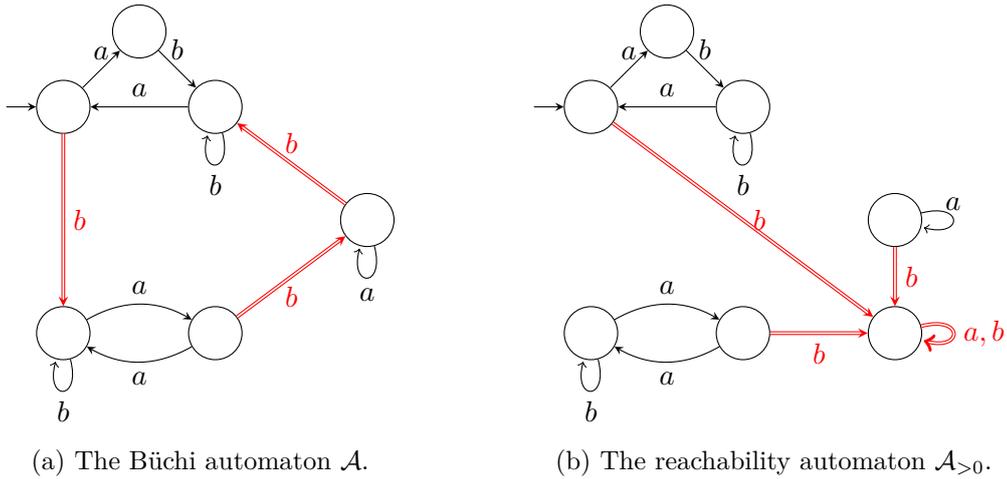

(a) The Büchi automaton $\mathcal{A}$.

(b) The reachability automaton $\mathcal{A}_{>0}$.

Figure 8: An illustration of a Büchi automaton and its 1-approximation. The double-arrowed transitions have priority 0, while the single-arrowed transitions have priority 1.

Analogous to how we defined 0-approximations for coBüchi or $[1, K+2]$ automata, we define the 1-*approximations* of Büchi automata. For a Büchi automaton $\mathcal{A}$, consider the automaton $\mathcal{A}_{>0}$ obtained by modifying $\mathcal{A}$ in the following way: transitions of priority 1 in $\mathcal{A}$ are kept as is, while transitions of priority 0 are redirected to an accepting sink state $q_{\top}$. The state $q_{\top}$ has self-loops of priority 0 on each letter in the alphabet. We say that $\mathcal{A}_{>0}$ is a 1-approximation of $\mathcal{A}$, and note that $\mathcal{A}_{>0}$ is a reachability automaton (see Fig. 8).

The following observation is easy to see.

**Proposition 8.1.** *For every Büchi automaton $\mathcal{A}$, $L(\mathcal{A}) \subseteq L(\mathcal{A}_{>0})$.*

*Proof.* If $\rho$ is an accepting run on some word $w$ in $\mathcal{A}$, then it must contain a priority 0 transition. Consider the run $\rho_{>0}$ of $\mathcal{A}_{>0}$ on $w$ that has the same transitions as the largest prefix of $\rho$ that contains only priority 1 transitions, and $\rho_{>0}$ has a transition to the accepting sink state at the position corresponding to the first priority 0 transition in $\rho$. Then $\rho_{>0}$ is an accepting run, as desired. □

Our proof of Theorem 3.8 is based on a property called reach covering that is based on the 1-token game relations between the states of 1-approximations of Büchi automata. This property is analogous to safe coverage for coBüchi automata. We show that any Büchi automaton that is semantically-deterministic— which is the case if Eve wins the 1-token game from everywhere (Lemma 5.8)—and has reach covering is history-deterministic. Then, we show that every Büchi automaton on which Eve wins the 1-token game from everywhere can be modified into a simulation-equivalent automaton that has reach covering. This proves Theorem 3.8.

## 8.1   Automata with reach covering

**Reach covering.**   We write that a Büchi automaton $\mathcal{A}$ has reach covering if for every state $q$, there is a state $p$ in $\mathsf{CR}^*(\mathcal{A}, q)$ such that Eve wins $G1(q; p)$ in $\mathcal{A}_{>0}$.

Let us compare the definitions of reach covering and safe coverage. Recall that for a coBüchi automaton $\mathcal{C}$, we say it has safe coverage if for each $q$ there is a $p$ such that Eve wins $G1(p; q)$ in $\mathcal{C}_{\mathtt{safe}}$. For reach covering, however, we note that the roles of $q$ and $p$ are reversed.

**Remark 3.** *The definition of reach covering is similar to the definition of sprint self-simulation introduced by Acharya, Jurdziński, and Prakash [AJP24] and to the self-dependency introduced by Kuperberg and Skrzypczak [KS15, Page 24 in full version], but there are subtle differences. They deal with G1 games on $\mathcal{A}$ where Eve's objective is to take a priority 0 transition on her token in no later round than the first time Adam takes a 0-transition. Our condition is weaker, however, and deals with G1 games that require Eve's token to eventually take an accepting transition if Adam's token takes an accepting transition.*

We proceed by showing the analogous statements of Lemmas 6.2 and 6.3 on coBüchi automata for Büchi automata. Due to transitivity of $G1$ (Lemma 2.4) and the fact that there are finitely many states in $\mathcal{A}$, we get the following result.

**Lemma 8.2.** *If $\mathcal{A}$ is a Büchi automaton with reach covering, then for each state $q$ in $\mathcal{A}$, there is a state $p$ in $\mathsf{CR}^*(\mathcal{A}, q)$ such that Eve wins $G1(q; p)$ and $G1(p; p)$ in $\mathcal{A}_{>0}$.*

*Proof.* This follows from the definition of reach covering, the transitivity of 1-token games (Lemma 2.4), and the fact that there are finitely many states in $\mathcal{A}$. Concretely, for each state $q$ in $\mathcal{A}$, consider the directed graph $G$ whose vertices are states in $\mathsf{CR}^*(\mathcal{A}, q)$. There is an edge in $G$ from vertices $r$ to $s$ if and only if Eve wins $G1(r; s)$ in $\mathcal{A}_{>0}$.

Since $\mathcal{A}$ has reach covering, every vertex in $G$ has outdegree at least 1. Thus, for every vertex $r$, there is a vertex $s$ such that there is a path from $r$ to $s$ and a cycle consisting of the vertex $s$ in $G$. Observe that if there is a path from $r$ to $s$ in $G$, then due to transitivity of $G1$ (Lemma 2.4), Eve wins $G1(r; s)$ in $\mathcal{A}_{>0}$. It follows that for each $r$ in $G$, there is an $s$ such that Eve wins $G1(r; s)$ and $G1(s; s)$ in $\mathcal{A}_{>0}$. This concludes our proof.  □

Recall however, that 1-token games characterise history-determinism on reachability automaton, and a reachability automaton is history-deterministic if and only if it is determinisable-by-pruning (Lemma 2.6). Call a state $q$ in $\mathcal{A}$ to be *reach-deterministic* if $(\mathcal{A}_{>0}, q)$ is determinisable by pruning. Then we can restate the above lemma as the following result.

**Lemma 8.3.** *If $\mathcal{A}$ is a Büchi automaton with reach covering, then for each state $q$ in $\mathcal{A}$, there is a state $p$ in $\mathsf{CR}^*(\mathcal{A}, q)$ such that Eve wins $G1(q; p)$ in $\mathcal{A}_{>0}$ and $p$ is reach-deterministic.*

We will use Lemma 8.3 above to show that every Büchi automaton that has reach covering and on which Eve wins the 1-token game from everywhere is history-deterministic.

**Lemma 3.9.** *Every Büchi automaton $\mathcal{A}$ on which Eve wins the 1-token game from everywhere and that has reach covering, is HD.*

*Proof.* Fix $\sigma_{>0}$ to be a positional strategy for Eve in the HD game on $\mathcal{A}_{>0}$ from states that are reach-deterministic. Let $\mathcal{D}_{>0}$ be the deterministic automaton obtained as the subautomaton of $\mathcal{A}_{>0}$ consisting of states that are reach-deterministic, and transitions that are in $\sigma_{>0}$. Note that for each state $p$ in $\mathcal{A}$ that is reach-deterministic, $L(\mathcal{D}_{>0}, p) = L(\mathcal{A}_{>0}, p)$.

Fix a uniform strategy $\tau_{G1}$ for Eve in the $G1$ game on $\mathcal{A}_{>0}$ from all pairs of states $q, p$ that are weakly coreachable in $\mathcal{A}$ such that Eve wins $G1(q; p)$ in $\mathcal{A}_{>0}$ and $p$ is reach-deterministic; we know such a strategy exists due to Lemmas 2.5 and 5.5. We will describe a winning strategy $\sigma_{HD}$ for Eve in the HD game using $\tau_{G1}$ and $\mathcal{D}_{>0}$.

Eve's strategy $\sigma_{HD}$ stores a token in her memory that takes transitions in $\mathcal{D}_{>0}$. When Eve's token is at some state $q$, her memory token will be at some state $p$ in $\mathcal{D}_{>0}$ such that either of the following two conditions hold.

1. Eve wins $G1(q; p)$ in $\mathcal{A}_{>0}$ and $p$ is the accepting sink state in $\mathcal{D}_{>0}$.

2. Eve Eve wins $G1(q; p)$ in $\mathcal{A}_{>0}$ and $p$ is in $\mathsf{CR}^*(\mathcal{A}, q)$.

When Adam chooses a letter $a$, the strategy $\sigma_{HD}$ in the HD game then picks the transition given by $\tau_{G1}$ on $G1(q; p)$ in $\mathcal{A}_{>0}$ if it is feasible to do so, i.e., the transition $\delta$ was not leading to the accepting sink state in $\mathcal{A}_{>0}$. The memory token $p$ takes the unique transition on $a$ in $\mathcal{D}_{>0}$.

Otherwise, if $\delta$ is a transition that leads to the accepting sink state in $\mathcal{A}_{>0}$, then there is a transition $q \xrightarrow{a:0} q'$ of priority 0 in $\mathcal{A}$, and Eve picks this transition in the HD game. Eve then *resets* her memory to be some $p' \in \mathcal{D}_{>0}$ such that $q'$ and $p'$ are weakly coreachable in $\mathcal{A}$, Eve wins $G1(q'; p')$ in $\mathcal{A}_{>0}$, and $p'$ is reach-deterministic.

We will show that the above strategy $\sigma_{HD}$ in winning for Eve in the HD game on $\mathcal{A}$. Suppose that in the HD game on $\mathcal{A}$, Adam plays a word $w \in L(\mathcal{A})$, and Eve builds the run $\rho_\mathcal{A}$ on her token using $\sigma_{HD}$. It suffices to prove that $\rho_\mathcal{A}$ is accepting, i.e., it contains infinitely many 0-transitions. We prove this inductively, by showing that there is at least one 0-transition in every suffix of $\rho_\mathcal{A}$.

To show this, suppose that after Adam has played a prefix word $u$ of $w$ such that $w = uv$, $\rho_\mathcal{A}$ is at state $q$, and that Eve's memory token is at state $p$. If the state $p$ is the accepting sink state in $\mathcal{A}_{>0}$, then the run on Eve's token from this point eventually takes a 0-transition, since $\tau_{G1}$ is a winning strategy.

Otherwise, $p$ is a state in $\mathsf{CR}^*(\mathcal{A}, q)$. Then note that

$$L(\mathcal{A}, q) = L(\mathcal{A}, p) = u^{-1}L(\mathcal{A}, q_0),$$

since $\mathcal{A}$ is semantically-deterministic (Lemma 5.8). Furthermore, due to Proposition 8.1,

$$L(\mathcal{A}, p) \subseteq L(\mathcal{A}_{>0}, p) = L(\mathcal{D}_{>0, p}).$$

Since $w = uv \in L(\mathcal{A})$, we note that, $v \in L(\mathcal{A}, p) \subseteq L(\mathcal{D}_{>0}, p)$. Thus the run on Eve's memory token in $\mathcal{D}_{>0}$ is accepting, and hence reaches the accepting sink state. Since Eve wins $G1(q; p)$ in $\mathcal{A}_{>0}$ using $\tau_{G1}$, it follows again that the run on Eve's token from this point takes at least one more 0-transition.

Thus, every suffix of $\rho_\mathcal{A}$ contains at least one 0-transition, and hence $\rho_\mathcal{A}$ infinitely many 0-transitions, as desired. $\qquad\square$

## 8.2 Iterating towards reach covering

We now describe a procedure to convert a nondeterministic Büchi automaton on which Eve wins the 1-token game from everywhere, into a simulation equivalent Büchi automaton that has reach covering. This involves considering the 1-token game as a $[0, 2]$ parity game, and then considering the *ranks* of the vertices in this game. This procedure is heavily inspired by a nearly-exact procedure presented in [AJP24], which in turn was inspired by a similar procedure of Kuperberg and Skrzypczak [KS15].

**The 1-token game as a parity game**

**Definition 8.4.** *For the Büchi automaton $\mathcal{A} = (Q, \Sigma, q_0, \Delta)$, define the $[0, 2]$ parity game $\mathcal{G}_1(\mathcal{A}) = (V, E)$ as follows.*

- *The set of vertices $V$ consists of the set $V = V_1 \cup V_2 \cup V_3$, where:*

    1. $V_1 = \{(q, p) \mid q, p \text{ are weakly coreachable}\}$
    2. $V_2 = \{(q, a, p) \mid (q, p) \in V_1 \text{ and } a \in \Sigma\}$
    3. $V_3 = \{(q', p, a) \mid (q, a, p) \in V_2 \text{ and } q \xrightarrow{a:c} q' \in \Delta\}$

    *Eve's vertices are $V_\exists = V_2$, while Adam's vertices are $V_\forall = V_1 \cup V_3$*

- *The set of edges $E$ is the union of following sets:*

    1. $E_1 = \{(q, p) \to (q, a, p) \mid a \in \Sigma\}$ *(Adam chooses a letter)*
    2. $E_2 = \{(q, a, p) \to (q', p, a) \mid q \xrightarrow{a:c\exists} q' \in \Delta\}$ *(Eve chooses a transition on her token)*
    3. $E_3 = \{(q', p, a) \to (q', p') \mid p \xrightarrow{a:c\forall} p' \in \Delta\}$ *(Adam chooses a transition on his token)*

- *The priorities on these edges are given as follows. All elements in $E_1$ have priority 2, while an edge $(q, a, p) \to (q', p, a)$ in $E_2$ has priority 0 if the transition $\delta = q \xrightarrow{a:c\exists} q'$ in $\mathcal{A}$ is accepting (or equivalently, $c\exists = 0$), and 2 otherwise. The edge $(q', p, a) \to (q', p')$ in $E_3$ has priority 1 if the priority of Adam's transition $p \xrightarrow{a:c\forall} p'$ is accepting, and 2 otherwise.*

We note that Eve wins the 1-token game $G1(p; q)$ in $\mathcal{A}$ if and only if Eve wins the above game $\mathcal{G}_1(\mathcal{A})$ from the vertex $(p, q)$ [BK18, Lemma 10].

**Lemma 8.5.** *For every Büchi automaton $\mathcal{A}$ and its states $q, p$ that are weakly coreachable, Eve wins $G1(q; p)$ in $\mathcal{A}$ if and only if Eve wins $\mathcal{G}_1(\mathcal{A})$ from the vertex $(q, p)$.*

*Proof.* We note that the game arena of the 1-token game on $\mathcal{A}$ and the game arena of $\mathcal{G}_1(\mathcal{A})$ (without its priorities) are syntactically equivalent. Thus, it suffices to prove that the winning plays in $G1(q; p)$ on $\mathcal{A}$ correspond to the winning plays in $\mathcal{G}_1(\mathcal{A})$ from $(q, p)$ and vice-versa.

To show this, we note that $\rho$ is a losing play for Eve in $\mathcal{G}_1(\mathcal{A})$ from $(q, p)$, if and only if, the priority 1 edges occur infinitely often and priority 0 edges occur finitely often in $\rho$, if and only if, the run on Adam's token contain infinitely many 0-transitions and is accepting while the run on Eve's token contain finitely many 0-transitions and is rejecting, if and only if, $\rho$ corresponds to a losing play in the 1-token game on $\mathcal{A}$. $\qquad\square$

**Ranks of parity games.** Our iterative procedure to obtain reach covering relies on modifying the automaton based on the *ranks* of the parity game $\mathcal{G}_1(\mathcal{A})$. These were introduced by Büchi in 1983 for a more general class of games [Büc83], and have been dubbed as signatures [Wal02], and progress measures [Jur00, JL17] over the years. Kuperberg and Skrzypczak called them ranks to give a determinisation procedure for history-deterministic Büchi automata [KS15], and we follow their precedent. We only require a restricted version of this concept for our purposes.

For a $[0, P]$ parity game $\mathcal{G}$ where $P \geq 1$, we define rank$(v)$ of a vertex $v$ as the largest number $k \in \mathbb{N} \cup \{\infty\}$ such that Adam has a strategy $\sigma$ in $\mathcal{G}$ that satisfies the following condition: in every play $\rho$ starting from $v$ where Adam is playing according to $\sigma$, at least $k$ many edges of priority 1 occur in $\rho$ before an edge of priority 0.

It is clear that if all vertices of $\mathcal{G}$ are in Eve's winning region, then rank$(v)$ is bounded for all vertices in $\mathcal{G}$. Furthermore, there is a positional winning strategy $\tau$ in $\mathcal{G}$ for Eve which ensures that any play from $v$ does not see more than rank$(v)$ 1-edges before a 0-edge.

**Lemma 8.6** ([Wal02, Lemma 8]). *Let $\mathcal{G}$ be a $[0, P]$ parity game such that Eve wins from all vertices in $\mathcal{G}$. Then there is a uniform strategy $\tau$ for Eve in $\mathcal{G}$ that ensures the following: in any play from $v$ where Eve is playing according to her strategy $\tau$, at most $rank(v)$ edges of priority 1 can occur before an edge of priority 0 occurs.*

We will call the strategy $\tau$ appearing in the above lemma *optimal strategy*. Let us observe that ranks, in a sense, are monotonic.

**Proposition 8.7** (Monotonicity of ranks). *Let $\mathcal{G}$ be a $[0, P]$ parity game where Eve wins from everywhere, and fix $\tau$ to be an optimal strategy in $\tau$. Then, for each vertex $u$ in $\mathcal{G}$, the following holds.*

1. *If $u \in V_\exists$ and $\tau(u) = u \xrightarrow{c} v$, then either $c = 0$, or $rank(u) \geq rank(v)$. This inequality is strict if $c = 1$.*

2. *If $u \in V_\forall$, then for all edges $u \xrightarrow{c} v$, either $c = 0$ or $rank(u) \geq rank(v)$. This inequality is strict if $c = 1$.*

Let us return to the game $\mathcal{G}_1(\mathcal{A})$, where $\mathcal{A}$ is a Büchi automaton such that Eve wins the 1-token game from everywhere. For each state $q$, define the optimal-rank of $q$, denoted by $opt(q)$ as

$$opt(q) = \min\{rank(q, p) \mid q \text{ and } p \text{ are weakly coreachable in } \mathcal{A}\}.$$

Suppose $rank(q, p) = 0$ for some weakly coreachable states $q$ and $p$ in $\mathcal{A}$. Then observe that Eve wins $G1(q; p)$ in $\mathcal{A}_{>0}$: indeed the optimal strategy $\tau$ ensures that Eve takes a priority 0 transition in her token before Adam does in $\mathcal{G}_1(\mathcal{A})$ from $(q, p)$. This corresponds to a strategy for Eve in $G1(q; p)$ in $\mathcal{A}_{>0}$ to reach the accepting sink state in $\mathcal{A}_{>0}$ no later than Adam. Thus, we obtain the following result.

**Proposition 8.8.** *If $opt(q) = 0$ for all states $q$ in $\mathcal{A}$, then $\mathcal{A}$ has reach covering.*

The iterative modification we carry out below, which we call the *rank-reduction procedure*, thus has the objective of making all optimal values 0.

**Rank-reduction procedure** Starting with a Büchi automaton $\mathcal{A}$ on which Eve wins the 1-token game from everywhere, we iteratively modify $\mathcal{A}$ to $\mathcal{A}_N$ so that $opt_N(q) = 0$ for each state $q$ in $\mathcal{A}_N$.

Set $\mathcal{A}_0 = \mathcal{A}$. For each $i \geq 0$, we perform the following three steps on $\mathcal{A}_i$ until $\mathcal{A}_{i+1} = \mathcal{A}_i$.

Step 1 For each state state $q$ in $\mathcal{A}_i$, compute the optimal rank of $q$ in $\mathcal{G}_1(\mathcal{A}_i)$, which we denote $opt_i(q)$.

Step 2 Obtain $\mathcal{A}'_i$ from $\mathcal{A}_i$ by removing all transitions $q \xrightarrow{a:1} q'$ with $opt_i(q) < opt_i(q')$.

Step 3 Obtain $\mathcal{A}_{i+1}$ from $\mathcal{A}'_i$, by changing priorities of transitions $q \xrightarrow{a:1} q'$ such that $opt_i(q) > opt_i(q')$ to 0.

Acharya, Jurdziński, and Prakash argue that these steps preserve the following two invariants (Lemmas 39 and 40 in the full version of [AJP24]). We will similarly argue that these invariants are preserved in Lemmas 9.14 and 9.15 for a similar rank-reduction procedure in Section 9.3, so we skip the details.

I1 Simulation-equivalence to the automaton $\mathcal{A}$.

I2 Eve winning the 1-token game from everywhere.

Observe that each iteration of the rank-reduction procedure either deletes transitions, or decreases the priority of certain transitions to 0. Thus, the rank-reduction procedure terminates after at most $|\Delta|$ many steps. Let $\mathcal{A}_N$ then be the automaton obtained by the rank-reduction procedure. We show that all states in $\mathcal{A}_N$ have optimal-rank 0.

**Lemma 8.9.** *For every state $q$ in $\mathcal{A}_N$, $opt_N(q) = 0$.*

*Proof.* Suppose that there exists a state $q$ such that $\mathrm{opt}_N(q) = \mathrm{rank}_N(q, p) > 0$. We will show that the rank-reduction procedure can then run for at least one more iteration on $\mathcal{A}_N$, which will be a contradiction to the fact that $\mathcal{A}_N$ is the automaton obtained after termination of the rank-reduction procedure.

Fix an optimal winning strategy $\tau$ for Eve in $\mathcal{G}_1(\mathcal{A}_N)$. Since $\mathrm{rank}_N(q, p) > 0$, there is a finite play $\rho$ of $\mathcal{G}_1(\mathcal{A}_N)$ starting from $(q, p)$ where Eve is playing according to $\tau$ such that $\rho$ contains an edge of priority 1 but no edge of priority 0. Note that in this play, Eve's token must take only priority 1 transitions since a priority 0 transition on her token corresponds to a priority 0 edge in $\mathcal{G}_1(\mathcal{A}_N)$.

By monotonicity of ranks (Proposition 8.7), we know that $\mathrm{rank}_N$ strictly decreases across $\rho$ at some point. Then there must be a transition on Eve's token in $\rho$ across which the quantity $\mathrm{opt}_N$ decreases as well, since $\rho$ started at $(q, p)$ such that $\mathrm{opt}_N(q) = \mathrm{rank}_N(q, p)$. But since such transitions are made accepting in Step 3 of the iteration, we get that the rank-reduction procedure can run for at least one more iteration on $\mathcal{A}_N$, as desired. $\square$

From Proposition 8.8, we obtain that $\mathcal{A}_N$ has reach covering, and hence is HD due to Lemma 3.9, assuming the invariants I1 and I2 hold. Since these invariants imply that $\mathcal{A}_N$ is simulation equivalent to $\mathcal{A}$, we obtain that $\mathcal{A}$ is HD, hence proving Theorem 3.8. We note that Theorem 3.8 together with Theorem 3.1 proves the 2-token theorem for Büchi automata.

**Corollary 3.10.** *For every Büchi automaton $\mathcal{A}$, Eve wins the 2-token game on $\mathcal{A}$ if and only if $\mathcal{A}$ is history-deterministic.*

**Remark 4.** *We remark that, for parity automata, Eve winning the 1-token game from everywhere does not imply history-determinism. A counterexample for this is the $[1, 3]$ automata presented in Fig. 1 [AJP24, Theorem 23].*

# 9 Climbing from Odd to Even

In this section, we show that if 2-token theorem holds for $[1, K]$ automata, then the 2-token theorem also holds for $[0, K]$ automata.

**Theorem 3.11.** *Let $K > 1$ be a natural number such that for every $[1, K]$ automaton $\mathcal{A}$, Eve wins the 2-token game on $\mathcal{A}$ if and only if $\mathcal{A}$ is HD. Then, for every $[0, K]$ automaton $\mathcal{A}$, Eve wins the 2-token game on $\mathcal{A}$ if and only if $\mathcal{A}$ is HD.*

Similar to how we argued at the start of Section 7, it suffices for us to show that under the hypothesis that the 2-token theorem holds for $[1, K]$ automata for some fixed $K > 1$, every $[0, K]$ automaton on which Eve wins the 2-token game is HD. We thus make the following assumption in the rest of this section.

**Hypothesis 9.1.** *For every $[1, K]$ automaton $\mathcal{A}$, Eve wins the 2-token game on $\mathcal{A}$ if and only if $\mathcal{A}$ is HD.*

Like for our even-to-odd induction step, we start by extending the property of reach covering used to prove the 2-token theorem on Büchi automata to 2-token games, which we call 0-reach covering, similar to how we extended the property of safe coverage for coBüchi automata to 1-safe coverage for the even-to-odd induction step. Then, showing that every semantically-deterministic $[0, K]$ automaton that has 0-reach covering is HD (Lemma 3.12) has a nearly identical proof to that for its Büchi analogue (Lemma 3.9). The crux in the proof of Theorem 3.11 is modifying $[0, K]$ automata on which Eve wins the 2-token game from everywhere into another simulation-equivalent automaton that additionally has 0-reach covering, and it requires more technical effort than the previous cases of coBüchi automata, Büchi automata, and our even-to-odd induction step.

The reason for this is that the 2-token game on $[0, K]$ automata are, unlike for Büchi automata, not parity games but Muller games. Our modification thus requires a careful analysis of the Zielonka tree of the 2-token game, which is a construction used to convert Muller games to winner-equivalent parity games [DJW97]. We then iteratively modify the automaton based on the ranks of the resulting parity game.

We start by defining 1-approximations for $[0, K]$ automata, similar to how we defined 1-approximations for $[0, K]$ automata. For a $[0, K]$ automaton $\mathcal{A}$, its *1-approximation*, denoted $\mathcal{A}_{>0}$, is the automaton that is obtained by preserving all transitions of $\mathcal{A}$ that have priority at least 1, and redirecting every priority 0 transition to an additional accepting sink state that has self loops with priority 2 on all letters in the alphabet of $\mathcal{A}$. Furthermore, we change the priority of the priority 0 transitions leading to the accepting sink state to be 2 (see Fig. 9). We note that $\mathcal{A}_{>0}$ is a $[1, K]$ automaton, since $K > 1$.

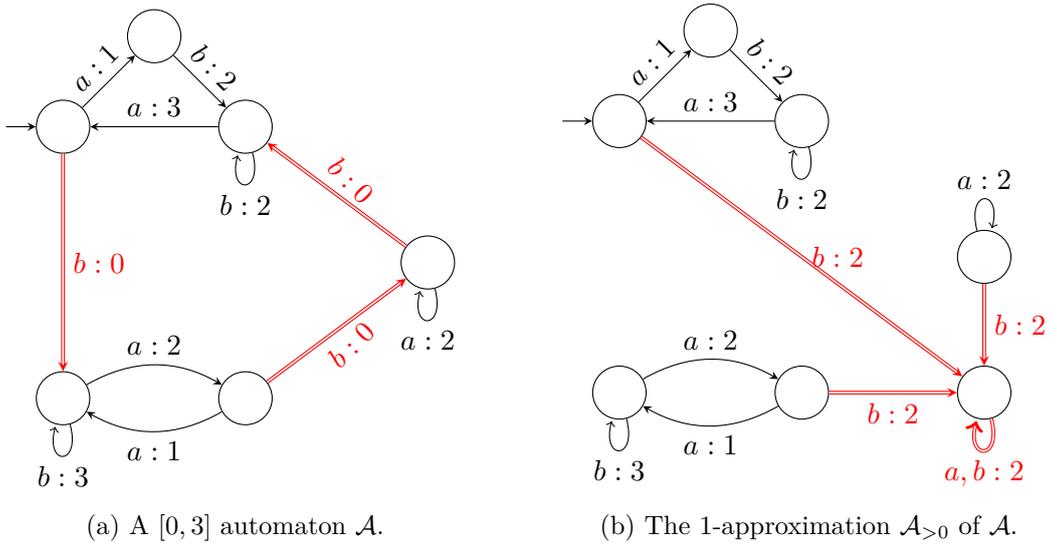

(a) A $[0, 3]$ automaton $\mathcal{A}$.

(b) The 1-approximation $\mathcal{A}_{>0}$ of $\mathcal{A}$.

Figure 9: An illustration of a $[0, 3]$ automaton and its 1-approximation. Note that $\mathcal{A}_{>0}$ is a $[1, 3]$ automaton.

The following observation is easy to see.

**Proposition 9.2.** *For every $[0, K]$ automaton $\mathcal{B}$, $L(\mathcal{B}) \subseteq L(\mathcal{B}_{>0})$.*

*Proof.* Let $\rho$ be an accepting run on some word $w$ in $\mathcal{B}$. Consider the run $\rho_{>0}$ of $\mathcal{B}_{>0}$ on $w$ that has the same transitions as the largest prefix of $\rho$ that contains only transitions of priority at least 1, and $\rho_{>0}$ has a transition to the accepting sink state at the position corresponding to the first priority 0 transition in $\rho$ (if such a transition exists). Then $\rho_{>0}$ is an accepting run in $\mathcal{B}_{>0}$, as desired. $\qquad \square$

**0-reach covering.** We say that a $[0, K]$ automaton $\mathcal{A}$ has 0-reach covering if for each state $p$, there is a state $q$ in $\mathsf{CR}^*(\mathcal{A}, p)$ such that Eve wins $G2(p; q, q)$ in $\mathcal{A}_{>0}$.

Observe that 0-reach covering is similar to 1-safe coverage for $[1, K+2]$ automata, but there is a crucial difference between the two. Namely, the roles of the states $q$ and $p$ in the $G2$ game has been reversed.

The rest of this section is organised as follows. In Section 9.1, we show that every $[0, K]$ automata that is semantically-deterministic and that has 0-reach covering is HD. Then, in Sections 9.2 to 9.6, we present our normalisation procedure to obtain automata with 0-reach covering. We start by detailing on Zielonka trees for 2-token games in Section 9.2. We then give an overview of the normalisation-procedure in Section 9.3, and describe the details of its subprocedures in Sections 9.4 and 9.5. Finally, in Section 9.6, we conclude with the correctness of our normalisation-procedure, which proves Theorem 3.11 and hence, the 2-token theorem.

## 9.1 Automata with 0-reach covering

Similar to Lemma 3.9 for Büchi automata, we show that every $[0, K]$ automaton that has 0-reach covering and on which Eve wins the 2-token game from everywhere is HD.

**Lemma 3.12.** *Every $[0, K]$ automaton on which Eve wins the 2-token game from everywhere and that has 0-reach covering is history-deterministic.*

We begin by observing that if Eve wins $G2(p; q, q)$ and $G2(q; r, r)$ in $\mathcal{A}_{>0}$, then Eve wins $G2(p; r, r)$ in $\mathcal{A}_{>0}$ (see Lemma 5.1). Since there are finitely many states in $\mathcal{A}$, we get that for each state $p$ in $\mathcal{A}$, there is a state $q$ in $\mathsf{CR}^*(\mathcal{A}, p)$ such that Eve wins $G2(p; q, q)$ Eve wins $G2(q; q, q)$ in $\mathcal{A}_{>0}$. Using Hypothesis 9.1, we obtain the following result that is proved analogously to Lemma 3.9 for Büchi automata.

**Lemma 9.3.** *Let $\mathcal{A}$ be a $[0, K]$ automaton with 0-reach covering. Then for every state $q$ in $\mathcal{A}$, there is a state $p$ in $\mathsf{CR}^*(\mathcal{A}, p)$ such that Eve wins $G1(q; p, p)$ in $\mathcal{A}_{>0}$ and $(\mathcal{A}_0, p)$ is HD.*

We will call a state $p$ *0-reach HD* if $(\mathcal{A}_{>0}, p)$ is HD. We now prove Lemma 3.12, which is very similar to the proof of Lemma 3.9 for Büchi automata.

*Proof of Lemma 3.12.* Fix $\sigma_{>0}$ to be a winning strategy for Eve in the HD game $\mathcal{A}_{>0}$ from all 0-reach HD states $p$, i.e., states $p$ such that $(\mathcal{A}_{>0}, p)$ is HD. We also fix a positional winning strategy $\sigma_{G1}$ for Eve in the $G1$ game on $\mathcal{A}_{>0}$ from all pairs of weakly coreachable states $q, p$ in $\mathcal{A}$ such that Eve wins $G1(q; p)$ in $\mathcal{A}_{>0}$ and $p$ is 0-reach HD (see Lemma 2.5). We will construct a winning strategy $\sigma_{HD}$ for Eve in the HD-game on $\mathcal{A}$ using $\sigma_{>0}$ and $\sigma_{G1}$.

In strategy $\sigma_{HD}$, when Eve's token is at a state $q$, she stores in her memory a token that will be at some state $p$ in $\mathcal{A}_{>0}$ such that Eve wins $G1(q; p)$ in $\mathcal{A}_{>0}$ and either of the following two conditions hold.

1. $p$ is 0-reach HD and in $\mathsf{CR}^*(\mathcal{A}, q)$.

2. $p$ is the accepting sink state in $\mathcal{A}_{>0}$.

Note that such a state $p$ exists for all states $q$ due to Lemma 9.3.

When Adam chooses a letter $a$, the strategy $\sigma_{HD}$ in the HD game on $\mathcal{A}$ then picks the transition given by $\sigma_{G1}$ on $G1(q; p)$ in $\mathcal{A}_{>0}$ if it is feasible to do so, i.e., the transition $\delta$ is not to the accepting sink state in $\mathcal{A}_{>0}$. The memory token $p$ takes the transition on $a$ given by $\sigma_{>0}$ in $\mathcal{A}_{>0}$.

Otherwise, if the transition $\delta$ leads to an accepting sink state in $\mathcal{A}_{>0}$, then we know there is a priority 0 transition from $q$ on the letter $a$ in $\mathcal{A}$. The strategy $\sigma_{HD}$ picks this transition $q \xrightarrow{a:0} q'$. We then reset the memory token to be at $p'$ such that $p'$ is weakly coreachable to $q'$ in $\mathcal{A}$, Eve wins $G1(q'; p')$ in $\mathcal{A}_{>0}$ and $p'$ is 0-reach HD.

We will show that the above strategy $\sigma_{HD}$ is winning for Eve in the HD game on $\mathcal{A}$. Indeed, let $\rho$ be a play of the HD game where Eve is playing according to $\sigma_{HD}$. Then, if the memory token is reset infinitely often in $\rho$, then the run on Eve's token contains infinitely many priority 0 transitions and is accepting.

Otherwise, suppose that the memory token is reset finitely often. Furthermore, suppose that after Adam has played a finite word $u$, Eve's token in the HD game is at state $q$, Eve's memory token is at state $p$, and the memory token is not reset after this point in the play. If the state $p$ is the accepting sink state in $\mathcal{A}_{>0}$, then the run on Eve's token is accepting since $\sigma_{G1}$ is a winning strategy. Otherwise if $p$ is in $\mathsf{CR}^*(\mathcal{A}, q)$, then

$$L(\mathcal{A}, q) = L(\mathcal{A}, p) = u^{-1}L(\mathcal{A}, q_0),$$

since $\mathcal{A}$ is semantically-deterministic (Lemma 5.8). Furthermore, observe that $L(\mathcal{A}, p) \subseteq L(\mathcal{A}_{>0}, p)$, due to Proposition 9.2. Thus, if Adam's word $w$ from this point is such that $uw \in L(\mathcal{A})$, or equivalently, $w \in L(\mathcal{A}, p) \subseteq L(\mathcal{A}_{>0}, p)$, then the run on Eve's memory token in $\mathcal{A}_{>0}$ is accepting. Since Eve's token is following transitions given by the winning strategy $\sigma_{G1}$, Eve's token either eventually takes a 0-transition and resets her memory token or she builds an accepting run on her token in $\mathcal{A}_{>0}$ that is also an accepting run in $\mathcal{A}$. Since Eve's memory token is not reset after Adam has played $u$ in the HD game on $\mathcal{A}$, we are in the latter case, and therefore, the run on Eve's token is accepting. We conclude that this strategy of Eve in the HD game on $\mathcal{A}$ is winning, and thus, $\mathcal{A}$ is HD. □

## 9.2 The Zielonka tree of the 2-token game

Having shown that $[0, K]$ automata on which Eve wins the 2-token game from everywhere and that have 0-reach covering are HD in the previous section (Lemma 3.12), we focus on showing the following result next.

**Lemma 9.4.** *Let $\mathcal{A}$ be a $[0, K]$ automaton on which Eve wins the 2-token game from everywhere. Then there is a $[0, K]$ automaton $\mathcal{A}^I$ such that $\mathcal{A}^I$ is simulation-equivalent to $\mathcal{A}$, Eve wins the 2-token game from everywhere on $\mathcal{A}^I$, and $\mathcal{A}^I$ has 0-reach covering.*

Lemmas 3.12 and 9.4 together would then prove Theorem 3.11. To prove Lemma 9.4, we will describe, in Section 9.3, a normalisation procedure for $[0, K]$ automata on which Eve wins the 2-token game from everywhere. This normalisation procedure is based on the ranks of the games obtained by converting 2-token games on $[0, K]$ automata into winner-equivalent parity games via Zielonka trees. In this section, we will focus on this conversion.

Unlike for Büchi automata, the 2-token game on a $[0, K]$ automaton $\mathcal{A}$ is not a parity game, but a *Muller game*. Similar to parity games, a Muller game $\mathcal{M}$ is a 2-player zero-sum turn-based game played on a finite graph, but instead of having priorities on edges, each edge is coloured by some *colour* from a finite set of colours $C$. The winning condition is specified by $\mathcal{F} \subseteq \mathcal{P}(C)$ as a set of non-empty subsets of colour from $C$. An infinite play in such a game is winning for Eve if the set of colours that occur infinitely often in the play belongs to $\mathcal{F}$. We call this winning condition as the $(C, \mathcal{F})$-Muller condition.

Every Muller game can be converted to a winner-equivalent parity game, as shown by Gurevich and Harrington [GH82]. We will use the conversion of Dziembowski, Jurdziński, and Walukiewikz that involve Zielonka trees [DJW97, CCF21].

**Definition 9.5** (Zielonka tree)**.** *Given a Muller condition $(C, \mathcal{F})$, the Zielonka tree of a Muller condition, denoted $\mathcal{Z}_{C, \mathcal{F}}$, is an ordered tree whose nodes are labelled by subsets of $C$, and is defined inductively. The root of the tree is labelled by $C$. For a node that is already constructed and labelled with the set $X$, its children are nodes labelled by distinct maximal non-empty subsets $Y \subsetneq X$ such that $X \in \mathcal{F} \Leftrightarrow Y \notin \mathcal{F}$. If there are no such $Y$, then the node labelled $X$ is a leaf of $\mathcal{Z}_{C, \mathcal{F}}$ and has no attached children.*

**Example 2.** *For a $(C, \mathcal{F})$ Muller condition where $C = \{1, 2, 3, 4\}$ and*

$$\mathcal{F} = \{\{1, 2, 3, 4\}, \{2, 3, 4\}, \{1, 2\}, \{2, 3\}, \{3, 4\}, \{1\}, \{2\}\},$$

*the Zielonka tree $\mathcal{Z}_{C, \mathcal{F}}$ is as shown in Fig. 10.*

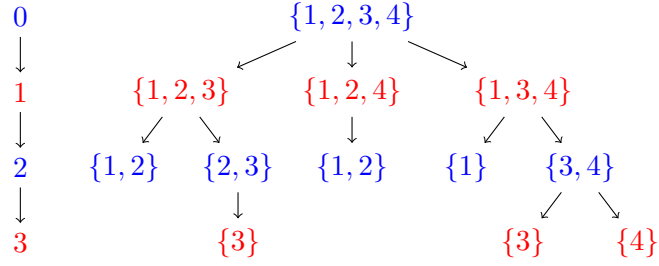

Figure 10: An example of a Zielonka tree

For a finite set of colours $C$ and a set of accepting subsets $\mathcal{F} \subseteq \mathcal{P}(\mathcal{C})$, the $(C, \mathcal{F})$ *Muller language* $L_{C,\mathcal{F}}$ is the language of words over $C^\omega$ in which the set of colours that appear infinitely often is a set in $\mathcal{F}$. We will use the Zielonka tree $Z_{C,\mathcal{F}}$ of the Muller condition $(C, \mathcal{F})$ to construct a deterministic parity automaton $\mathcal{D}_{C,\mathcal{F}}$ that recognises $L_{C,\mathcal{F}}$. We will then use this automaton $\mathcal{D}_{C,\mathcal{F}}$ to describe a conversion from Muller games to winner-equivalent parity games.

For a $(C, \mathcal{F})$-Muller condition, the states of $\mathcal{D}_{C,\mathcal{F}}$ are the *branches* of $\mathcal{Z}_{C,\mathcal{F}}$; a branch is a path in $\mathcal{Z}_{C,\mathcal{F}}$ from the root to a leaf. The initial state can be any branch of $\mathcal{Z}_{C,\mathcal{F}}$.

We now describe the transitions in $\mathcal{D}_{C,\mathcal{F}}$. On the branch $\beta$ of $\mathcal{Z}_{C,\mathcal{F}}$ and the letter $c$, the outgoing transition from $\beta$ on $c$ is given via the following steps.

1. We find the node $\nu = support(\beta, c)$ in $\beta$ that has the largest $\mathsf{prioritydepth}$ amongst nodes whose label contains $c$. We call $\nu$ as the support of $c$ in $\beta$.

2. If $\nu$ is a leaf, then we have the transition $\beta \xrightarrow{c:d} \beta$ in $\mathcal{D}_{C,\mathcal{F}}$ where $d = \mathsf{prioritydepth}(\nu)$.

3. Otherwise, let $\chi$ be the child of $\nu$ in $\beta$. Recall that the children of $\nu$ are ordered from left to right. Define $rightsibling(\chi)$ to the be the leftmost child of $\nu$ that is strictly to the right of $\chi$. If $\chi$ is the rightmost child of $\nu$, then $rightsibling(\chi)$ is the leftmost child of $\nu$.

4. Let $\beta'$ be the leftmost branch in $\mathcal{Z}_{C,\mathcal{F}}$ that contains the node $rightsibling(\chi)$. Then, we have the transition $\beta \xrightarrow{c:d} \beta'$ in $\mathcal{D}_{C,\mathcal{F}}$ where $d = \mathsf{prioritydepth}(\nu)$ is the priority-depth of $\nu$.

This concludes our description of $\mathcal{D}_{C,\mathcal{F}}$.

**Example 3.** *We illustrate a transition of $\mathcal{D}_{C,\mathcal{F}}$ for the Muller condition whose Zielonka tree is given in Fig. 10. Suppose that the current state is the leftmost branch $\beta$ in $\mathcal{Z}_{C,\mathcal{F}}$. On seeing the colour 4, we find the highest node in $\beta$ whose label contains 4: this is the support of 4 in $\beta$. In our example, this node happens to be the root of $\mathcal{Z}_{C,\mathcal{F}}$. The node labelled $\{1, 2, 3\}$ is the child of the root in $\beta$, whose rightsibling is the node labelled $\{1, 2, 4\}$. The new state, therefore, is the leftmost branch that contains $\{1, 2, 4\}$, i.e., the branch $\beta' = \{1, 2, 3, 4\} \rightarrow \{1, 2, 4\} \rightarrow \{1, 2\}$. Thus, we have the transition $\beta \xrightarrow{4:0} \beta'$, where we see priority 0 because 0 is the $\mathsf{prioritydepth}$ of support of 4 in $\beta$.*

**Lemma 9.6** ([CCF21, Proposition 3.9])**.** *For every Muller condition $(C, \mathcal{F})$ and a word $w$ in $C^\omega$, $w$ satisfies the $(C, \mathcal{F})$ Muller condition if and only if $w$ is accepted by $\mathcal{D}_{C,\mathcal{F}}$.*

The correctness of Lemma 9.6 does not depend on what we set the initial state of $\mathcal{D}_{C,\mathcal{F}}$ to be. The automaton $\mathcal{D}_{C,\mathcal{F}}$ allows us to convert a Muller game $\mathcal{M}$ with the winning condition as $(C, \mathcal{F})$-Muller condition to an equivalent parity game $\mathcal{G}$ by a product construction as follows. The vertices of $\mathcal{G}$ are given by $(m, \beta)$, where $m$ is a vertex in $\mathcal{M}$ and $\beta$ is a state in $\mathcal{D}_{C,\mathcal{F}}$. The game $\mathcal{G}$ contains the edge $(m, \beta) \xrightarrow{p} (m', \beta')$ if and only if $m \xrightarrow{c} m'$ is an edge in $\mathcal{M}$ and $\beta \xrightarrow{c:p} \beta'$ is the transition of $\mathcal{D}_{C,\mathcal{F}}$ from $\beta$ on $c$. The ownership of vertices in $\mathcal{G}$ is inherited from $\mathcal{M}$.

**Lemma 9.7.** *For games $\mathcal{M}$ and $\mathcal{G}$ as above, the following conditions are equivalent.*

1. *Eve wins from $m$ in $\mathcal{M}$.*

2. *Eve wins from $(m, \beta)$ in $\mathcal{G}$ for some branch $\beta$.*

3. *Eve wins from $(m, \beta)$ in $\mathcal{G}$ for all branches $\beta$.*

Let us return to 2-token games on $[0, K]$ automata, which when represented explicitly, is a Muller game. The colours of this game are represented by $C = [0, K] \times [0, K] \times [0, K]$, where the first component represent the priorities that occur along the transitions of Eve's token, while the second and the third component represent the priorities that occur along the transitions of Adam's two tokens. The winning condition of the 2-token game can then be given by $(C, \mathcal{F})$ where $\mathcal{F}$ consists of subsets $S$ of $C$ in which the following conditions holds: if the least priority occurring amongst the second component in elements of $S$ is even or if the least priority occurring amongst the third component in elements of $S$ is even, then the least priority occurring amongst the first component in elements of $S$ is even as well.

We will denote the Zielonka tree of the condition $(C, \mathcal{F})$ above as $\mathcal{Z}_{[0,K]}$, and the corresponding automaton recognising the $(C, \mathcal{F})$ Muller condition as $\mathcal{D}_{[0,K]}$. We will use $\mathcal{D}_{[0,K]}$ to construct parity games that have the same winner as the 2-token games on $[0, K]$ automata, obtained by taking the product of $\mathcal{Z}_{[0,K]}$ with the 2-token game, which itself is a game on a product of three copies of automata.

**Definition 9.8.** *For a $[0, K]$ automaton $\mathcal{A} = (Q, \Sigma, q_0, \Delta)$, we define the parity game $\mathcal{G}_2(\mathcal{A}) = (V, E)$ as follows.*

- *The set of vertices $V$ consists of the set $V = V_1 \cup V_2 \cup V_3$, where:*

  1. *$V_1 = \{(q, p_1, p_2, \beta) \mid q, p_1, p_2$ are weakly coreachable states and $\beta$ is a branch in $\mathcal{Z}_{[0,K]}$, or equivalently, a state in $\mathcal{D}_{[0,K]}\}$*
  2. *$V_2 = \{(q, a, p_1, p_2, \beta) \mid (q, p_1, p_2, \beta) \in V_1, a \in \Sigma\}$;*
  3. *$V_3 = \{(\delta, p_1, p_2, a, \beta) \mid (q, a, p_1, p_2, \beta) \in V_2$ and $\delta = q \xrightarrow{a:c_1} q' \in \Delta\}$.*

  *Eve's vertices are $V_\exists = V_2$, while Adam's vertices are $V_\forall = V_1 \cup V_3$*

- *The set of edges $E$ is the union of following sets:*

  1. *$E_1 = \{(q, p_1, p_2, \beta) \to (q, a, p_1, p_2, \beta) \mid a \in \Sigma\}$ (Adam chooses a letter)*
  2. *$E_2 = \{(q, a, p_1, p_2, \beta) \to (\delta, p_1, p_2, a, \beta) \mid q \xrightarrow{a:c_1} q' \in \Delta\}$ (Eve chooses a transition on her token)*
  3. *$E_3 = \{(\delta, p_1, p_2, a, \beta) \xrightarrow{d} (q', p'_1, p'_2, \beta') \mid \delta = q \xrightarrow{a:c_1} q', p_1 \xrightarrow{a:c_2} p'_1, p_2 \xrightarrow{a:c_3} p'_2$ are transitions in $\Delta$ and $\beta \xrightarrow{(c_1, c_2, c_3):d} \beta'$ is a transition in $\mathcal{D}_{[0,K]}\}$ (Adam chooses transitions on his tokens)*

- *The priority of edges is as follows. All elements in $E_1$ and $E_2$ are assigned priority $h + 1$, where $h$ is the height of the Zielonka tree $\mathcal{Z}_{[0,K]}$. The edges in $E_3$ as above are assigned priorities according to the transition from the current branch in $\mathcal{D}$ on the colour of the three transitions on Eve's token and Adam's two tokens.*

The correctness of the conversion from Muller to parity games via Zielonka trees (Lemma 9.7) implies that for every $[0, K]$ automaton $\mathcal{A}$ and states $q, p, r$ that are weakly coreachable, Eve wins the 2-token game on $\mathcal{A}$ from $(q; p, r)$, if and only if, Eve wins $\mathcal{G}_2(\mathcal{A})$ from $(q, p, r, \beta)$ for some branch $\beta$, if and only if, Eve wins $\mathcal{G}_2(\mathcal{A})$ from $(q, p, r, \beta)$ for all branches $\beta$. In particular, Eve wins the 2-token game on $\mathcal{A}$ from everywhere if and only if Eve wins $\mathcal{G}_2(\mathcal{A})$ from all vertices.

**Lemma 9.9.** *For every $[0, K]$ automaton $\mathcal{A}$, Eve wins the 2-token game on $\mathcal{A}$ from everywhere if and only if Eve wins $\mathcal{G}_2(\mathcal{A})$ from all vertices.*

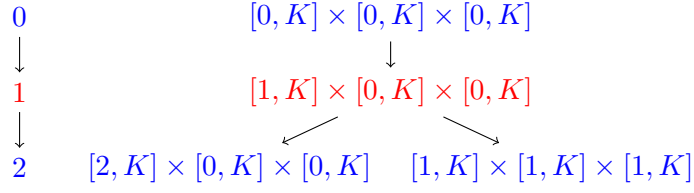

Figure 11: The first three layers of the Zielonka tree $\mathcal{Z}_{[0,K]}$

For our purposes, we only require to understand the top three layers of $\mathcal{Z}_{[0,K]}$, which is as shown in the figure below.

We say that a branch $\beta$ in $\mathcal{Z}_{[0,K]}$ is a *right-branch* if $\beta$ contains the node labelled $[1, K] \times [1, K] \times [1, K]$. Otherwise, we say that $\beta$ is a *left-branch* if $\beta$ contains the node labelled $[2, K] \times [0, K] \times [0, K]$. Note that every branch in $\mathcal{Z}_{[0,K]}$ is either a right-branch or a left-branch, and right-branches and left-branches do not overlap. The following observation follows from the construction of $\mathcal{D}_{[0,K]}$ and $\mathcal{Z}_{[0,K]}$.

**Proposition 9.10.** *Let* $\beta \xrightarrow{(c_e, c_{a1}, c_{a2}):d} \beta'$ *be a transition in* $\mathcal{D}_{[0,K]}$. *Then the following statements are true.*

1. *If the priority $d$ is at least 2 then $\beta$ and $\beta'$ are either both right branches, or are both left branches.*

2. *Let $\beta$ be a left branch. Then $d$ is 0 if and only if $c_e$ is 0, and $d$ is 1 if and only if $c_e$ is 1.*

3. *Let $\beta$ be a right branch. Then $d$ is 0 if and only if $c_e$ is 0, and $d$ is 1 if and only if at least one of $c_{a1}$ and $c_{a2}$ is 0.*

For the rest of this section, let us fix a $[0, K]$ automaton $\mathcal{A}$. Recall the concept of ranks for parity games, which we we described in Section 8.2, that measure how many 1 edges Adam can force Eve to see before a 0 edge. Specifically, note that if $\text{rank}(q, p, r, \beta) = 0$ for some vertex $(p, q, r, \beta)$ in $\mathcal{G}_2(\mathcal{A})$, then Eve's optimal winning strategy $\sigma$ ensures that a 0 priority edge occurs before the first occurrence of a 1 priority edge in any play of $\mathcal{G}_2(\mathcal{A})$ that starts at $(q, p, r, \beta)$ (Lemma 8.6). Thus, the following result is true due to Proposition 9.10 and the fact that the first component in the elements of $C = [0, K] \times [0, K] \times [0, K]$ corresponds to the priorities of Eve's token, while the second and third components correspond to the priorities of Adam's first and second tokens, respectively.

**Proposition 3.13.** *Suppose* $\text{rank}(q, p, r, \beta) = 0$ *in* $\mathcal{G}_2(\mathcal{A})$ *for some weakly coreachable states $q, p, r$ in $\mathcal{A}$ and a branch $\beta$ of the Zielonka tree. Fix $\sigma$ to be an optimal winning strategy for Eve in* $\mathcal{G}_2(\mathcal{A})$. *Then the following two statements are true.*

1. *If $\beta$ is a left-branch, then the strategy $\sigma$ from $\mathcal{G}_2(q, p, r, \beta)$ in $\mathcal{A}$ ensures that a priority 0 transition occurs before the first priority 1 transition in the run of Eve's token.*

2. *If $\beta$ is a right-branch, then the strategy $\sigma$ from $\mathcal{G}_2(q, p, r, \beta)$ in $\mathcal{A}$ ensures that a 0-priority transition in Eve's token occurs earlier or in the same round as the first occurrence of a 0-priority transition in any of Adam's tokens.*

A consequence of the observations in Proposition 3.13 above is that if $\text{rank}(q, p, r, \beta) = 0$ for some right-branch $\beta$, then Eve wins $G2(q; p, r)$ in $\mathcal{A}_{>0}$. Indeed the optimal winning strategy for Eve in $\mathcal{G}_2(\mathcal{A})$ from $(q, p, r, \beta)$ ensures that a 0-transition occurs in Eve's token no later than the first occurrence of a 0-transition in any of Adam's tokens, and Eve can use this strategy to play $G2(q; p, r)$ in $\mathcal{A}_{>0}$ keeping the branch of the corresponding vertices in $\mathcal{G}_2(\mathcal{A})$ in her memory to reach the accepting sink state in no later round than Adam. Otherwise if the run on Eve's token does not reach the accepting sink state in $\mathcal{A}_{>0}$, then neither of Adam's token reach the accepting sink state, and then the run on Eve's token in $\mathcal{A}_{>0}$ is accepting if any of the runs on Adam's token in $\mathcal{A}_{>0}$ is accepting since $\sigma$ is a winning strategy.

**Lemma 9.11.** *If $rank(q, p, r, \beta) = 0$ for weakly coreachable states $q, p,$ and $r$ and some right-branch $\beta$, then Eve wins $G2(q; p, r)$ in $\mathcal{A}_{>0}$.*

As for Büchi automata, we define the optimal-rank of a state $q$ as

$$\mathrm{opt}(q) = \min\{\mathrm{rank}(q, p, r, \beta) \mid q, p, r \in \mathsf{CR}^*(\mathcal{A}) \text{ and } \beta \text{ is a branch in } \mathcal{Z}_{0,K}\}.$$

If $\mathrm{opt}(q) = 0$ and there is a right-branch $\beta$ such that $\mathrm{rank}(q, p, r, \beta) = 0$, then we say that $q$ is a *right state*. We define the *non-right states* as states $q$ for which $\mathrm{opt}(q) = 0$, but that are not right states.

Inspired by Lemma 9.11, the objective of the normalisation procedure we present next is to make all states right states (while preserving simulation equivalence). Once this is the case, then we use Lemma 9.11 with Lemma 5.1 to conclude that the automaton obtained after normalisation has 0-reach covering (Lemma 9.18). Combining this with Lemma 3.12, we will deduce that $\mathcal{A}$ is HD.

### 9.3 Overview of the normalisation procedure

For the rest of this section, let us fix a $[0, K]$ automaton $\mathcal{A}$ on which Eve wins the 2-token game. By Theorem 3.1, we assume without loss of generality, that Eve wins the 2-token game from everywhere on $\mathcal{A}$.

The objective of the normalisation procedure on $\mathcal{A}$ we present below is to ensure that all states in the automaton are right states. In each step of the normalisation procedure, we will ensure that the automaton we construct are simulation equivalent to $\mathcal{A}$, and Eve wins the 2-token game from everywhere. Thus, if the automaton we get at the end of the procedure is HD, so is the automaton $\mathcal{A}$ that we started with (Lemma 2.8).

**The normalisation procedure.** Set $\mathcal{A}_0 = \mathcal{A}$. For each $i \geq 0$, we do the following three steps on $\mathcal{A}_i$ till $\mathcal{A}_i = \mathcal{A}_{i+1}$ via the following three subprocedures.

1. *Rank-reduction.* We modify the automaton $\mathcal{A}_i$ into a simulation equivalent automaton $\mathcal{B}_i$ so that $\mathrm{opt}(q) = 0$ for each state $q$ in $\mathcal{G}_2(\mathcal{B}_i)$. This is done similarly to the rank-reduction procedure for Büchi automata (Section 8.2), and we will elaborate upon the details of this subprocedure later.

2. *Branch-separation.* For the automaton $\mathcal{B}_i$, we remove the following transitions to obtain $\mathcal{C}_i$.

   (a) Transitions $q \xrightarrow{a:c} q'$ from a right state $q$ to a non-right state $q'$ in which the priority $c$ is at least 1.

   (b) Transitions of priority 1 that are outgoing from non-right states.

3. *Priority-modification.* The branch separation step might change the fact that all states have their optimal rank as 0. We nevertheless use the right (resp. non-right) states of $\mathcal{C}_i$ to refer to the states that were originally right (resp. non-right) in $\mathcal{B}_i$. In automaton $\mathcal{C}_i$, for any transition $p \xrightarrow{a:c} p'$ where $p$ is an non-right state, if the priority $c$ is not 0 or 1, then decrease the priority of that transition by 2. We let $\mathcal{A}_{i+1}$ be the automaton thus obtained.

Each of the above subprocedures, as we will detail on them in Sections 9.4 and 9.5, only involve either removal of certain transitions or decreasing the priorities of certain transitions. Thus, the normalisation procedure eventually stabilises to an automaton $\mathcal{A}_I$. We will show that the automaton $\mathcal{A}_I$ has 0-reach covering in Section 9.6. Additionally, we will argue that $\mathcal{A}_I$ is simulation-equivalent to $\mathcal{A}$ and Eve wins the 2-token game from everywhere on $\mathcal{A}_I$ by showing that each of the above subprocedures preserve the following invariants.

I1 Simulation-equivalence to the original automaton.

I2 Eve winning the 2-token game from everywhere.

## 9.4 Rank-reduction procedure

Our rank-reduction step is iterative in itself, and is similar to the rank-reduction procedure we presented in Section 8. Starting with a $[0, K]$ automaton $\mathcal{B}$ on which Eve wins the 2-token game from everywhere and is simulation-equivalent to $\mathcal{A}$, we iteratively modify $\mathcal{B}$ to obtain a $[0, K]$ automaton $\mathcal{B}_N$ in which the optimal-ranks of all states are 0.

Set $\mathcal{B}_0 = \mathcal{B}$. For each $i \geq 0$, we perform the following three steps on $\mathcal{B}_i$ until $\mathcal{B}_{i+1} = \mathcal{B}_i$.

Step 1 For each state state $q$ in $\mathcal{B}_i$, compute the optimal rank of $q$ in $\mathcal{G}_2(\mathcal{B}_i)$, which we denote by $\mathrm{opt}_i(q)$.

Step 2 Obtain $\mathcal{B}_i'$ from $\mathcal{B}_i$ by removing all transitions $q \xrightarrow{a:c} q'$ with $\mathrm{opt}_i(q) < \mathrm{opt}_i(q')$ and $c > 0$.

Step 3 Obtain $\mathcal{B}_{i+1}$ from $\mathcal{B}_i'$, by changing priorities of transitions $q \xrightarrow{a:c} q'$ with $c > 0$ and $\mathrm{opt}_i(q) > \mathrm{opt}_i(q')$ to $c = 0$.

We will argue in Lemmas 9.14 and 9.15 that the invariants I1 and I2 are preserved through Steps 2 and 3 respectively. Assuming that these invariants hold, observe that in Steps 2 and 3, we are either removing transitions of priority greater than 0, or changing the priority of some transitions with priority greater than 0 to be 0. Thus, this procedure terminates after at most $|\Delta|$ many iterations. Let $\mathcal{B}_N$ be the automaton obtained after this procedure terminates. We argue that in $\mathcal{B}_N$, all opt values of states $q$, denoted $\mathrm{opt}_N(q)$, are all 0. The proof is nearly identical to that of Lemma 8.9.

**Lemma 9.12.** *For all states $q$ in $\mathcal{B}_N$, we have that $\mathit{opt}_N(q) = 0$.*

*Proof.* Assume, towards a contradiction, that there is a state $q$ such that

$$\mathrm{opt}_N(q) = \mathrm{rank}_N(c_0) > 0$$

for some $c_0 = (q, p_1, p_2, \beta) \in \mathcal{G}_2(\mathcal{B}_N)$, where $\mathrm{rank}_N$ is the function that assigns vertices in $\mathcal{G}_2(\mathcal{B}_N)$ their ranks in $\mathcal{G}_2(\mathcal{B}_N)$. We will show that the rank-reduction procedure then can run for at least one more iteration on $\mathcal{B}_N$, which contradicts our assumption that the rank-reduction procedure has stabilised on $\mathcal{B}_N$.

Fix an optimal winning strategy $\sigma$ for Eve in $\mathcal{G}_2(\mathcal{B}_N)$. Since $\mathrm{rank}_N(c_0) > 0$, there is a finite play $\rho$ of $\mathcal{G}_2(\mathcal{B}_N)$ in which Eve is playing according to $\sigma$ and that contains a 1 priority edge but no 0 priority edge. Since a 0 priority edge in $\mathcal{G}_2(\mathcal{B}_N)$ corresponds to a 0-transition on Eve's token, it follows that Eve's token only takes transitions of priority at least 1 in $\rho$.

Since $\mathrm{rank}_N$ does not increase across edges of priority at least 1 and strictly decreases when an edge of priority 1 is seen (Proposition 8.7), we get that $\mathrm{rank}_N$ strictly decreases across $\rho$. Then there must be a transition on Eve's token in $\rho$ across which the quantity $\mathrm{opt}_N$ decreases as well, since we started with $\mathrm{opt}_N(q) = \mathrm{rank}_N(c_0)$. This implies that there must be a transition in $\mathcal{B}_N$ that has priority greater than 0, and across which the quantity $\mathrm{opt}_N$ strictly decreases. But since Step 3 of the rank-reduction procedure changes the priority of such transitions to 0, we get a contradiction to the fact that the rank-reduction procedure has stabilised. □

We now shift our focus towards proving the invariants for Steps 2 and 3. Step 2 removes certain transitions from the automaton. To show that the invariants are preserved during Step 2, we will use the following result, which will also be useful later in Section 9.5 for Branch-separation procedure.

**Lemma 9.13.** *Let $\mathcal{P}$ be a nondeterministic parity automaton, and $\mathcal{P}'$ a subautomaton of $\mathcal{P}$ such that for every three weakly-coreachable states $q, p, r$ in $\mathcal{P}$, Eve wins $G2((\mathcal{P}', q); (\mathcal{P}, p), (\mathcal{P}, r))$. Then the following statements hold.*

1. *$\mathcal{P}$ and $\mathcal{P}'$ are simulation equivalent.*

2. *Eve wins the 2-token game from everywhere in $\mathcal{P}'$.*

*Proof.* Proof of (1). Note that since $\mathcal{P}'$ is a subautomaton of $\mathcal{P}$, $\mathcal{P}$ simulates $\mathcal{P}'$: Eve can win the simulation game of $\mathcal{P}'$ by $\mathcal{P}$ by simply copying the transitions of Adam's token in $\mathcal{P}'$ in her token in $\mathcal{P}$. For the other direction, note that Eve wins $G1(\mathcal{P}'; \mathcal{P})$, and hence $\mathcal{P}'$ simulates $\mathcal{P}$ (Lemma 5.1). Thus $\mathcal{P}$ and $\mathcal{P}'$ are simulation-equivalent.

Proof of (2). Note that if states $q, p, r$ are weakly coreachable in $\mathcal{P}'$, then they are also weakly coreachable in $\mathcal{P}$. Therefore Eve wins $G2((\mathcal{P}', q); (\mathcal{P}, p), (\mathcal{P}, r))$, and since $\mathcal{P}'$ is a subautomaton of $\mathcal{P}$, Eve wins $G2(q; p, r)$ in $\mathcal{P}'$. It follows that Eve wins the 2-token game from everywhere in $\mathcal{P}'$. □

**Lemma 9.14** (Invariants for Step 2). *If Eve wins the 2-token game from everywhere on $\mathcal{B}_i$ and $\mathcal{B}_i$ is simulation-equivalent to $\mathcal{A}$, then the same holds for $\mathcal{B}_i'$.*

*Proof.* Fix $\sigma_{G2}$ to be an optimal positional strategy for Eve in the parity game $\mathcal{G}_2(\mathcal{B}_i)$, and fix $\sigma_{G1}$ to be a winning positional strategy for Eve in $G1$ from all states $q, p$ that are weakly coreachable in $\mathcal{A}$. Consider the strategy $\sigma_{G2}'$ in $\mathcal{G}_2(\mathcal{B}_i)$ that takes a 0 priority transition on Eve's token whenever Adam's letter $a$ is such that there is an outgoing transition on $a$ with priority 0 from Eve's token, and otherwise follows $\sigma_{G2}$. Note that $\sigma_{G2}'$ is a winning strategy since Eve wins $G2$ from everywhere, and is an optimal positional strategy since 0 priority transitions on Eve's token corresponds to 0 priority edges in $\mathcal{G}_2(\mathcal{B}_i)$. Thus we assume, without loss of generality, that the strategy $\sigma_{G2}$ takes a 0-transition on Eve's token whenever possible.

We will describe a winning strategy $\sigma'$ for Eve in $G2((\mathcal{B}_i', q); (\mathcal{B}_i, p^1), (\mathcal{B}_i, p^2))$, for all triplets of states $q, p^1, p^2$ are weakly coreachable in $\mathcal{A}_i$. It will then follow from Lemma 9.13 that the invariants I1 and I2 are preserved during Step 2.

At a high level, the strategy $\sigma'$ will require Eve to store as memory two additional tokens in $\mathcal{B}_i$, and a branch $\beta$ of $\mathcal{Z}_{[0,K]}$—or equivalently a state in $\mathcal{D}_{[0,K]}$. Eve's two memory tokens will each choose transitions according to $\sigma_{G1}$ by playing the 1-token game against the two Adam's tokens. Eve's token will select transitions according to $\sigma_{G2}$ by playing in $\mathcal{G}_2(\mathcal{B}_i)$ against her memory tokens and memory branch, till she takes a 0-transition or the transition given by $\sigma_{G2}$ had been deleted, in which case she *resets* her memory.

We will describe the strategy $\sigma'$ for Eve inductively, as rounds of $G2((\mathcal{B}_i', q); (\mathcal{B}_i, p^1), (\mathcal{B}_i, p^2))$ proceed along.

At the start of the 2-token game, let $q_0 = q, p_0^1 = p^1, p_0^2 = p_2$. We let Eve's memory tokens be at states $r_0^1$ and $r_0^2$ and her memory branch be at $\beta_0$ such that

$$\mathrm{opt}(q_0) = \mathrm{rank}(q_0, r_0^1, r_0^2, \beta_0).$$

Throughout the play, we will preserve the invariant that the states of Eve's token, Eve's memory tokens, and Adam's tokens are all weakly coreachable in $\mathcal{B}_i$.

After $j$ rounds in the 2-token game for some $j \geq 0$, suppose that the 2-token game is at the position $(q_j, p_j^1, p_j^2)$; i.e., Eve's token is at $q_j$, and Adam's tokens are at $p_j^1$ and $p_j^2$. Suppose that Eve's memory is $(r_j^1, r_j^2, \beta_j)$. Eve then plays as follows. Adam chooses the letter $a_j$, and let $\delta_j = q_j \xrightarrow{a_j : c_j} q'_j$ be the transition given by $\sigma_{G2}$ from $(q_j, a_j, r_j^1, r_j^2, \beta_j)$. We distinguish between the following three cases.

**Case 1.** $\delta_j$ is a 0-transition.

Then Eve takes this transition on her token, thus setting $q_{j+1}$ to be $q'_j$, and she updates her memory to be $(r_{j+1}^1, r_{j+1}^2, \beta_{j+1})$ so that

$$\mathrm{rank}(q_{j+1}, r_{j+1}^1, r_{j+1}^2, \beta_{j+1}) = \mathrm{opt}(q_{j+1}).$$

**Case 2.** If $\delta_j$ is a transition of priority at least 1 that hasn't been removed in $\mathcal{B}'_i$.

Then Eve moves her token to $q_{j+1}$. She updates her memory tokens at $r_{j+1}^1, r_{j+1}^2$ to take the transitions given by her strategy $\sigma_{G1}$ in $G1(\mathcal{B}_i)$ against Adam's tokens at $p_j^1, p_j^2$, respectively. The branch $\beta_{j+1}$ is updated according to the automata $\mathcal{D}_{[0,K]}$, based on the priorities that Eve's token and Eve's memory tokens take in the 2-token game of Eve's token against Eve's memory tokens.

**Case 3.** If $\delta_j$ is no longer a transition in $\mathcal{B}'_i$.

Eve then finds states $s_j^1, s_j^2 \in \mathsf{CR}^*(\mathcal{B}_i, q_j)$ and a branch $\beta'_j$ so that

$$\mathrm{opt}(q_j) = \mathrm{rank}(q_j, s_j^1, s_j^2, \beta'_j).$$

She picks the transition given by $\sigma_{G2}$ from $(q_j, a, s_j^1, s_j^2, \beta'_j)$. Eve's memory tokens are then reset to $s_j^1$ and $s_j^2$, and she then takes the $a$-transitions on her memory tokens given by her strategy $\sigma_{G1}$ in the 1-token game against Adam's tokens at $p_j^1$ and $p_j^2$, respectively. The branch $\beta_{j+1}$ is obtained by the unique outgoing transition in $\mathcal{D}_{[0,K]}$ from $\beta'_j$, on the priorities of the transitions that Eve's token and Eve's memory tokens took (from $s_j^1$ and $s_j^2$).

This concludes the description of Eve's strategy $\sigma'$. Let $\rho$ be a play for Eve in the game $G2((\mathcal{B}'_i, q); (\mathcal{B}_i, p^1), (\mathcal{B}_i, p^2))$, where Eve is playing according to $\sigma'$. We will show that $\rho$ is a winning play for Eve.

Firstly, suppose that eventually only moves from Case 2 occur in $\rho$. Let $w$ be the word that Adam plays in $\rho$, and suppose that the run of his token $t_j$ on $w$ is accepting, for some $j = 1$ or 2. Then, since Eve's memory tokens are eventually not reset, and the suffix of the moves on the Eve's memory token $j$ constitute a run on some suffix of $w$ in $\mathcal{A}_i$, which is accepting since $\sigma_{G1}$ is a winning strategy. Because Eve's token is taking transitions according to a winning strategy in the 2-token game against her memory tokens, the run of Eve's token then is accepting as well. Thus, Eve wins $\rho$ in this case.

If moves from Case 1 are taken infinitely often in $\rho$, then the run on Eve's token contains infinitely many 0-transitions and it is accepting, implying Eve wins $\rho$.

We will therefore conclude from the next claim that $\rho$ is a winning play for Eve, which will finish our proof of Lemma 9.14.

**Claim 3.** *In the play $\rho$, either moves from Case 1 are taken infinitely often, or eventually only moves from Case 2 are taken.*

In order to prove the claim, it suffices to prove that at most $m$ many moves from Case 3 can be taken before a move from Case 1 is taken, where $m$ is the number of vertices in $\mathcal{G}_2(\mathcal{B}_i)$. Note that the ranks of vertices in $\mathcal{G}_2(\mathcal{B}_i)$ are bounded by $m$. We will show that the rank of the configuration between the state of Eve's token, the state of Eve's memory tokens, and Eve's memory branch strictly decreases whenever a move from Case 3 is taken; it is clear from Proposition 8.7 that a move from Case 2 does not increase the rank.

Suppose at the start of round $j$ in $\rho$, Eve's token is initially at $q_j$ and her memory is $(r_j^1, r_j^2, \beta_j)$, and Adam chooses a letter $a_j$ such that the strategy $\sigma'$ takes a move from Case 3.

Suppose that the new Eve's state and memory then is $q_{j+1}$ and $(r_{j+1}^1, r_{j+1}^2, \beta_{j+1})$, respectively. Let $\delta_j = q_j \xrightarrow{a_j} q_j'$ be the transition given by $\sigma_{G2}$ from $(q_j, r_j^1, r_j^2, \beta_j)$. Since $\delta_j$ is a deleted transition, we note that $\mathrm{opt}(q_j) < \mathrm{opt}(q_j')$. We therefore have the following series of inequalities

$$\begin{aligned}
\mathrm{rank}(q_j, r_j^1, r_j^2, \beta_j) &\geq \mathrm{opt}(q_j') \\
&> \mathrm{opt}(q_j) \\
&\geq \mathrm{rank}(q_{j+1}, r_{j+1}^1, r_{j+1}^2, \beta_{j+1}).
\end{aligned}$$

Here, the first inequality holds due to monotonicity of ranks (Proposition 8.7). For the third inequality, note that the transition $\delta_j = q_j \xrightarrow{a_j} q_{j+1}$ cannot have priority 0 since otherwise $\sigma_{G2}$ would have picked a 0-transition, as we assumed in the start of this proof. Thus, the third inequality also follows from monotonicity of ranks (Proposition 8.7). Thus, the quantity $\mathrm{rank}(q_l, r_l^1, r_l^2, \beta_l)$ strictly decreases from round $l$ to round $(l+1)$ whenever a move from Case 3 is taken, and is non-increasing on moves from Case 2 due to Proposition 8.7. Since the ranks of vertices in $\mathcal{G}_2(\mathcal{B}_i)$ are bounded by $m$, we conclude the proof of the claim and hence of Lemma 9.14. $\qquad\square$

Showing that invariants are preserved during Step 3 is easier.

**Lemma 9.15** (Invariants for Step 3)**.** *If Eve wins the 2-token game from everywhere on $\mathcal{B}_i'$ and $\mathcal{B}_i'$ is simulation-equivalent to $\mathcal{A}$, then the same holds for $\mathcal{B}_{i+1}$.*

*Proof.* Recall that $\mathcal{B}_{i+1}$ is obtained by relabelling priorities of certain transitions in $\mathcal{B}_i'$. Hence, it suffices to show that a run is accepting in $\mathcal{B}_{i+1}$ if and only if that same run is accepting in $\mathcal{B}_i'$.

One direction is clear: if a run in $\mathcal{B}_i'$ is accepting, then the same run must be accepting in $\mathcal{B}_{i+1}$, since we only changed the priority of certain transitions to 0.

For the other direction, suppose that $\rho$ is an accepting run in $\mathcal{B}_{i+1}$, and consider the same run $\rho$ in $\mathcal{B}_i'$.

If $\rho$ in $\mathcal{B}_{i+1}$ is such that it contains finitely many transitions of priority 0 in $\mathcal{B}_{i+1}$, then because $\rho$ in $\mathcal{B}_{i+1}$ is accepting, it eventually only contains transitions of priority at least 2. Then note that the set of priorities visited infinitely often in $\rho$ in $\mathcal{B}_i'$ is the same as that in $\mathcal{B}_{i+1}$, and hence, $\rho$ is also accepting in $\mathcal{B}_{i+1}$.

Otherwise, suppose that $\rho$ contains infinitely many priority 0 transitions in $\mathcal{B}_{i+1}$. Note that due to Step 2, the $\mathrm{opt}_i$ values are non-increasing across transitions with priority greater than 0, and, due to Step 3, for every transition $\delta = q \xrightarrow{a:c} q'$ that has priority 0 in $\mathcal{B}_{i+1}$ but not in $\mathcal{B}_i'$, the $\mathrm{opt}_i$ value strictly decreases, i.e., $\mathrm{opt}_i(q) > \mathrm{opt}_i(q')$. Since $\mathrm{opt}_i$ is bounded by the size of the arena of $\mathcal{G}_2(\mathcal{B}_i)$, it follows that $\rho$ must also have infinitely many 0-transitions in $\mathcal{B}_i'$, and therefore is accepting. $\qquad\square$

We have thus shown that the first subprocedure of an iteration of the normalisation step, the rank-reduction subprocedure, preserves the invariants I1 and I2. By Lemma 9.12, we conclude that in the automaton obtained after the rank-reduction procedure, all states have their optimal rank 0.

## 9.5 Branch-separation and Priority-reduction.

We next prove that the subprocedures of branch-separation and priority-reduction involved in the normalisation procedure also preserve the invariants I1 and I2.

**Branch-separation.** Let $\mathcal{B}$ be a $[0, K]$ automaton such that $\mathrm{opt}(q) = 0$ for all states $q$, and on which Eve wins the 2-token game from everywhere and is simulation equivalent to $\mathcal{A}$. The branch-separation procedure on automaton $\mathcal{B}$ removes the following transitions from $\mathcal{B}$, to obtain $\mathcal{C}$.

1. Transitions of priority at least 1 from right states to non-right states.

2. Transitions of priority 1 that are outgoing from non-right states.

Recall that we call a state $q$ as a right state if there is a right-branch $\beta$ and states $p, r \in \mathsf{CR}^*(\mathcal{B}, q)$ such that $\mathrm{rank}(q, p, r, \beta) = 0$, and *non-right states* as states $q$ for which $\mathrm{opt}(q) = 0$, but that are not right states. Since $\mathrm{opt}(q) = 0$ for all states $q$ in $\mathcal{B}$, every state is either a right state or a non-right state.

We next show that the invariants I1 and I2 are preserved during the branch-separation procedure.

**Lemma 9.16.** *The branch separation procedure preserves the invariants I1 and I2.*

*Proof.* Let $\mathcal{B}$ be a $[0, K]$ automaton on which Eve wins the 2-token game from everywhere, and in which all states have optimal rank 0. Let $\mathcal{C}$ be the $[0, K]$ automaton obtained by the branch-separation procedure on $\mathcal{B}$. We will show that for all states $q, p, r$ that are weakly coreachable in $\mathcal{B}$, Eve wins $G2((\mathcal{C}, q); (\mathcal{B}, p), (\mathcal{B}, r))$. Due to Lemma 9.13, we conclude the proof of Lemma 9.16.

Fix an optimal strategy $\sigma_{G2}$ for Eve in $\mathcal{G}_2(\mathcal{B})$ from all triplets of weakly coreachable states (Lemma 8.6), as well as a winning positional strategy $\sigma_{G1}$ for Eve on the 1-token game from all pairs of weakly coreachable states (Lemma 2.5). We will describe a winning strategy $\sigma$ for Eve in $G2((\mathcal{C}, q); (\mathcal{B}, p), (\mathcal{B}, r))$ using $\sigma_{G2}$ and $\sigma_{G1}$.

We describe $\sigma$ inductively, as rounds of $G2$ proceed along. Suppose that after $i$ rounds, the game is at the position $((\mathcal{C}, q_i); (\mathcal{B}, p_i), (\mathcal{B}, r_i))$. Eve will store, in her memory, a branch $\beta_i$ of $\mathcal{Z}_{[0, K]}$ and two additional tokens that are at states $s_i$ and $t_i$ such that $s_i, t_i \in \mathsf{CR}^*(\mathcal{B}, q)$ and $\mathrm{rank}(q_i, s_i, t_i, \beta_i) = 0$ in $\mathcal{G}_2(\mathcal{B})$. Additionally, the branch $\beta_i$ is a right (resp. left) branch whenever $q_i$ is a right (resp. non-right) state. Since $\mathrm{opt}(q_i) = 0$ in $\mathcal{B}$, we know that such states $s_i, t_i$ and $\beta_i$ exist.

Informally, similar to the proof of Lemma 9.14, Eve's token will play the 2-token game against her memory tokens, and her memory tokens will play the 1-token game against Adam tokens.

When Adam picks a letter $a_i$, let $\delta = q_i \xrightarrow{a_i : c_i} q'$ be the transition that is given by $\sigma_{G2}$ at the position $(q_i, a_i, s_i, t_i, \beta_i)$ in $\mathcal{G}_2(\mathcal{B})$. We distinguish between the following four cases.

**Case** 1. The priority of $\delta = q_i \xrightarrow{a_i : c_i} q'$ is not 0, and $q$ is a right state.

We first observe that $q'$ must be a right state too due to the structure of $\mathcal{D}_{[0, K]}$ and $\sigma_{G2}$ being an optimal strategy. Indeed, from the position $v = (\delta, s_i, t_i, \beta_i)$ in $\mathcal{G}_2(\mathcal{A})$ that is an Adam vertex, every outgoing edge from $v$ to $(q', s', t', \beta')$ must be such that $\mathrm{rank}(q', s', t', \beta') = 0$ (Proposition 8.7), and note that if $\beta$ is a right branch, and $\beta \xrightarrow{(c_i, c', c'') : d} \beta'$ is the corresponding transition in $\mathcal{D}$, then $\beta'$ is a left branch only if $d$ is 0 or 1 (Proposition 9.10). Note that $d$ cannot be 1 because $\sigma_{G2}$ is an optimal strategy, and $d$ is 0 if and only if $c_i$ is 0, which we assumed not to be the case. Thus, $q'$ is a right state, and $\beta'$ is a right branch.

It follows the transition $\delta$ has not been deleted in $\mathcal{C}$. Eve thus picks the transition $\delta$ on her token, and she updates her memory tokens at $s_i$ and $t_i$ to take the transitions on $a_i$ given by the strategy $\sigma_{G1}$ against Adam's tokens at $p_i$ and $r_i$, respectively. The memory branch $\beta_i$ is updated to $\beta_{i+1}$ in $\mathcal{D}_{[0, K]}$, according to the priorities of the transitions that Eve's tokens and Eve's memory tokens took in the 2-token game of Eve's token against Eve's memory tokens. Note that $\beta_{i+1}$ is a right branch, due to our argument in the previous paragraph.

**Case** 2. The priority of $\delta = q_i \xrightarrow{a_i : c_i} q'$ is not 0, and both $q_i$ and $q'$ are non-right states.

If $q_i$ is a non-right state, then $\delta$ cannot have priority 1 since $\sigma_{G2}$ is a rank-optimal strategy (Proposition 3.13), and thus, $\delta$ is a transition in $\mathcal{C}$. We let Eve move her token along the transition $\delta$. She updates her memory tokens at $s_i$ and $t_i$ to take transitions given by $\sigma_{G1}$ against Adam's tokens at $q_i$ and $r_i$ respectively. The memory $\beta_i$ is updated to $\beta_{i+1}$ in $\mathcal{D}_{[0,K]}$, according to the priorities of the transitions that Eve's tokens and Eve's memory tokens took in the $\mathcal{G}_2$ game between Eve's token against Eve's memory tokens.

Observe that $\beta_{i+1}$ is a left branch. The corresponding transition $\delta_D$ from $\beta_i$ to $\beta_{i+1}$ does not have priority 0 since $\delta$ does not have priority 0, and since we argued that $\delta$ does not have priority 1, the transition $\delta_D$ does not have priority 1 too (Proposition 9.10). Since transitions of priority at least 2 from left states lead to left states (Proposition 9.10) in $\mathcal{D}_{[0,K]}$, we get that $\beta_{i+1}$ is a left branch.

**Case 3.** The priority of $\delta = q_i \xrightarrow{a_i : c_i} q'$ is not 0, $q_i$ is a non-right state, and $q'$ is a right state. Since $\sigma_{G2}$ is a rank-optimal strategy, $\delta$ does not have priority 1 (Proposition 3.13). Eve thus takes the transition $\delta$ on her token to $q'$. Eve then *resets* her memory tokens to be at the state $s_{i+1}, t_{i+1}$ that are weakly coreachable to $q'$ in $\mathcal{B}$, and her memory branch to be a right branch $\beta_{i+1}$ such that $\mathrm{rank}(q', s_{i+1}, t_{i+1}, \beta_{i+1}) = 0$ in $\mathcal{G}_2(\mathcal{B})$.

**Case 4.** The priority of $\delta = q_i \xrightarrow{a_i : c_i} q'$ is 0.
Then Eve takes the transition $\delta$ on her token. Eve then *resets* her memory tokens to be at $s_{i+1}, t_{i+1} \in \mathrm{CR}^*(\mathcal{B}, q')$ and her memory branch to be $\beta_{i+1}$ such that $\mathrm{rank}(q', s_{i+1}, t_{i+1}, \beta_{i+1}) = 0$ in $\mathcal{G}_2(\mathcal{B})$ and $\beta_{i+1}$ is a right (resp. left) branch if $q'$ is a right (resp. non-right) state.

Note that at the end of the round, in each of the above four cases, the configuration of the state of Eve's token, the states of Eve's memory token, and her memory branch has rank 0 in $\mathcal{G}_2(\mathcal{B})$. For moves from Case 1 and 2, this follows from the monotonicity of ranks (Proposition 8.7), and for moves from Cases 3 and 4, this is true because that is how we reset Eve's memory. Additionally, observe that if Eve's token is at a right (resp. non-right) state, then her memory branch is a right (resp. left) branch. Thus, Eve can continue playing similarly in the next round, and this concludes our inductive description of Eve's strategy.

We claim that the above strategy is winning for Eve. Indeed, if her token visits both right states and non-right states infinitely often, then since the only transitions from right states to non-right states have priority 0, Eve's run on her token contains infinitely many priority 0 transitions and hence is accepting.

Otherwise, Eve's token eventually stays only in right states, or only in non-right states. If this is the case, then eventually, Eve's memory tokens are never reset. Then eventually, Eve is choosing transitions on her token according to a $G2$ winning strategy against her memory tokens, which in turn are playing according to a $G1$ winning strategy against Adam's tokens. Thus, if the run of some Adam's token is accepting, then some suffix of the moves on the corresponding Eve's memory token constitutes an accepting run, and therefore, the run on Eve's token is accepting as well. It follows that we have described a winning strategy for Eve, as desired. $\qquad\square$

We now proceed to show that the invariants **I1** and **I2** are preserved during the priority-reduction subprocedure.

**Priority-reduction.** We note that the removal of transitions from $\mathcal{B}$ to $\mathcal{C}$ in the branch separation procedure described above might change the fact that all states had their optimal rank as 0 in $\mathcal{B}$. We still use the terminology of right states and non-right states to denote the states in $\mathcal{C}$ that were originally right states and non-right states in $\mathcal{B}$, respectively.

Due to branch-separation, we know that there are no transitions of priority at least 1 from right states to non-right states. We do the following priority relabelling on $\mathcal{C}$ to obtain $\mathcal{A}'$: for every transition $\delta = p \xrightarrow{a:c} p'$ in $\mathcal{C}$ where $p$ is a non-right state and $c$ has priority greater than 0, change the priority of $\delta$ to be $c-2$. Observe that $c$ cannot be 1, since we removed outgoing transitions of priority 1 from non-right states in the branch-separation procedure. We claim that relabelling priorities this way does not change the acceptance of each run.

Indeed, let $\rho$ be an infinite run. If $\rho$ eventually only stays in right states or only in non-right states, then it is clear that $\rho$ is accepting in $\mathcal{C}$ if and only if $\rho$ is accepting in $\mathcal{A}'$. Otherwise, suppose $\rho$ visits both right and non-right states infinitely often. Then $\rho$ is accepting in $\mathcal{C}$, since every transition from a right state to non-right state must have priority 0. For the same reasoning, $\rho$ is also accepting in $\mathcal{A}'$. Thus, we conclude that a run in $\mathcal{A}'$ is accepting if and only if that same run is accepting in $\mathcal{C}$. We obtain that $\mathcal{A}'$ and $\mathcal{C}$ are simulation equivalent, and Eve wins the 2-token game from everywhere in $\mathcal{A}'$.

We have thus shown that the subprocedures of our normalisation procedure: rank-reduction, branch-separation, and priority-reduction each preserve the invariants I1 and I2.

## 9.6 Stabilisation

Observe that each of the three subprocedures of rank-reduction, branch separation, and priority-reduction either decrease the priorities of some transitions, or remove certain transitions. Thus, the normalisation procedure terminates after at most $(K+1) \times |\Delta|$ many iterations on $\mathcal{A}$, where $\Delta$ is the set of transitions of $\mathcal{A}$. Suppose that the automaton then obtained is $\mathcal{A}^I$. Since we have shown that each of the subprocedures of the normalisation procedure preserves the invariants I1 and I2, so does the entire normalisation procedure, and thus $\mathcal{A}^I$ is simulation-equivalent to $\mathcal{A}$ and Eve wins the 2-token game from everywhere in $\mathcal{A}^I$. We will now show that $\mathcal{A}^I$ has 0-reach covering (Lemma 9.18), and therefore it is HD (Lemma 3.12). We start by showing that all states in $\mathcal{A}^I$ are right states.

**Lemma 9.17.** *Every state in the automaton $\mathcal{A}_I$ has optimal rank 0 and is a right state.*

*Proof.* If the optimal rank of $q$ is not 0 for some state $q$ in $\mathcal{A}^I$, then we can run the normalisation procedure for one more step to obtain a different automaton, since the rank-reduction procedure would then modify the automaton so that all states have optimal ranks 0. But since the normalisation procedure terminates at $\mathcal{A}^I$, we deduce that the optimal ranks of all states in $\mathcal{A}^I$ is 0.

We next show that each state $q$ is a right state in $\mathcal{A}^I$. Assume, towards a contradiction, that $q$ is a non-right state for some state $q$ in $\mathcal{A}^I$. If $q$ had an outgoing transition of priority at least 1, then either the branch-separation or the priority-modification step would have changed the automaton $\mathcal{A}^I$. Thus, all transitions outgoing from $q$ have priority 0. But then it is clear that $q$ is a right-state, which is a contradiction. $\square$

We now show that the automaton $\mathcal{A}^I$ has 0-reach covering.

**Lemma 9.18.** *For each state $q$ in $\mathcal{A}^I$, there is a state $p$ weakly coreachable to $q$ in $\mathcal{A}^I$ such that Eve wins $G2(q; p, p)$ in $(\mathcal{A}^I)_{>0}$.*

*Proof.* From Lemmas 9.11 and 9.17, we know that for each state $q$, there are weakly coreachable states $p_1, p_2$ such that Eve wins $G2(q; p_1, p_2)$ in $(\mathcal{A}^I)_{>0}$. We need to show that for each $q$, there is a $p$ weakly coreachable to $q$ such that Eve wins $G2(q; p, p)$ in $(\mathcal{A}^I)_{>0}$. We prove the existence of such a $p$ as follows. Construct a directed graph $G$ whose vertices are states of $\mathcal{A}^I$. For each state $q$, we add two edges $q \to p_1$ and $q \to p_2$ such that Eve wins $G2(q; p_1, p_2)$ in $(\mathcal{A}^I)_{>0}$, and $p_1$ and $p_2$ are weakly coreachable to $q$ in $\mathcal{A}^I$. We will then find a $p$ such that there are two distinct paths from $q$ to $p$ in $\mathcal{G}$, which would imply that Eve wins $G2(q; p, p)$ in $(\mathcal{A}^I)_{>0}$ due to the following claim.

**Claim 4.** *Suppose there are two distinct paths from $q$ to $p_1, p_2$. Eve then wins $G2(q; p_1, p_2)$ in $(\mathcal{A}^I)_{>0}$.*

We begin by recalling the following observations from Lemma 5.1, which we will use repeatedly to show the claim.

**Observation 0.** If Eve wins $G2(\mathcal{B}; \mathcal{C}_1, \mathcal{C}_2)$, then Eve wins $G2(\mathcal{B}; \mathcal{C}_2, \mathcal{C}_1)$.

**Observation 1.** If Eve wins $G2(\mathcal{B}; \mathcal{C}_1, \mathcal{C}_2)$, then Eve wins $G1(\mathcal{B}; \mathcal{C}_1)$ and $G1(\mathcal{B}; \mathcal{C}_2)$.

**Observation 2.** If Eve wins $G2(\mathcal{B}; \mathcal{C}_1, \mathcal{C}_2)$ and $G1(\mathcal{C}_1; \mathcal{C}_1')$ (resp. $G1(\mathcal{C}_2; \mathcal{C}_2')$), then Eve wins $G2(\mathcal{B}; \mathcal{C}_1', \mathcal{C}_2)$ (resp. $G2(\mathcal{B}; \mathcal{C}_1, \mathcal{C}_2')$).

**Observation 3.** If Eve wins $G1(\mathcal{B}; \mathcal{B}')$ and $G2(\mathcal{B}'; \mathcal{C}_1, \mathcal{C}_2)$, then Eve wins $G2(\mathcal{B}; \mathcal{C}_1, \mathcal{C}_2)$.

Let $P_1$ and $P_2$ be paths from $q$ to $p_1, p_2$ respectively, and let $P_0$ be the longest common prefix of $P_1$ and $P_2$. Let $r$ be the vertex at the end of $P_0$, and let $P_1', P_2'$ be paths such that $P_1 = P_0 P_1'$ and $P_2 = P_0 P_2'$. From Observation 1, it is clear that if there is an edge from $q_u$ to $q_v$, then Eve wins $G1(q_u; q_v)$ in $(\mathcal{A}^I)_{>0}$. By transitivity of $G1$ (Lemma 2.4), this extends to paths, i.e., for $q$ and $r$ as above,

$$\text{Eve wins } G1(q; r) \text{ in } (\mathcal{A}^I)_{>0}. \tag{5}$$

The paths $P_1'$ and $P_2'$ are both distinct and start at $r$. Let $r_1, r_2$ be states such that there are $P_1' = e_1 P_1''$ starts with the edge $e_1 = r \rightarrow r_1$ and $P_2' = e_2 P_2''$ starts with the edge $e_2 = r \rightarrow r_2$. Then

$$\text{Eve wins } G2(r; r_1, r_2) \text{ in } (\mathcal{A}^I)_{>0}. \tag{6}$$

The transitivity of $G1$ across paths $P_1''$ and $P_2''$ implies Eve wins $G1(r_1; p_1)$ and $G1(r_2; p_2)$ in $(\mathcal{A}^I)_{>0}$. Combining this with Eq. (6) using Observation 2, we get that Eve wins $G2(r; p_1, p_2)$ in $(\mathcal{A}^I)_{>0}$. This together with the fact that Eve wins $G1(q; r)$ in $(\mathcal{A}^I)_{>0}$ (Eq. (5)), we get using Observation 3 that Eve wins $G2(q; p_1, p_2)$ in $(\mathcal{A}^I)_{>0}$, as desired. This concludes the proof of our claim.

Due to Claim 4, it suffices to show that for each $q$ there is a $p$ such that there are two distinct paths from $q$ to $p$ in $G$. Consider the SCC-decomposition of $G$, and consider the SCCs of $G$ that do not have an outgoing edge to another SCC—call these SCCs *end SCCs*. For each state $p$ in an end SCC, there are edges from $p$ to $p_1$ and $p_2$, and there are paths from $p_1$ to $p$ and $p_2$ to $p$. Thus, each state in an end SCC has two distinct paths from itself to itself. Since every other state $q$ has a path to a vertex in an end SCC, we get the desired result. $\qquad\square$

We have thus proved Lemma 9.4.

**Lemma 9.4.** *Let $\mathcal{A}$ be a $[0, K]$ automaton on which Eve wins the 2-token game from everywhere. Then there is a $[0, K]$ automaton $\mathcal{A}^I$ such that $\mathcal{A}^I$ is simulation-equivalent to $\mathcal{A}$, Eve wins the 2-token game from everywhere on $\mathcal{A}^I$, and $\mathcal{A}^I$ has 0-reach covering.*

The proof of Theorem 3.11 now follows easily.

**Theorem 3.11.** *Let $K > 1$ be a natural number such that for every $[1, K]$ automaton $\mathcal{A}$, Eve wins the 2-token game on $\mathcal{A}$ if and only if $\mathcal{A}$ is HD. Then, for every $[0, K]$ automaton $\mathcal{A}$, Eve wins the 2-token game on $\mathcal{A}$ if and only if $\mathcal{A}$ is HD.*

*Proof.* We will show that if $\mathcal{A}$ is a $[0, K]$ automaton on which Eve wins the 2-token game, then $\mathcal{A}$ is HD under the assumption of Hypothesis 9.1 (that the 2-token game characterised history determinism of $[1, K]$ automata). By Theorem 3.1, we know that $\mathcal{A}$ has a simulation-equivalent subautomaton $\mathcal{B}$ on which Eve wins the 2-token game from everywhere. From Lemma 9.4, $\mathcal{B}$ has a simulation-equivalent $[0, K]$ automaton $\mathcal{B}^I$ on which Eve wins the 2-token game from everywhere and that has 0-reach covering. By Lemma 3.12, $\mathcal{B}^I$ is HD. Since $\mathcal{A}$ is simulation-equivalent to $\mathcal{B}^I$, it follows from Lemma 2.8 that $\mathcal{A}$ is HD as well. $\qquad\square$

Combining Corollary 3.3 and Theorems 3.5 and 3.11, we obtain 2-token theorem.

**The 2-Token Theorem.** *For every nondeterministic parity automaton $\mathcal{A}$, Eve wins the 2-token game on $\mathcal{A}$ if and only if $\mathcal{A}$ is history-deterministic. Thus, the problem of deciding history-determinism is in* **P** *for parity automata with a fixed number of priorities, and in* **PSPACE** *if the number of priorities is part of the input.*

**Acknowledgements.**

We are grateful to Udi Boker, Denis Kuperberg, Michał Skrzypczak and the anonymous reviewers of STOC for their feedback and suggestions. We are also grateful to Dmitry Chistikov, Marcin Jurdziński, K. S. Thejaswini, and Patrick Totzke for their feedback on the PhD thesis of the second author, many of which have also improved the quality of the current paper. The first author is funded by ANR QuaSy 23-CE48-0008-01.